\documentclass[fleqn,usenatbib]{mnras}

\usepackage{newtxtext,newtxmath}

\usepackage[T1]{fontenc}

\DeclareRobustCommand{\VAN}[3]{#2}
\let\VANthebibliography\thebibliography
\def\thebibliography{\DeclareRobustCommand{\VAN}[3]{##3}\VANthebibliography}

\usepackage{graphicx}	
\usepackage{graphbox}
\usepackage{float}
\usepackage{amsmath}	
\usepackage{bm}
\usepackage{caption}
\usepackage{subcaption}
\usepackage{enumerate}

\title[\texttt{Gmunu}: Paralleled, grid-adaptive, GRMHD code in curvilinear geometries]{\texttt{Gmunu}: Paralleled, grid-adaptive, general-relativistic magnetohydrodynamics in curvilinear geometries in dynamical spacetimes}

\author[Cheong et al.]{
Patrick Chi-Kit Cheong,$^{1}$\thanks{chi-kit.cheong@link.cuhk.edu.hk}
Alan Tsz-Lok Lam,$^{1,2}$\thanks{alantllam@link.cuhk.edu.hk}
Harry Ho-Yin Ng$^{1,3}$\thanks{hoyin.ng@ligo.org}
and Tjonnie Guang Feng Li$^{1,4,5}$\thanks{tgfli@cuhk.edu.hk}
\\
$^{1}$Department of Physics, The Chinese University of Hong Kong, Shatin, N. T., Hong Kong \\ 
$^{2}$Max Planck Institute for Gravitational Physics (Albert Einstein Institute), Am Mühlenberg 1, Postdam-Golm 14476, Germany \\
$^{3}$Institut f\"ur Theoretische Physik, Goethe Universit\"at, Max-von-Laue-Str. 1, 60438 Frankfurt am Main, Germany \\
$^{4}$Institute for Theoretical Physics, KU Leuven, Celestijnenlaan 200D, B-3001 Leuven, Belgium \\
$^{5}$Department of Electrical Engineering (ESAT), KU Leuven, Kasteelpark Arenberg 10, B-3001 Leuven, Belgium 
}

\date{Accepted XXX. Received YYY; in original form ZZZ}

\pubyear{2021}

\begin{document}
\label{firstpage}
\pagerange{\pageref{firstpage}--\pageref{lastpage}}
\maketitle

\begin{abstract}
	We present an update {on} the \texttt{G}eneral-relativistic \texttt{mu}ltigrid \texttt{nu}merical (\texttt{Gmunu}) code, a parallelised, multi-dimensional curvilinear, general relativistic magnetohydrodynamics code with an {efficient non-linear cell-centred multigrid elliptic solver, which is fully coupled with an efficient block-based adaptive mesh refinement module}.
	{To date, as described in this paper, } \texttt{Gmunu} is able to solve the elliptic metric equations in the conformally flat condition approximation with the multigrid approach and the equations of ideal general-relativistic magnetohydrodynamics by means of high-resolution shock-capturing finite-volume method with {reference metric formularised} multi-dimensionally in {Cartesian}, cylindrical or spherical geometries.
	To guarantee the absence of magnetic monopoles during the evolution, we have developed an elliptical divergence cleaning method by using {the} multigrid solver.
	In this paper, we present the methodology, full evolution equations and implementation details of \texttt{Gmunu} and its properties and performance in some benchmarking and challenging relativistic magnetohydrodynamics problems.
\end{abstract}




\section{Introduction}
{Many astrophysical scenarios involving neutron stars and black holes such as core-collapse supernovae, mergers of compact objects are the most important events in gravitational wave physics or multimessenger astrophysics.}
In order to have a better {understanding} of such detected events and gain our understanding of the physics at nuclear densities in the postmerger remnant of binary neutron mergers {(e.g. \citet{2015IJMPD..2430012R})} and neutron star-black hole mergers {(e.g. \citet{2017LRR....20....3M})}, accurate general relativistic (magneto-)hydrodynamic (GR(M)HD) simulations are essential.

Depending on the configuration and focus of the problems, the computational cost can be significantly reduced if some symmetries can be imposed {or simulating the problems in certain geometries, e.g.} core-collapse supernovae \citet{2007PhR...442...38J, 2013RvMP...85..245B}, mangetars \citet{2015RPPh...78k6901T,2015SSRv..191..315M,2017ARA&A..55..261K}, pulsars \citet{2005LRR.....8....7L}, compact binary merger remnants \citet{2011LRR....14....6S,2012LRR....15....8F,2017RPPh...80i6901B,2019RPPh...82a6902D,2020ARNPS..70...95R}, and self-gravitating accretion disks \citet{2013LRR....16....1A}.
While these problems can be simulated in three-dimensional {Cartesian} coordinate, these systems with approximate symmetries are better captured in spherical or cylindrical coordinates due to better angular momentum conservation. 
Furthermore, the dimensionality of the problems and the computational cost can be reduced significantly if any symmetry can be imposed.
{Moreover, the simulation code is required not only to be robust in the highly relativistic region but also be able to resolve different scales accurately since most of such astrophysical systems are usually highly relativistic, include multi-time scale and multi-length scale physics.}
For instance, in stellar core collapse problem, the length scale could vary from the pre-supernova stellar core (thousands of kilometres) and down to a small length scale such as the turbulence in the postbounce flow {(}on the order of meters), and a typical time step size is of the order $\mathcal{O}(10^{-6})$ s and {one needs} to evolve such systems up to $1 \sim 2$ s for the development of a full explosion or for black hole formation \citet{2009CQGra..26f3001O}.
Thus, to numerically model these systems accurately within a reasonable time and affordable computational resources, a multi-scale, multi-dimensional, fully parallelised, support different geometries general relativistic (magneto-)hydrodynamics code is desired.

{Several} GRMHD codes are developed recently \citet{2017ComAC...4....1P, 2019A&A...629A..61O, 2019ApJS..244...10R, 2019arXiv191210192L, 2020PhRvD.101j4007M, 2020CQGra..37m5010C}.
However, most of them are either designed for particular coordinates, or {does not allow for a dynamical evolution of spacetime}.
In our previous work \citet{2020CQGra..37n5015C}, we presented {an} axisymmetric general relativistic hydordynamics code \texttt{Gmunu} (\texttt{G}eneral-relativistic \texttt{mu}ltigrid \texttt{nu}merical solver) and {show} that cell-centred multigrid method is an efficient and robust approach of solving the elliptic metric equations in the conformally flat condition (CFC) approximation \citet{CoCoA, XCFC}.
{
However, the previous version of \texttt{Gmunu} has limitations such as it has no GRMHD solver, not parallelised, not grid adaptive, supports two-dimensional spherical coordinates only.
The aim of this work is to extend the capabilities of \texttt{Gmunu} to overcome these difficulties and enable us to apply it to more generic astrophysical problems.
}
The key updates of \texttt{Gmunu} are the following:
{
\begin{itemize} 
        \item one-, two- and three- dimensional {Cartesian}, cylindrical and spherical coordinates are supported;
	\item general relativistic magnetohydrodynamics (GRMHD) solver is implemented;
	\item multigrid based elliptic divergence cleaning is implemented {for magnetic fields divergenceles handling};
	\item fully parallelised with {Message Passing Interface (MPI)};
        \item block-based adaptive mesh refinement module is included.
\end{itemize}
}
The parallelization and the adaptive mesh refinement module of current \texttt{Gmunu} are provided by coupling with \texttt{MPI-AMRVAC} PDE toolkit \citet{2018ApJS..234...30X, 2020arXiv200403275K}, a Message Passing Interface (MPI) based parallelised toolkit with a block-based quadtree-octree (in 2D-3D) Adaptive Mesh Refinement (AMR) module.
In this paper, we present the methodology and the implementation details of the code and valid our code though some benchmarking tests.

The paper is organised as follows.
In section \ref{sec:formulations} we outline the formalism we used in this work.
The details of the numerical settings and, the methodology, implementation of our magneto-hydrodynamics solver and our multigrid solver are presented in respectively.
The code tests and results are presented in section \ref{sec:numerical_tests}.
This paper ends with a discussion section in section \ref{sec:conclusions}.
{Unless} explicitly stated, we work in geometrized {Heaviside}-Lorentz units, {for} which the speed of light $c$, gravitational constant $G$, solar mass $M_{\odot}$, vacuum permittivity $\epsilon_0$ and vacuum permeability $\mu_0$ are all {equal} to one ( $c=G=M_{\odot}=\epsilon_0 = \mu_0 = 1$ ).
Greek indices, running from 0 to 3, are used for 4-quantities while the Roman indices, running from 1 to 3, are used for 3-quantities.

\section{Formulation and Numerical methods}\label{sec:formulations} 
\subsection{GR(M)HD in the reference-metric formalism}\label{sec:GRMHD_equations}
We use the standard ADM (Arnowitt-Deser-Misner) 3+1 formalism \citet{2007gr.qc.....3035G,2008itnr.book.....A}. 
The metric can be written as
\begin{equation}
	ds^2 = g_{\mu\nu}dx^\mu dx^\nu = -\alpha^2  dt^2 + \gamma_{ij} \left( dx^i + \beta^i dt \right) \left( dx^j + \beta^j dt \right)
\end{equation}
where $\alpha$ is the lapse function, $\beta^i$ is the spacelike shift vector and $\gamma_{ij}$ is the spatial metric.
We adopt a conformal decomposition of the spatial metric $\gamma_{ij}$ with the conformal factor $\psi$:
\begin{equation}
\gamma_{ij} = \psi^4 \bar{\gamma}_{ij},
\end{equation}
where $\bar{\gamma}_{ij}$ is the conformally related metric.

The evolution equations for matter are derived from the local conservations of the rest-mass and energy-momentum and the homogeneous Faraday's law:
\begin{align}
	&\nabla_\mu \left( \rho u^\mu \right) = 0, \\
	&\nabla_\mu T^{\mu\nu} = 0, \\
	&\nabla_\mu {^*F^{\mu\nu}} = 0 \label{eq:maxwellseq}, 
\end{align}
where $\rho$ is the rest-mass density of the fluid, $u^\mu$ is the fluid four-velocity, $T^{\mu\nu}$ is the total energy-momentum tensor and $^*F^{\mu\nu}$ is the dual Faraday tensor.
From Faraday tensor, we define the magnetic field four-vector (the projection of the Faraday tensor parallel to the fluid four-velocity):
\begin{equation}
	b^{\mu} \equiv {^*F^{\mu\nu}} u_\nu.
\end{equation}
With $b^\mu$ and $u^\mu$, the {total} energy-momentum tensor can be expressed as:
\begin{equation}
	T^{\mu\nu} = \rho h^* u^{\mu} u^{\nu} + p^* g^{\mu\nu} - b^{\mu} b^{\nu},
\end{equation}
where we further define the square of the fluid frame magnetic field strength $b^2 \equiv b^\mu b_\mu = B^2 - E^2$, the magnetically modified specific enthalpy $ h^* \equiv 1 + \epsilon + (p + b^2) / \rho$ and the magnetically modified isotropic pressure $ p^* \equiv p + b^2/2 $.

The reference-metric formalism was originally presented in \citet{2014PhRvD..89h4043M} for GRHD and is recently extended to GRMHD \citet{2020PhRvD.101j4007M}.
By introducing a time-independent reference metric $\hat{\gamma}_{ij}$, the Valencia formulation can be generalized as the following form:
\begin{equation}\label{eq:GRMHD}
	\partial_t( \bm{q}) + \hat{\nabla}_i(\bm{f^i}) = \bm{s}, 
\end{equation}
\begin{equation}
	\bm{q} = \begin{bmatrix}
           q_D \\
           q_{S_j} \\
           q_{\tau} \\
           q_{{B}^j}
        \end{bmatrix}, 
	\bm{f^i} = 
        \begin{bmatrix}
           \left(f_D\right)^i \\
           \left(f_{S_j}\right)^i \\
           \left(f_\tau\right)^i \\
           \left(f_{B^j}\right)^i \\
        \end{bmatrix} ,
	\bm{s} = 
        \begin{bmatrix}
           s_D \\
           s_{S_j} \\
           s_{\tau} \\
           s_{{B}^j}
        \end{bmatrix}
	,
\end{equation}
where the $\hat{\nabla}_i$ here is the covariant derivatives associated with the \emph{time-independent} reference metric $\hat{\gamma}_{ij}$.
Here, $\bm{q}$ are the conserved quantities:
\begin{align}
	q_D &\equiv \psi^6 \sqrt{\bar{\gamma}/\hat{\gamma}} D =  \psi^6 \sqrt{\bar{\gamma}/\hat{\gamma}} \left[ \rho W  \right] ,\\
	q_{S_j} &\equiv  \psi^6 \sqrt{\bar{\gamma}/\hat{\gamma}} S_j =  \psi^6 \sqrt{\bar{\gamma}/\hat{\gamma}} \left[  \rho h^* W^2 v_j - \alpha b^0 b_j \right] ,\\
	q_{\tau} &\equiv  \psi^6 \sqrt{\bar{\gamma}/\hat{\gamma}} \tau =  \psi^6 \sqrt{\bar{\gamma}/\hat{\gamma}} \left[  \rho h^* W^2 - p^* - ( \alpha b^0 )^2 - D \right] ,\\
	q_{{B}^j} &\equiv \psi^6 \sqrt{\bar{\gamma}/\hat{\gamma}} B^j  ,
\end{align}
where $v^i = u^i/W + \beta^i/\alpha$ is the 3-velocity seen by an Eulerian observer at rest in current spatial 3-hypersurface, $W \equiv 1/\sqrt{1-v^i v_i}$ is the Lorentz factor.
{The} magnetic field in fluid's rest frame can be obtained by: 
\begin{align}
& b^0 = \frac{W B^k v_k}{\alpha}, \qquad b^i = \frac{B^i}{W} + b^0 \hat{v}^i, \\
& b^2 = \frac{B^iB_i}{W^2} + (B^k v_k)^2 ,
\end{align}
where $\hat{v}^i \equiv \left( \alpha v^i - \beta^i \right)$.
Note that $b_i = b^\mu g_{\mu i} = B_i / W + \alpha b^0 v_i$.
The corresponding fluxes $\bm{f}^i$ are given by:
\begin{align}
           \left(f_D\right)^i  &\equiv \psi^6 \sqrt{\bar{\gamma}/\hat{\gamma}} \left[  D \hat{v}^i \right] , \\
           \left(f_{S_j}\right)^i  &\equiv \psi^6 \sqrt{\bar{\gamma}/\hat{\gamma}} \left[ S_j \hat{v}^i  + \delta^i_j \alpha p^* - \alpha b_j B^i / W  \right] , \\
           \left(f_\tau\right)^i  &\equiv \psi^6 \sqrt{\bar{\gamma}/\hat{\gamma}}  \left[  \tau \hat{v}^i + \alpha p^* v^i - \alpha^2 b^0 B^i / W \right] , \\
           \left(f_{B^j}\right)^i  &\equiv \psi^6 \sqrt{\bar{\gamma}/\hat{\gamma}}  \left[ \hat{v}^i {B}^j -  \hat{v}^j {B}^i  \right].
\end{align}
Finally, the corresponding source terms $\bm{s}$ are given by:
\begin{align}
	s_D = &0 , \\
        s_{S_i} = &\alpha \psi^6 \sqrt{\bar{\gamma}/\hat{\gamma}} \Big\{ 
                  - T^{00} \alpha \partial_i \alpha + T^{0}_k \hat{\nabla}_i \beta^k \\ \nonumber
                  & + \frac{1}{2} \left( T^{00} \beta^j \beta^k + 2 T^{0j}\beta^k + T^{jk} \right) \hat{\nabla}_i \gamma_{jk} \Big\} , \\
        s_\tau = &\alpha \psi^6 \sqrt{\bar{\gamma}/\hat{\gamma}} \Big\{ T^{00} \left( K_{ij} \beta^i \beta^j - \beta^k \partial_k \alpha \right) \\ \nonumber
        	&+ T^{0j} \left( 2 K_{jk} \beta^k - \partial_j \alpha \right) + T^{ij}K_{ij} \Big\} , \\
	s_{B^i} = & 0 ,
\end{align}
where $K_{ij}$ is the extrinsic curvature.

In order to solve eq.~\eqref{eq:GRMHD} with {the} finite volume formulation, we further express the equations in the following form:
\begin{equation} \label{eq:GRMHD_ref}
	\partial_t \bm{q} + \frac{1}{\sqrt{\hat{\gamma}}}\partial_j\left[\sqrt{\hat{\gamma}} \bm{f}^j\right] = \bm{s} + \bm{s}_{\text{geom}} ,
\end{equation}
where $\bm{s}_{\text{geom}}$ are so-called geometrical source terms which contain the 3-Christoffel symbols $ \hat{\Gamma}^l_{ik} $ associated with the reference metric $\hat{\gamma}_{ij}$.
Explicitly, eq.~\eqref{eq:GRMHD_ref} can be expressed as:
\begin{align}
&\partial_t (q_D) + \frac{1}{\sqrt{\hat{\gamma}}}\partial_j\left[\sqrt{\hat{\gamma}} (f_D)^j\right] = 0 ,\\
&\partial_t (q_{S_i}) + \frac{1}{\sqrt{\hat{\gamma}}}\partial_j\left[\sqrt{\hat{\gamma}} (f_{S_i})^j\right] = s_{S_i} + \hat{\Gamma}^l_{ik}(f_{S_l})^k ,\\
&\partial_t (q_\tau) + \frac{1}{\sqrt{\hat{\gamma}}}\partial_j\left[\sqrt{\hat{\gamma}} (f_\tau)^j\right] = s_\tau ,\\
&\partial_t (q_{B^i}) + \frac{1}{\sqrt{\hat{\gamma}}}\partial_j\left[\sqrt{\hat{\gamma}} (f_{B^i})^j\right] = 0 \label{eq:induction}.
\end{align}
Note that the momentum conservation in this expression {are satisfied} to \emph{machine precision} rather than to the level of truncation error {because} the geometrical source terms $\bm{s}_{\text{geom}}$ are identically vanishing for the components associated with ignorable coordinates in the metric.
For example, in spherical {coordinates} $(r,\theta,\phi)$, since the coordinate $\phi$ does not explicitly enter into the metric, the corresponding geometrical source term vanish for the $q_{S_\phi}$ equations.
Physically, unlike in the expression in \citet{2014PhRvD..89h4043M, 2020PhRvD.101j4007M} where the angular momentum is conserved to the level of truncation error due to the explicit expression of the covariant derivatives, in our expression, the angular momentum conservation {is numerically satisfied} to machine precision since the corresponding geometrical source term is identically {equal} to zero.
{Similar implementations that minimize coordinate-dependent part of the code can be found in \citet{2003ApJ...589..444G} and in a recent work \citet{2019ApJS..241....7S} }

We then discretize the volume averages of eq.~\eqref{eq:GRMHD_ref}.
{Using} divergence theorem and some {algebra}, the discretized version of eq.~\eqref{eq:GRMHD_ref} in the cell $(i,j,k)$ can be expressed as
\begin{equation}
\begin{aligned}
	\frac{ d }{dt}\left<\bm{q}\right>_{\texttt{i,j,k}} = 
	& \frac{1}{\Delta V_{\texttt{i,j,k}}} \times \\
	& \Bigg\{ 
	\left[ \left( \left<\bm{f}\right>^1\Delta A^1\right)\Big|_{\texttt{i+1/2,j,k}} -  \left( \left<\bm{f}\right>^1\Delta A^1 \right) \Big|_{\texttt{i-1/2,j,k}} \right] \\
	& + \left[ \left( \left<\bm{f}\right>^2\Delta A^2\right)\Big|_{\texttt{i,j+1/2,k}} -  \left(\left<\bm{f}\right>^2\Delta A^2\right) \Big|_{\texttt{i,j-1/2,k}} \right]  \\
	& + \left[ \left( \left<\bm{f}\right>^3\Delta A^3\right)\Big|_{\texttt{i,j,k+1/2}} -  \left(\left<\bm{f}\right>^3\Delta A^3\right) \Big|_{\texttt{i,j,k-1/2}} \right] 
	\Bigg\} \\ 
	& + \left<\bm{s}\right>_{\texttt{i,j,k}} + \left<\bm{s}_{\text{geom}}\right>_{\texttt{i,j,k}}, 
\end{aligned}
\end{equation}
where the cell volume and volume-average are defined as
\begin{align}
&\Delta V \equiv \int_{\text{cell}} \sqrt{\hat{\gamma}} dx^1 dx^2 dx^3 ,\\
&\left<\bullet\right> \equiv \frac{1}{\Delta V} \int_{\text{cell}} \bullet \sqrt{\hat{\gamma}} dx^1 dx^2 dx^3 ,
\end{align}
while the surface area and surface-average is defined as 
\begin{align}
&\Delta A^i \equiv \int_{\text{surface}} \sqrt{\hat{\gamma}} dx^{j,j\neq i} ,\\
&\left<\bullet\right>^i \equiv \frac{1}{\Delta A^i} \int_{\text{surface}} \bullet^i \sqrt{\hat{\gamma}} dx^{j,j\neq i}.
\end{align}

Here we note that, as the reference metric $\hat{\gamma}_{ij}$ is \emph{time-independent}, the volume-averaged 3-Christoffel symbols $ \left< \hat{\Gamma}^l_{ik} \right>$ in the geometrical source terms, cell volume $\Delta V$ and surface area $\Delta A$ are fixed once the coordinate {system} is {chosen}.
For completeness, we included these quantities in both cylindrical and spherical coordinates in Appendix \ref{appendix:coordinates}.

\subsection{Divergenceless handling and elliptic divergence cleaning}\label{sec:elliptic_cleaning}
The time-component of eq.~\eqref{eq:maxwellseq} implies that the divergence of the magnetic field is zero, namely:
\begin{align}
	&\nabla \cdot \vec{B} \equiv \frac{1}{\sqrt{\gamma}} \partial_i \left( \sqrt{\gamma} B^i \right) = 0 \\ \nonumber
        \Rightarrow & \hat{\nabla}_i q_{B^i} = \frac{1}{\sqrt{\hat{\gamma}}} \partial_i \left( \sqrt{\hat{\gamma}} q_{B^i} \right) = 0.
\end{align}
In practice, this condition is not {satisfied} if we evolve the induction equation \eqref{eq:induction} directly without any treatment due to the accumulating numerical error.
As a result, non-vanishing monopoles are introduced and {the code returns} non-physical results.
Various treatments are introduced to enforce this constraint in (GR)MHD calculations.
The most common approaches recently are
(i) hyperbolic divergence cleaning through a generalized Lagrange multiplier (GLM) {(e.g. \citet{2017ComAC...4....1P})};
(ii) constrained transport (CT) scheme which updates the magnetic fields while controlling the divergence-free constraint to numerical round-off accuracy {(e.g. \citet{2017ComAC...4....1P, 2019A&A...629A..61O})};
and (iii) evolving the vector potentials directly and compute the magnetic {field} by taking the curl of the vector potential {(e.g. \citet{2020PhRvD.101j4007M})}.
Here, we adopt a different approach, the so-called \emph{elliptic} divergence cleaning, by solving Poisson's equation and enforce the magnetic field is divergence-free:
\begin{align}
        & \hat{\nabla}^2 \Phi = \hat{\nabla}_i q_{B^i}^{\text{old}} , \label{eq:poisson}\\
	& q_{B^i}^{\text{new}} = q_{B^i}^{\text{old}} - \left( \hat{\nabla}\Phi \right)^i \label{eq:update_bfield}.
\end{align}
The \texttt{BHAC} code \citet{2017ComAC...4....1P}, elliptic divergence cleaning is available to be used only for the magnetic fields initialization \citet{2019CoPhC.24506866T}.

In the current implementation of \texttt{Gmunu} with elliptic divergence cleaning, the magnetic field is defined at cell centres.
Whenever the conserved magnetic field $q_{B^i}$ is updated at each timestep, we first solve Poisson's equation in eq.~\eqref{eq:poisson} through the multigrid solver (see section \ref{sec:MG_solver}), then we update the magnetic field with the solution $\Phi$ as shown in eq.~\eqref{eq:update_bfield}.

In addition to the elliptic cleaning mentioned above, generalized Lagrange multiplier (GLM), constrained transport (CT) and the vector potential schemes {are planned for} \texttt{Gmunu}.
The implementations and comparisons of these divergence-free treatments will be presented in future work.
Here, we will only focus on the elliptic divergence cleaning approach as our main divergence-free treatment for evolution.

\subsection{\label{sec:characteristic_speed}Characteristic speed}
In relativistic magnetohydrodynamics, one has to solve a quartic equation if we wish to obtain the exact form of the characteristic wave speeds $\lambda_{\pm}$ (e.g.~\citet{1990rfmf.book.....A}).
To reduce the computational cost and complexity of the implementation, instead of obtaining the exact characteristic speeds, we follow the approach presented in \citet{2003ApJ...589..444G}.
In this approach, the upper bound $a$ for the fast wave speed is
\begin{equation}
	a^2 = c^2_s + c^2_a - c^2_s c^2_a,
\end{equation}
where $c_s$ is the sound speed and $c_a$ is the Alfven speed which can be obtained by
\begin{equation}
	c^2_a = \frac{b^2}{\rho h + b^2} = \frac{b^2}{\rho h^*}.
\end{equation}
The characteristic velocities can then be calculated by
\begin{align}
	&\lambda^i_{\pm} = \alpha \bar{\lambda}^i_{\pm} - \beta^i, \\
	&\bar{\lambda}^i_{\pm} = \frac{ \left( 1 - a^2 \right) v^i \pm \sqrt{a^2 \left( 1 - v^2 \right) \left[ \left( 1 - v^2 a^2 \right) \gamma^{ii} - \left( 1 - a^2 \right)\left( v^i \right)^2 \right]} }{\left( 1 - v^2 a^2 \right)}.
\end{align}

\subsection{\label{sec:PPlimiter}Positivity Preserving Limiter}
Positivity preserving limiter was originally introduced in \citet{HU2013169} for Newtonian hydrodynamics and was successfully applied on GR(M)HD \citet{Radice_2014, 2017ComAC...4....1P}.
Here we will discuss the basic concept of positivity preserving limiter and its implementation in \texttt{Gmunu}.
For simplicity, let us consider the evolution equation of conserved density in one-dimensional case:
\begin{equation}\label{eq:1D}
\partial_t (u) + \frac{1}{\sqrt{\hat{\gamma}}}\partial_1\left[\sqrt{\hat{\gamma}} (f(u))\right] = 0 .
\end{equation}
Note that if the positivity of $u$ is guaranteed over one first-order Euler timestep, then the positivity is also guaranteed for any strong-stability preserving Runge-Kutta (SSPRK) time integrator since the time integrator is always constructed as a convex combination of Euler steps.
So, we discretized eq.~\eqref{eq:1D} as the following form:
\begin{equation}
\begin{aligned}
&\frac{ u^{\texttt{n+1}}_{\texttt{i}} - u^{\texttt{n}}_{\texttt{i}} }{\Delta t} = - \frac{1}{\Delta V_{\texttt{i}}} \Big\{ f_{\texttt{i+1/2}} \Delta A_{\texttt{i+1/2}}  - f_{\texttt{i-1/2}} \Delta A_{\texttt{i-1/2}} \Big\} \\ 
\Rightarrow & u^{\texttt{n+1}}_{\texttt{i}} = \frac{1}{2}\left[ u^-_{\texttt{i}} + u^+_{\texttt{i}}\right],
\end{aligned}
\end{equation}
where
\begin{equation}
\begin{aligned}
u^-_{\texttt{i}} &\equiv \left( u^{\texttt{n}}_{\texttt{i}} - 2 \frac{\Delta t}{\Delta V_{\texttt{i}}}f_{\texttt{i+1/2}} \Delta A_{\texttt{i+1/2}}\right) , \\
u^+_{\texttt{i}} &\equiv \left( u^{\texttt{n}}_{\texttt{i}} + 2 \frac{\Delta t}{\Delta V_{\texttt{i}}}f_{\texttt{i-1/2}} \Delta A_{\texttt{i-1/2}}\right) .
\end{aligned}
\end{equation}
To ensure both $u^+_{\texttt{i}}$ and $u^-_{\texttt{i}}$ are positive, we modify the flux as
\begin{equation}\label{eq:pplimiter_flux}
f_{\texttt{i+1/2}} = \theta f^{\text{HO}}_{\texttt{i+1/2}} + (1-\theta)f^{\text{LF}}_{\texttt{i+1/2}},
\end{equation}
where $f^{\text{HO}}_{\texttt{i+1/2}}$ is the high-order flux of the original scheme while $f^{\text{LF}}_{\texttt{i+1/2}}$ is the first order Lax-Friedrichs flux {and} $\theta \in [0,1]$ is the maximum value such that both $u^+_{\texttt{i+1}}$ and $u^-_{\texttt{i}}$ are positive.
Since the Lax-Friedrichs scheme is positivity preserving, it is always possible to choose some $\theta$ such that positivity is guaranteed.
In \texttt{Gmunu}, we implemented this limiter to preserve the positivity of conserved density $D$ and energy density $\tau$, to preserve the positivity of density {$\rho$} and pressure {$p$}.
For multi-dimensional cases, we apply the limiter {component by component}.

\subsubsection{Implementation of positivity preserving limiter}
After calculating $f^{\text{HO}}_{\texttt{i+1/2}}$ and $f^{\text{LF}}_{\texttt{i+1/2}}$, we check if the {the relationship} $f_{\texttt{i+1/2}} = f^{\text{HO}}_{\texttt{i+1/2}}$ {needs} to be modified with a small value $\epsilon$ (which is set as $10^{-16}$ in \texttt{Gmunu}), i.e., we check if the following relations hold:
\begin{equation}
\begin{aligned}\label{eq:pplimiter_check}
u^-_{\texttt{i}} &= \left( u^{\texttt{n}}_{\texttt{i}} - 2 \frac{\Delta t}{\Delta V_{\texttt{i}}}f_{\texttt{i+1/2}} \Delta A_{\texttt{i+1/2}}\right) > \epsilon\\
u^+_{\texttt{i}} &= \left( u^{\texttt{n}}_{\texttt{i}} + 2 \frac{\Delta t}{\Delta V_{\texttt{i}}}f_{\texttt{i-1/2}} \Delta A_{\texttt{i-1/2}}\right) > \epsilon.
\end{aligned}
\end{equation}
If relations above do not hold, we then work out $\theta$s by substituting \eqref{eq:pplimiter_flux} into \eqref{eq:pplimiter_check}:
\begin{equation}
\begin{aligned}
\theta^-_{\texttt{i}} &= \frac{ \frac{\Delta V_{\texttt{i}}}{2\Delta t}\left( u^{\texttt{n}}_{\texttt{i}} - \epsilon \right) - f^{\text{LF}}_{\texttt{i+1/2}}\Delta A_{\texttt{i+1/2}}}{f^{\text{HO}}_{\texttt{i+1/2}}\Delta A_{\texttt{i+1/2}} - f^{\text{LF}}_{\texttt{i+1/2}}\Delta A_{\texttt{i+1/2}}} ,\\
\theta^+_{\texttt{i}} &= \frac{ \frac{\Delta V_{\texttt{i}}}{2\Delta t}\left( u^{\texttt{n}}_{\texttt{i}} - \epsilon \right) + f^{\text{LF}}_{\texttt{i-1/2}}\Delta A_{\texttt{i-1/2}}}{f^{\text{LF}}_{\texttt{i-1/2}}\Delta A_{\texttt{i-1/2}} - f^{\text{HO}}_{\texttt{i-1/2}}\Delta A_{\texttt{i-1/2}}} , \\
\theta_{\texttt{i}} &= \min\left(\theta^+_{\texttt{i}},\theta^-_{\texttt{i}}\right) , \\
\theta_{\texttt{i}} &= \min\left(\max\left(\theta_{\texttt{i}},0\right),1\right).
\end{aligned}
\end{equation}
After obtaining $\theta$s for both continuity and energy equations, we pick the maximum one and substitute the resulting $\theta$ into \eqref{eq:pplimiter_flux} to modify the \emph{all} the flux terms at that particular grid point.

\subsection{Conserved to Primitive variables conversion}
Recovery of the primitive variables $(\rho, Wv^i, p)$ from the conservative variables $(D, S_i, \tau)$ in GR(M)HD is non-trivial, one has to solve non-linear equations numerically.
Most of the root-finding methods used in GR(M)HD simulations are Newton-Raphson method which works fine with analytic equations of state.
However, it might return inaccurate results with tabulated equations of state {because} it requires the partial derivatives $\partial \hat{p}/\partial \rho$ and $\partial \hat{p}/\partial \epsilon$.
In \texttt{Gmunu}, two conserved to primitive variables conversions which do not require derivatives are implemented for GRHD and GRMHD respectively.
For the GRHD cases, the implementation basically follows the formulation presented in Appendix C in \citet{2013PhRvD..88f4009G} while for the GRMHD cases we mainly follow a recent work \citet{2021PhRvD.103b3018K}.
Although GRHD can be reduced from GRMHD by letting all magnetic field $B^i = 0$ and the recovery of primitive variables method presented in \citet{2021PhRvD.103b3018K} actually works well for vanishing magnetic fields, for different applications and development {purposes} {(e.g. for the systems which have no magnetic fields, it is better to use the GRHD module to lower the computational cost)}, we implemented two separate modules called \texttt{grhd} and \texttt{grmhd} correspondingly.
For completeness, we included the implementation details of both GRHD and GRMHD here.

\subsubsection{ {Implementation of} recovery of primitive variables in GRHD }
\begin{enumerate}[{Step} 1:]
	\item Calculate the rescaled variables and {the following} useful relations which are fixed during the iterations.
		\begin{align}
			& S \equiv \sqrt{S^i S_i}, \\
			r \equiv \frac{S}{D}, \qquad & q \equiv \frac{\tau}{D}, \qquad k \equiv \frac{S}{\tau + D}, 
		\end{align}
	\item Determine the bounds of the root {$z_{-}$ and $z_{+}$}, where
		\begin{equation}
			z_{-} \equiv \frac{k / 2}{\sqrt{1-k^{2} / 4}}, \qquad z_{+} \equiv \frac{k}{\sqrt{1-k^{2}}}
		\end{equation}
	\item In the inverval $\left[ z_{-}, z_{+} \right]$, solve:
		\begin{equation}\label{eq:f_z}
			f (z) = z - \frac{r}{\hat{h}(z)},
		\end{equation}
		where
		\begin{align}
			\hat{h} (z) & = (1+\hat{\epsilon}(z))(1+\hat{a}(z)), \\
			\hat{p} (z) & = p\left(\hat{\rho} (z),\hat{\epsilon} (z)\right) , \qquad \hat{a} (z) = \frac{\hat{p} (z)}{\hat{\rho} (z) \left( 1 + \hat{\epsilon} (z) \right) } ,\\
			\hat{\rho} (z) & = \frac{D}{\hat{W}(z)}, \\
			\hat{\epsilon} (z) & = \hat{W}(z)q - zr + \frac{z^2}{1+\hat{W}(z)}, \\
			\hat{W} (z) & = \sqrt{1+z^2} .
		\end{align}
		In \texttt{Gmunu}, we numerically solve eq.~\eqref{eq:f_z} with {the} Illinois algorithm \citet{dowell1971modified}, which is an improved version of Regula-Falsi method.
		Note that during the iterations, we enforce {that} the density $\rho$ and the specific energy $\epsilon$ {fall within} the validity region of the EOS, i.e., 
		we {evaluate} the updated $\rho$ and $\epsilon$ with $\hat{\rho} = \max \left( \min \left( \rho_{\max} ,\hat{\rho} \right), \rho_{\min} \right)$ and $\hat{\epsilon} = \max \left( \min \left( \epsilon_{\max}(\hat{\rho}) ,\hat{\epsilon} \right), \epsilon_{\min}(\hat{\rho}) \right)$.
	\item With the root $z_0$ of eq.~\eqref{eq:f_z}, we can then work out the primitive variables $[\rho, \epsilon, p]$ respectively with the equations used in step 3.
		{The} velocity $v^i$ can be obtained with $z$ by:
		\begin{equation}
			\hat{v}^i (z) = \frac{ S^i / D }{ \hat{h}(z)\hat{W}(z)}.
		\end{equation}
\end{enumerate}

\subsubsection{ {Implementation of} recovery of primitive variables in GRMHD }
\begin{enumerate}[{Step} 1:]
	\item Calculate the rescaled variables and the {following} useful relations which are fixed during the iterations.
		\begin{align}
			q \equiv \frac{\tau}{D}, \qquad r_i \equiv \frac{S_i}{D}, \qquad \mathcal{B}^i \equiv \frac{B^i}{\sqrt{D}}, 
		\end{align}
		and then calculate
		\begin{align}
			{	r^2 \equiv r^i r_i, \quad \mathcal{B}^2 \equiv \mathcal{B}^i\mathcal{B}_i \quad \text{and} \quad \mathcal{B}^2r^2_\perp \equiv \mathcal{B}^2 r^2 - (r^l\mathcal{B}_l)^2.}
		\end{align}
	\item In the interval $\left( 0, h_0^{-1} \right]$, solve:
		\begin{equation}\label{eq:f_mu_plus}
			f_a (\mu) = \mu \sqrt{ h_0^2 + \bar{r}^2(\mu)} - 1,
		\end{equation}
		where $h_0$ is the relativistic enthalpy lower bound over {the} entire validity region of the EOS and 
		\begin{align}
			&\bar{r}^2 (\mu) = r^2 \chi^2(\mu) + \mu \chi(\mu) \left( 1 + \chi(\mu) \right) \left( r^l \mathcal{B}_l \right)^2\label{eq:r_bar_sqr}, \\
			&\chi (\mu) = \frac{1}{1 + \mu \mathcal{B}^2}\label{eq:chi} .
		\end{align}
		Here the root of $f_a$ {in eq.\eqref{eq:f_mu_plus}} is denoted as $\mu_+$.
		{Since $f_a$ is smooth and its derivative can be expressed analytically, we numerically solve eq.~\eqref{eq:f_mu_plus} with Newton-Raphson method, which is usually more efficient than bracketing methods.}
		In case the Newton-Raphson method {fails} to converge, we use the Illinois algorithm to solve this equation.
	\item In the interval $\left( 0, \mu_+ \right]$, solve:
		\begin{equation}\label{eq:f_mu}
			f (\mu) = \mu - \frac{1}{\hat{\nu} + \mu \bar{r}^2(\mu)},
		\end{equation}
		where
		\begin{align}
			&\hat{\nu} (\mu) = \max(\nu_A (\mu),\nu_B (\mu)), \\
			&\nu_A (\mu) = \left( 1 + \hat{a} (\mu) \right) \frac{1 + \hat{\epsilon} (\mu)}{\hat{W} (\mu)}, \\
			&\nu_B (\mu) = \left( 1 + \hat{a} (\mu) \right) \left( 1 + \bar{q} (\mu) - \mu \bar{r}^2 (\mu) \right) \\
			&\hat{p} (\mu) = p\left(\hat{\rho} (\mu),\hat{\epsilon} (\mu)\right) , \qquad \hat{a} (\mu) = \frac{\hat{p} (\mu)}{\hat{\rho} (\mu) \left( 1 + \hat{\epsilon} (\mu) \right) } ,\\
			&\hat{\rho} (\mu) = \frac{D}{\hat{W} (\mu)}, \quad \hat{\epsilon} (\mu) = \hat{W} (\mu) \left( \bar{q} (\mu) - \mu \bar{r}^2 (\mu) \right) + \hat{v}^2 (\mu) \frac{\hat{W}^2 (\mu)}{1 + \hat{W} (\mu)} ,\\
			&\hat{v}^2 (\mu) = \min( \mu^2 \bar{r}^2 (\mu), v_0^2 ), \quad \hat{W} (\mu) = \frac{1}{\sqrt{1-\hat{v}^2 (\mu)}}, \\
			&\bar{q} (\mu) = q - \frac{1}{2} \mathcal{B}^2 - \frac{1}{2} \mu^2\chi^2 (\mu) \left( \mathcal{B}^2 r_\perp^2 \right),
		\end{align}
		$\bar{r}^2 (\mu)$ and $\chi(\mu)$ {are} defined {in} eq.~\eqref{eq:r_bar_sqr} and eq.~\eqref{eq:chi}, and the upper velocity limit square $v_0^2$ is defined as $v_0^2 \equiv r^2 / (h_0^2 + r^2) < 1$.
		In \texttt{Gmunu}, we numerically solve eq.~\eqref{eq:f_mu} with the Illinois {algorithm}.
		Note that during the iterations, we enforce the density $\rho$ and the specific energy $\epsilon$ {fall within} the validity region of the EOS, i.e., 
		we {evaluate} the updated $\rho$ and $\epsilon$ with $\hat{\rho} = \max \left( \min \left( \rho_{\max} ,\hat{\rho} \right), \rho_{\min} \right)$ and $\hat{\epsilon} = \max \left( \min \left( \epsilon_{\max}(\hat{\rho}) ,\hat{\epsilon} \right), \epsilon_{\min}(\hat{\rho}) \right)$.
	\item With the root $\mu$ of eq.~\eqref{eq:f_mu}, we can then work out the primitive variables $[\rho, \epsilon, p]$ respectively with the equations used in step 3.
		{The velocity $v^i$ can be obtained with $\mu$ by}:
		\begin{equation}
			\hat{v}^i (\mu) = \mu \chi (\mu) \left( r^i + \mu \left( r^l \mathcal{B}_l\right) \mathcal{B}^i \right).
		\end{equation}
\end{enumerate}
{
Note that, as in \cite{2021PhRvD.103b3018K}, we assume positive baryon number density, positive total energy density and positive pressure.
For the case of the classical ideal-gas equation of state, the specific energy is also non-negative, i.e. $\epsilon \geq 0$.
It is worth to point out that, since the definitions of the specific energy $\epsilon$ and the relativistic enthalpy depend on the arbitrary choice of the mass constant $m_B$, relations such as $\epsilon > 0$ or $h \geq 1$ may not hold in general.
For example, negative specific energy $\epsilon$ is possible in nuclear physics equations of state.
}

{
	The recovery scheme described here has been shown to be robust and efficient in typical scenarios such as binary neutron star mergers and core-collapse supernovae, where the rescaled magnetic field $\mathcal{B}^i := {B^i}/{\sqrt{D}}$ should below $10^2$ \cite{2021PhRvD.103b3018K}. 
Although $\mathcal{B}^i \sim 10^4$ is far beyond practical uses, this scheme still converges in this case with around 40 iterations \cite{2021PhRvD.103b3018K}.
}

\subsubsection{{Error handling}}
{
Although some primitive variables are enforced to fall within the validity region during the iterations, this conserved to primitive variables conversion occasionally return unphysical results, especially at the surfaces of neutron stars.
These errors are mostly harmless and can be corrected.
After the primitive variables are obtained, we check whether any correction is needed.
By following \cite{2013PhRvD..88f4009G}, some corrections are allowed only for low density region or at black hole centre.
Low density region is defined as $\rho < \rho_{\rm{low}}$ and inside a black hole is defined as $\alpha < \alpha_{\rm{BH}}$.
In this work, we set $\rho_{\rm{low}} = \rho_{\rm{atmo}} \times 10^{2}$ while $\alpha_{\rm{BH}} = 10^{-2}$.
The error handling processes are the following:
\begin{enumerate}
	\item $\rho < \rho_{\min}$: set everything to atmosphere.
	In particular, we enforce the rest-mass density to be $\rho_\text{atmo}$ and the velocity is set to be zero, then update the rest of the primitive variables such as pressure $p$ and specific energy density $\epsilon$ by using polytropic equation of state.
	\item $\rho > \rho_{\max}$: a fatal error, stop the code.
	\item $\epsilon < \epsilon_{\min}$: set $\epsilon = \epsilon_{\min}$.
	\item $\epsilon > \epsilon_{\max}$: set $\epsilon = \epsilon_{\max}$ for low density or at black hole centre, otherwise it is a fatal error.
	\item $v > v_{\max}$: adjust for low density or at black hole centre, otherwise it is a fatal error.
	In particular, we rescale the velocity such that $v = v_{\max}$ as well as the Lorentz factor $W_{\max}$.
	Here we keep conserved density $D$ fixed, the rest-mass density $\rho$ increase slightly.
	We limit the rest-mass density and the specific energy $\epsilon$ again.
	\item The electron fraction $Y_e$ out of range: adjust for low density or at black hole centre, otherwise it is a fatal error.
\end{enumerate}
}

\subsection{Metric equations and Conformal flatness approximation}
In this work, we adopt conformal flatness approximation and solve the Einstein field equations with xCFC scheme as in \citet{2020CQGra..37n5015C}.
For the details of CFC/ xCFC schemes and how to numerically solve the metric equations, we refer readers to \citet{CoCoA, XCFC, XECHO, 2020CQGra..37n5015C}.
Here we briefly outline the basic equations and the formulations.

In a CFC approximation \citet{CoCoA, XECHO}, the three metric $\gamma_{ij}$ is assumed to be decomposed according to   
\begin{equation}
	\gamma_{ij} := \psi^4 f_{ij}, 
\end{equation}
where $f_{ij}$ is a time-independent flat background metric and $\psi$ is the conformal factor which is a function of space and time.
In the {updated} implementation, we let the flat background metric $f_{ij}$ equals the reference metric $\hat{\gamma}_{ij}$.
Another assumption is the maximal slicing condition of foliations $K = 0$.
For the matter sources, we define: $U\equiv n_\mu n_\nu T^{\mu\nu}$, $S^{i} \equiv - n_\mu \gamma^i_\nu T^{\mu\nu}$ and $S^{ij} \equiv \gamma^i_\mu \gamma^j_\nu T^{\mu\nu}$, where $T^{\mu\nu}$ is the energy-momentum tensor.
In the xCFC scheme, one introduces a vector potential $X^i$, and the metric can be solved by the following equations:   
\begin{align}
	&\tilde{\Delta} X^i + \frac{1}{3}\tilde{\nabla}^i \left( \tilde{\nabla}_j X^j \right) = 8 \pi f^{ij} \tilde{S_j} \label{eq:X} ,\\
	&\tilde{\Delta} \psi = -2\pi \tilde{U}\psi^{-1} - \frac{1}{8}f_{ik}f_{jl}\tilde{A}^{kl}\tilde{A}^{ij} \psi^{-7}  \label{eq:psi} ,\\
	&\tilde{\Delta} (\alpha\psi) = (\alpha\psi)\left[ 2\pi \left( \tilde{U} + 2\tilde{S} \right) \psi^{-2} + \frac{7}{8}f_{ik}f_{jl}\tilde{A}^{kl}\tilde{A}^{ij} \psi^{-8} \right] \label{eq:alpha} ,\\
	&\tilde{\Delta} \beta^i + \frac{1}{3}\tilde{\nabla}^i \left( \tilde{\nabla}_j \beta^j \right) = 16 \pi \alpha \psi^{-6} f^{ij} \tilde{S_i} + 2 \tilde{A}^{ij}\tilde{\nabla}_j\left(\alpha \psi^{-6}\right) \label{eq:beta},
\end{align}
where $\tilde{\nabla_i}$ and $\tilde{\Delta}$ are the covariant derivative and the Laplacian with respect to the flat three metric $f_{ij}$, respectively, and $\tilde{U}:=\psi^6 U$, $\tilde{S_i}:=\psi^6 S_i$ and $\tilde{S}:=\psi^6 S = \psi^6 \gamma_{ij} S^{ij}$ are the rescaled fluid source terms.
The tensor field $\tilde{A}^{ij}$ can be approximated on the CFC approximation level by (see the Appendix of \citet{XCFC}): 
\begin{align}\label{eq:Aij}
	&\tilde{A}^{ij} \approx \tilde{\nabla}^i X^j + \tilde{\nabla}^j X^i - \frac{2}{3}\tilde{\nabla}_k X^k f^{ij}.
\end{align}

Once the conformally rescaled hydrodynamical conserved variables $(q_{D}, q_{S_i}, q_{\tau})$ (for their definitions, see {section} \ref{sec:GRMHD_equations}) are given, the metric can be solved by the following steps:
\begin{enumerate}[{Step} 1:]
	\item Solve eq.~\eqref{eq:X} for the vector potential $X^i$ from the conserved variables $q_{S_i}$.
	\item Calculate the tensor $\tilde{A}^{ij}$ in eq.~\eqref{eq:Aij} from the vector potential $X^i$.
	\item Solve eq.~\eqref{eq:psi} for the conformal factor $\psi$.
	\item With the updated conformal factor $\psi$, calculate the conserved variables $({D}, {S}_i, {\tau})$ and thus convert the conserved variables to the primitive variables $({\rho}, W{v}^i, {P})$.
		Then $\tilde{S}$ can be worked out consistently.
	\item Solve eq.~\eqref{eq:alpha} for the lapse function $\alpha$.
	\item Solve eq.~\eqref{eq:beta} for the shift vector $\beta^i$.
\end{enumerate}

\subsubsection{{Boundary conditions}}
As in \citet{2020CQGra..37n5015C}, in the simulations of spheric-like astrophysical systems (e.g. isolated neutron star and core-collapse supernova), we set the Schwarzschild solution as the outer boundary condition.
In particular, we impose the following boundary conditions:
\begin{align}\label{eq:bc}
	&\frac{\partial \psi}{\partial r} \Big|_{r_{\text{max}}} = \frac{1-\psi}{r}, \\ 
	&\frac{\partial \alpha}{\partial r} \Big|_{r_{\text{max}}} = \frac{1-\alpha}{r}, \\
	&\beta^i \Big|_{r_{\text{max}}} = 0, \\ 
	&X^i \Big|_{r_{\text{max}}} = 0 ,
\end{align}
Note that due to the non-linearity of the scalar equations eq.~\eqref{eq:psi} and eq.~\eqref{eq:alpha}, instead of solving $\psi$ and $\alpha$ directly, we solve for its deviation, e.g. $\delta_\psi \equiv \psi - 1$ as in \citet{2020CQGra..37n5015C, XECHO}.
The boundary conditions {to eq.~\eqref{eq:bc} that} we implemented in \texttt{Gmunu} for the equation of the conformal factor $\psi$ is
\begin{align}\label{eq:robin}
	\frac{\partial}{\partial r} \left( r \delta_\psi \right) = 0.
\end{align}
In spherical {coordinates} $(r, \theta, \phi)$, the implementation of this Robin boundary condition eq.~\eqref{eq:robin} on the cell-face is straightforward.
However, this is not the case when we are working {in Cartesian coordinates} $(x,y,z)$ or cylindrical {coordinates} $(R,z,\varphi)$.
In these particular cases, we define the outer boundary at the outermost cell-centre, the boundary condition eq.~\eqref{eq:robin} can then be implemented as
\begin{align}
	\begin{cases}
		\delta_\psi + x \frac{\partial \delta_\psi}{\partial x} + y \frac{\partial \delta_\psi}{\partial y} + z \frac{\partial \delta_\psi}{\partial z} = 0 	&	\text{in {Cartesian} coordinates $(x,y,z)$},\\
		\delta_\psi + R \frac{\partial \delta_\psi}{\partial R} + z \frac{\partial \delta_\psi}{\partial z} = 0 	&	\text{in cylindrical coordinates $(R,z,\varphi)$}.
	\end{cases}
\end{align}

\subsubsection{{Frequency of solving the metric}}
{
In most of the applications, the metric quantities do not change too rapidly with time, it is in general not necessary to solve the metric equations at every time step in order to reduce the computational time.
In practice, the metric equations are solved for every $\Delta n$ time steps, and extrapolation could also be used to obtain the metric quantities in between, e.g., \citet{CoCoA}.
The number of time steps between solving the metric $\Delta n$ could vary from case to case, typically vary from 10 to 100 in spherical coordinates \citet{CoCoA, 2006MNRAS.368.1609D, XECHO, 2020CQGra..37n5015C}.
In our previous study \citet{2020CQGra..37n5015C}, we found that $\Delta n \sim 50$ is sufficient to extract the oscillation modes of isolated neutron stars correctly, and extrapolation has negligible effects on the results.
Since the time step $\Delta t$ is usually determined by Courant-Friedrichs-Lewy condition (see section~\ref{sec:cfl_condition}), we empirically found that the choice of the number of time steps between solving the metric $\Delta n \sim 10-50$ usually works well for isolated neutron star simulations even in Cartesian coordinates $(x,y,z)$ or cylindrical coordinates $(R,z,\varphi)$.
}

{
For more dynamical situations, the metric variation timescale may differ from time to time, keeping $\Delta n$ fixed during the entire simulation may be problematic.
For instance, the evolution of the metric is not correct if $\Delta n$ is too large while the computational power is wasted if $\Delta n$ is too small.
To have an adaptive $\Delta n$, we use the metric equation of $\psi$ (equation~\ref{eq:psi}) to monitor numerical errors.
In particular, at each time step, the metric will be updated if the $L^\infty$ norm of the residual of equation~\ref{eq:psi} (i.e., the maximum value of the absolute of equation~\ref{eq:psi}) below a threshold $\epsilon_{\rm{residual}}$.
It is worth to point out that, the metric equation of $\psi$ (equation~\ref{eq:psi}) is actually originated from Hamiltonian constraint equation (see, for example, \citet{doi:10.1142/9692}), which is widely used to monitor numerical errors in dynamical numerical relativity simulations.
We experimentally found out that, with this approach only (i.e., set $\Delta n$ as an extremely large number), the choice of $\epsilon_{\rm{residual}} = 10^{-3}$ (the tolerance of the metric solver is typically set as $10^{-6}$) is sufficient to obtain correct results in both stable neutron star evolution and migration tests (e.g., see section~\ref{sec:GRMHD_tests}).
}

{
Unless explicitly stated, in this work, we set $\Delta n = 50$, $\epsilon_{\rm{residual}} = 10^{-3}$ and the tolerance of the metric solver is set as $10^{-6}$.
That is, the metric variables are updated at every $\leq 50$ time steps, and keep them fixed in between.
}

\subsection{\label{sec:MG_solver}Non-linear cell-centred multigrid solver}
To solve the \emph{elliptical} metric equations \eqref{eq:X} - \eqref{eq:beta}, as in the {previous} version of \texttt{Gmunu}, we use the non-linear cell-centred multigrid (CCMG) elliptic solver \citet{2020CQGra..37n5015C}.
Since the current version of \texttt{Gmunu} is developed on top of \texttt{MPI-AMRVAC 2.0} framework \citet{2018ApJS..234...30X, 2020arXiv200403275K}, it is natural to couple {\texttt{Gmunu}} to the existing open-source geometric multigrid library \texttt{octree-mg}\footnote{As the authors did not name their code in \citet{2019CoPhC.24506866T}, here we use the name of the git repository, \texttt{octree-mg}, as the name of the library.} \citet{2019CoPhC.24506866T}.
This library is parallelised with MPI, {supports} coupling with quadtree/octree AMR grids and provides Dirichlet, Neumann and periodic boundary conditions.

However, the library has its limitations, e.g. {polar} and spherical grids are not supported, {supports} only simple and {non-varying} source terms and {has} no Robin boundary conditions and thus cannot be applied directly on the metric equations or on spherical polar coordinates.
Although the convergence rate is reduced when using point-wise smoothers directly on spherical polar/3D-cylindrical coordinates \citet{DBLP:books/daglib/0002128}, in the current implementation, we still adopt point-wise smoothers, and extend the library based on our previous implementation \citet{2020CQGra..37n5015C} so that the extended multigrid library can be applied to solve the elliptical metric equations on cylindrical and spherical coordinates.
The extension of supporting curvilinear coordinates also benefits us when handling divergenceless constraint of the magnetic field in different geometries.
{In the following, we outline the key elements of our multigrid solver.}

\subsubsection{{Cell-centred discretization and operators}}
{
To solve a non-linear elliptic equation $\mathcal{L}(u) = f$, where $\mathcal{L}$ is an elliptic operator, $u$ is the solution and $f$ is the source term, we can discretize the equation on a grid with resolution $h$ as
\begin{equation}\label{eq:elliptic_op}
	\mathcal{L}_h(u_h) = f_h,
\end{equation}
where all the solution $u_h$ and the right-hand side $f_h$ are defined at the cell centres.
The elliptic operators are discretized with a standard 5/7-point (in 2D/3D) second-order accurate discretization.
These operators contain Laplacian operators, first- and second-order derivatives etc.
Here, we list some discretized operators used in \texttt{Gmunu}.
}

{
The Laplacian of a scalar function $u(x,y,z)$ in Cartesian coordinate is 
\begin{equation}
	{\nabla^2} u = 
		\frac{\partial^2 u}{\partial x^2}
		+ \frac{\partial^2 u}{\partial y^2}
		+ \frac{\partial^2 u}{\partial z^2},
\end{equation}
which is discretized as
\begin{equation}
	\begin{aligned}
		\left({\nabla^2} u\right)_\texttt{i,j,k} := 
		&\frac{1}{\Delta x^2}\left( u_\texttt{i+1,j,k} - 2u_\texttt{i,j,k} + u_\texttt{i-1,j,k} \right) \\
		& + \frac{1}{\Delta y^2}\left( u_\texttt{i,j+1,k} - 2u_\texttt{i,j,k} + u_\texttt{i,j-1,k} \right) \\
		& + \frac{1}{\Delta z^2}\left( u_\texttt{i,j,k+1} - 2u_\texttt{i,j,k} + u_\texttt{i,j,k-1} \right) ,
	\end{aligned}
\end{equation}
where $\Delta x$, $\Delta y$ and $\Delta z$ are the grid spacing in the $x$, $y$ and $z$ directions.
On the other hand, the Laplacian of a scalar function $u(R,z,\varphi)$ in cylindrical coordinate is 
\begin{equation}
	\nabla^2 u = \frac{1}{R}\frac{\partial}{\partial R}\left(R\frac{\partial u}{\partial R}\right) 
	+ \frac{\partial^2 u}{\partial z^2}
	+ \frac{1}{R^2}\frac{\partial^2 u }{\partial \varphi^2},
\end{equation}
which is discretized as
\begin{equation}
	\begin{aligned}
		\left({\nabla^2} u\right)_\texttt{i,j,k} := & \frac{1}{R_\texttt{i,j,k}} \frac{1}{\Delta R}\Big( R_\texttt{i+1/2,j,k} \frac{u_\texttt{i+1,j,k} - u_\texttt{i,j,k} }{\Delta R} \\
		& - R_\texttt{i-1/2,j,k} \frac{u_\texttt{i,j,k} - u_\texttt{i-1,j,k} }{\Delta R}\Big) \\
		& + \frac{1}{\Delta z^2}\left( u_\texttt{i,j+1,k} - 2u_\texttt{i,j,k}- u_\texttt{i,j-1,k} \right) \\
		& + \frac{1}{R_\texttt{i,j,k}^2} \frac{1}{\Delta \varphi^2}\left( u_\texttt{i,j,k+1} - 2u_\texttt{i,j,k}- u_\texttt{i,j,k-1} \right).
	\end{aligned}
\end{equation}
Finally, in spherical coordinate, the Laplacian of a scalar function $u(r,\theta, \phi)$ reads
\begin{equation}
	\nabla^2 u = \frac{1}{r^2}\frac{\partial}{\partial r}\left(r^2\frac{\partial u}{\partial r}\right) 
	+ \frac{1}{r^2 \sin\theta}\frac{\partial }{\partial \theta} \left(\sin\theta \frac{\partial u}{\partial \theta}\right)
	+ \frac{1}{r^2 \sin^2\theta}\frac{\partial^2 u }{\partial \phi^2} ,
\end{equation}
which is discretized as
\begin{equation}
	\begin{aligned}
		\left({\nabla^2} u\right)_\texttt{i,j,k} := & \frac{1}{r_\texttt{i,j,k}^2} \frac{1}{\Delta r}\Big( r^2_\texttt{i+1/2,j,k} \frac{u_\texttt{i+1,j,k} - u_\texttt{i,j,k} }{\Delta r} \\
		& - r^2_\texttt{i-1/2,j,k} \frac{u_\texttt{i,j,k} - u_\texttt{i-1,j,k} }{\Delta r}\Big) \\
		& + \frac{1}{r^2_\texttt{i,j,k} \sin\theta_\texttt{i,j,k}} \frac{1}{\Delta \theta}\Big( \sin\theta_\texttt{i,j+1/2,k} \frac{u_\texttt{i,j+1,k} - u_\texttt{i,j,k} }{\Delta \theta} \\
		& - \sin\theta_\texttt{i,j-1/2,k} \frac{u_\texttt{i,j,k} - u_\texttt{i,j-1,k} }{\Delta \theta}\Big) \\
		& + \frac{\left( u_\texttt{i,j,k+1} - 2u_\texttt{i,j,k}- u_\texttt{i,j,k-1} \right)}{r_\texttt{i,j,k}^2 \sin^2\theta_\texttt{i,j,k} \Delta \phi^2}.
	\end{aligned}
\end{equation}
The first- and second-order derivatives are discretized as, for example, 
\begin{align}
	\left( \frac{\partial u}{ \partial x}\right)_\texttt{i,j,k} &= \frac{u_\texttt{i+1,j,k} - u_\texttt{i-1,j,k}}{2 \Delta x},\\
	\left( \frac{\partial^2 u}{ \partial x^2}\right)_\texttt{i,j,k} &= \frac{u_\texttt{i+1,j,k} -2u_\texttt{i,j,k} + u_\texttt{i-1,j,k}}{ \Delta x^2},\\
	\left( \frac{\partial^2 u}{ \partial x \partial y}\right)_\texttt{i,j,k} &= \frac{u_\texttt{i+1,j+1,k} - u_\texttt{i+1,j-1,k}- u_\texttt{i-1,j+1,k} + u_\texttt{i-1,j-1,k}}{ 4 \Delta x \Delta y}.
\end{align}
Note that the diagonal ghost cells {(i.e. layers of cells around every grid blocks, which is used to contain data from neighbouring blocks for parallel communication)} are not passed when different processors are communicating as in \citet{2019CoPhC.24506866T}, to calculate some mixed differentiation such as $\frac{\partial^2 }{\partial x \partial y}$ in equations \eqref{eq:X} and \eqref{eq:beta} without a large amount of communication between processors, at the \emph{block corner}, we adopt the following discretization which requires not all diagonal elements:   
\begin{equation}
	\begin{aligned}
		\left( \frac{\partial^2 f }{ \partial x \partial y} \right)_{\texttt{i,j,k}} \approx \frac{1}{2 \Delta x_{\texttt{i}} \Delta y_{\texttt{j}}} \Big( 
		&- f_{\texttt{i+1,j-1,k}} - f_{\texttt{i-1,j+1,k}} \\
		&+ f_{\texttt{i+1,j,k}} + f_{\texttt{i-1,j,k}} \\ 
		&+ f_{\texttt{i,j+1,k}} + f_{\texttt{i,j-1,k}} \\
		&- 2f_{\texttt{i,j,k}}	\Big).
	\end{aligned}
\end{equation}
}

\subsubsection{{Smoothers and solvers}}
{
Relaxation method can be used as a smoother since it smooths the error in the solution.
In this work, Gauss-Seidel type point-wise smoothers is included, in which the solution $u_{\texttt{i,j,k}}$ is solved while keeping the neighbours values fixed.
To deal with the case where the operator $\mathcal{L}$ is non-linear in $u$, instead of implementing the traditional Gauss-Seidel iteration, we implement a Newton Gauss-Seidel iteration \citet{Press:1992:NRF:141273}:
\begin{equation}\label{eq:ngs_iteration}
	u^{\text{new}}_\texttt{i,j,k} = u^{\text{old}}_\texttt{i,j,k} - \Bigg( \mathcal{L}\left(u^{\text{old}}_\texttt{i,j,k}\right) - f_\texttt{i,j,k} \Bigg) \Bigg/ \left( \frac{ \partial \mathcal{L}}{\partial u_\texttt{i,j,k}} \bigg|_{u = u^{\text{old}}_\texttt{i,j,k}} \right).
\end{equation}
Note that equation~\eqref{eq:ngs_iteration} reduces to the standard Gauss-Seidel iteration if $\mathcal{L}$ is linear in $u$.
}

{
The update ordering of the indices $\texttt{i,j,k}$ affects the smoothing behaviour.
Two orderings are available:
\begin{itemize}
	\item Linear ordering: all the indices $\texttt{i,j,k}$ are looped linearly (in the order they are stored in the computer's memory);
	\item Red-black ordering: also known as odd-even ordering, the solution at the points where $\texttt{i+j+k}$ is even will be updated first, and then update the rest where the points $\texttt{i+j+k}$ is odd.
\end{itemize}
As mentioned, while the convergence rate is reduced when using point-wise smoothers directly on spherical polar or three-dimensional cylindrical coordinates \citet{DBLP:books/daglib/0002128}, in the current implementation, we still adopt point-wise smoothers for all coordinates.
}

{
Although the computational cost for solving elliptic equations at the coarsest grid is low even with other robust direct solver, for simplicity, in this work, the Newton Gauss-Seidel relaxation is used as a direct solver.
}

\subsubsection{{Intergrid transfer operators: Prolongation and restriction}}
{
Grids at different levels are connected by inter grid transfer operators.
The operators that map the values from a fine to a coarse grid are called \emph{restriction} and the mapping from a coarse to the fine grid are called \emph{prolongation}.
Although there are many possible choices of restriction and prolongation operators, they cannot be chosen arbitrarily \citet{Mohr2004}.
Kwak prolongation is adopted in this work, the data at finer grids (in 2D, for example) can be written as:
\begin{equation}
	\begin{aligned}
		&u_{x+\Delta x/4, y+\Delta y/4} = \frac{1}{4}\left( 2 u_{x,y} + u_{x+\Delta x, y} + u_{x, y+\Delta y}\right) ,\\
		&u_{x-\Delta x/4, y+\Delta y/4} = \frac{1}{4}\left( 2 u_{x,y} + u_{x-\Delta x, y} + u_{x, y+\Delta y}\right) .
	\end{aligned}
\end{equation}
This can also be shown in the stencil notation, as shown in figure~\ref{fig:kwak}.
The communication costs can be saved significantly since this prolongation requires no diagonal ghost cells, thus, this is used by default.
On the other hand, for the restriction, we adopt the first-order accurate piecewise constant restriction (figure~\ref{fig:pc}).
}
\begin{figure}
	\centering
	\begin{subfigure}{0.4\columnwidth}
		\centering
\begin{equation*}
	\frac{1}{4} \left]\begin{array}{ccccc}
		  &   &   &   &  	\\
		  & 1 &   & 1 &  	\\
		  &   & * &   &  	\\
		  & 1 &   & 1 &  	\\
		  &   &   &   &  
	\end{array}\right[^h_{2h}
\end{equation*}
		\caption{piecewise constant}
		\label{fig:pc}
	\end{subfigure}
	\begin{subfigure}{0.4\columnwidth}
		\centering
\begin{equation*}
	\frac{1}{4} \left]\begin{array}{ccccc}
		\cdot  & 1 &   & 1 & \cdot	\\
		1 & 2 &   & 2 & 1	\\
		  &   & * &   &  	\\
		1 & 2 &   & 2 & 1	\\
		\cdot & 1 &   & 1 & \cdot
	\end{array}\right[^h_{2h}
\end{equation*}
		\caption{\citet{doi:10.1137/S1064827597327310}}
		\label{fig:kwak}
	\end{subfigure}
	\caption{ 
	{
	The stencil notation of the interpolation operators in 2D.
	The ``*'' denotes the location of the coarse grid node.
	The notation shows the weighting of the value which are the neighbours of the coarse grid node ``*''.
	Kwak interpolations here is second-order accurate while the price-wise constant here is first-order accurate.
	}
	}
	\label{fig:interpolation}	
\end{figure}
{
It is worth to point out that the choice of restriction and prolongation operators affect the convergence rate but not the order of accuracy of the solution, the latter is determined by the accuracy of the smoothers and solvers.
}

\subsection{\label{sec:amr}{Adaptive Mesh Refinement}}
{
As mentioned, the simulation code is required to resolve different scales accurately since most of astrophysical systems include multi-time scale and multi-length scale physics.
}

{
Adaptive mesh refinement (AMR) algorithms enable us to resolve different length scale accuracy and significantly reduce the computational costs.
In particular, AMR algorithms change the grid spacing and the structure of the computational domains during the numerical calculations.
One of the widely adopted AMR algorithms in grid-based codes is the \emph{patch-based} AMR, which is based on overlapping patches, was introduced by Berger and Oliger in 1984 \citet{1984JCoPh..53..484B}.
Some examples can be found, for example, \citet{1989JCoPh..82...64B}, \texttt{PLUTO} \citet{2007ApJS..170..228M, 2012A&A...545A.152M, 2012ApJS..198....7M}, \texttt{Athena} \citet{2008ApJS..178..137S} and \texttt{Enzo} \citet{2014ApJS..211...19B}.
Although the patch-based AMR strategy has been successfully applied in various astrophysical studies, it was found not to perform well on modern highly parallel architectures.
Alternatively, the so-called \emph{block-based} AMR (e.g., \citet{10.1145/509593.509650}) has great performance and scaling on parallel architectures, and the implementation of which is much simpler.
Notable examples are \texttt{MPI-AMRVAC} \citet{2018ApJS..234...30X, 2020arXiv200403275K} and its sister code \texttt{BHAC} \citet{2017ComAC...4....1P}, \texttt{ECHO} \citet{2015CoPhC.188..110Z}, an updated version of \texttt{Athena++} \citet{2020ApJS..249....4S}, \texttt{RAM} \citet{2006ApJS..164..255Z}, \texttt{FLASH} \citet{2010ascl.soft10082F} which is based on \texttt{PARAMESH} AMR library \citet{2000CoPhC.126..330M}, and a GPU-accelerated code called \texttt{GAMER} \citet{2010ApJS..186..457S, 2018MNRAS.481.4815S}.
}

{
The parallelization and the adaptive mesh refinement module of current \texttt{Gmunu} are provided by coupling with \texttt{MPI-AMRVAC} PDE toolkit \citet{2018ApJS..234...30X, 2020arXiv200403275K}, a open-source Message Passing Interface (MPI) based parallelised toolkit with a pure block-tree (i.e., block-based quadtree-octree (in 2D-3D), as well as their 1D equivalent) Adaptive Mesh Refinement (AMR) module.
Since \texttt{MPI-AMRVAC} is a stand alone parallelised block-grid adaptive framework and provides user-defined physics interface modules, only minor modifications are needed to use this library.
In this section, we briefly summarise the essential elements of the adaptive mesh refinement module from \texttt{MPI-AMRVAC}.
For more details of implementations, we refer readers to \citet{2012JCoPh.231..718K}.
}

\subsubsection{\label{sec:block-tree_amr}{Block-tree AMR}}
{
The computational domain is considered to be logically rectangular region, i.e., it is bounded by $\left[ x^i_{\min}, x^i_{\max} \right]$ in each dimension $i \in \left\{ 1, \cdots, N_\text{dim} \right\}$, where $N_\text{dim}$ is the number of dimensions.
The domain decomposition on the lowest grid level $l = 1$ is determined by specifying the total number of grid cells per dimension $i$, which is denoted as $N^i_{l=1}$, and the number of grid cells per dimension per block $N_\text{grid}$, where $N^i_{l=1}$ must be an integer multiple of $N_\text{grid}$.
Note that the number of grid cells per dimension $N_\text{grid}$ is independent of the grid level $l$ and also the direction $i$.
Based on this decomposition, by using the refinement strategies discussed in section~\ref{sec:refinement_criteria}, the code check whether refinement is needed for each block at each level $l < l_{\max}$.
When a block at level $l < l_{\max}$ is identified for refinement, $2^{N_{\text{dim}}}$ child blocks with the resolution $\Delta x^i_{l+1}$ are activated.
Currently, the refinement ratios between grid levels is fixed to 2, namely, $\Delta x^i_{l+1} = \Delta x^i_{l} / 2$.
The newly generated blocks will be marked as ``active grid leafs'' while their parent blocks will be removed from the active grid leafs.
Obviously, the total number of active grid leafs $N_\text{leaf}$ may change after the regridding.
}

{
Overall, the computational domain is decomposed into $N_{\text{block}}$ blocks, each block contains $\left(N_\text{grid}\right)^{N_\text{dim}}$ grid cells.
The decomposition of the computational domain into blocks is arbitrary, and the total cell number must be the number of blocks times the block size.
To minimize times for inter-processor communications and improve data locality, all the blocks are connected with a Morton-ordered space-filling curve (also known as Z-order curve) \citet{2018ApJS..234...30X}.
}

\subsubsection{{Prolongation and restriction}}
{
The prolongation-restriction formulae adopted in this work follow \citet{2007JCoPh.226..925V, 2012JCoPh.231..718K}, which can be used in curvilinear coordinate systems for which the Jacobian is separable \citet{2007JCoPh.226..925V}.
For simplicity, in this section, we express the formula for two-dimensional case, which can be generalized straightforwardly to any dimensional cases.
The prolongation adopted in this work (for two-dimensional case) is
\begin{equation}\label{eq:amr_prolongation}
	\left<q\right>^{l+1}_{\texttt{I,J}} = \left<q\right>^{l}_{\texttt{i,j}} + 
	\sum\limits_{k=1}^{N_{\text{dim}} = 2} \overline{\Delta_k q}_\texttt{i,j}^{\; l} \times 2 \frac{x^{l+1}_{\texttt{I,J}} - x^{l}_{\texttt{i,j}}}{\Delta x^l_{k}} \left( 1 - \frac{\Delta V^{l+1}_{\texttt{I,J}}}{\sum_{(k)} \Delta V^{l+1}_{{\texttt{I,J}}}}\right),
\end{equation}
where $\texttt{i,j}$ are the location indices for the variable $\left<q\right>$ at the level $l$ while $\texttt{I,J}$ is used at level $l+1$, and the volume summation indicated with $\sum_{(k)}$ sums up the 2 fine cell volumes along direction $k$, e.g., for $k = 2$, $\sum_{(k=2)} \Delta V^{l+1}_{{\texttt{I,J}}} = \Delta V^{l+1}_{{\texttt{I,J}}} + \Delta V^{l+1}_{{\texttt{I,J+1}}}$.
Here, $\overline{\Delta_k q}_\texttt{i,j}^{\; l}$ denotes the slope limited linear reconstruction of $\left<q\right>_\texttt{i,j}^{l}$ along $k$ direction.
The Total-Variation-Diminishing (TVD) slope limiters such as minmod or MC limiters can be used.
}

{
For the restriction formula, since the coarse cell values $\left<q\right>^{l}_{\texttt{i,j}}$ can be obtained by using the prolongation formula (equation~\eqref{eq:amr_prolongation}) as well, the prolongation formula is also used for restriction.
}

\subsubsection{\label{sec:refinement_criteria}{Refinement criteria}}
{
The criteria of controlling grid refinement significantly affects the efficiency and accuracy of adaptive mesh refinement calculations.
With block-tree data structure, the needs of refining or coarsening are determined on a block-by-block basis.
The regridding process for each block can be summarised as follows:
\begin{enumerate}[{Step} 1:]
	\item Check weather the block level $l$ is in the valid range, i.e., $1 \leq l < l_{\max}$, where $l_{\max}$ is the maximal grid level;
	\item At \emph{each} grid point $\bm{x}$, compute the local error $E_{\bm{x}}$, and compare with a user-set tolerance $\epsilon_l$;
	\item If \emph{any} point has the local error $E_{\bm{x}}$ larger than a user-set tolerance $\epsilon_l$, namely, if $E_{\bm{x}} > \epsilon_l$, refine this block;
	\item If \emph{all} points has the local error below the user-set tolerance with a user-defined fraction $f^{\epsilon}_l < 1$, namely, if $E_{\bm{x}} < f^{\epsilon}_l \epsilon_l$, coarsen the block if the level $l$ is larger than 1 $(l > 1)$.
\end{enumerate}
The key of the refinement criteria here is to compute the local error $E_{\bm{x}}$ at each point, and with a properly set tolerance $\epsilon_l$ and a fraction $f^{\epsilon}_l < 1$.
Depending on different application, the local error needed to be estimated on various primitive or auxiliary variables.
For example, consider a set of variables $q_k$, the final local error $E_{\bm{x}}$ are calculated by the formula
\begin{equation}
	E_{\bm{x}} = \sum_k \sigma_{k} E_{\bm{x}, k}^{\text{rel}} ,
\end{equation}
where $k$ is a variable index for $q_k$, $E_{\bm{x}, k}^{\text{rel}}$ is the local relative variable errors of variable $q_k$ at grid $\bm{x}$, and $\sigma_{k}$ is the corresponding weighting which is defined by users and obey $\sum_k \sigma_{k} = 1$.
The local relative variable errors $E_{\bm{x}, k}^{\text{rel}}$ can be obtained by the so-called error estimators. 
In the following, we describe various possible estimators.
}

\paragraph{{Historical estimator}}
{
Historical estimator requests the information of previous time steps such as $t^{n-1}$ and $t^{n}$.
Richardson extrapolation can be used to compute the local error by the following: 
\begin{equation}
	E^{\text{rel}}_{\bm{x},k} = \frac{\lvert q^{\text{CI}}_k - q^{\text{IC}}_k \rvert}{\sum_k \sigma_{k} \lvert q^{\text{IC}}_k \rvert} ,
\end{equation}
where $q^{\text{CI}}_k$ is the coarsened-integrated solution which can be obtained by coarsening it at time $n-1$ from a $\Delta x_i$ grid to a $2 \Delta x_i$ grid, then time integrating it with time step $2\Delta t^{n-1}_l$;
while $q^{\text{IC}}_k$ is the integrated-coarsened solution which can be obtained by time integrating $\left(q_k\right)^{n}_l$ with time step $\Delta t^{n-1}_l$.
To be specific, $q^{\text{CI}}_k$ and $q^{\text{IC}}_k$ can be obtained by
\begin{equation}
	\begin{aligned}
		&\left(q_k\right)^{n-1}_l \xrightarrow{\text{coarsening}} \left(q_k\right)^{n-1}_{2 \Delta x_i} 
		\xrightarrow[2\Delta t^{n-1}_l]{\text{advance}} q^{\text{CI}}_k , \\
		&\left(q_k\right)^{n}_l \xrightarrow[\Delta t^{n-1}_l]{\text{advance}} q^{\text{IC}}_k .
	\end{aligned}
\end{equation}
The integrator used in this Richardson-based estimator can be low-order (e.g., first-order), dimensionally unsplit, incorporating unsplit source terms.
}

{
A simpler and computationally cheaper variant of historical estimator is the local comparison of the solution between $t^{n-1}$ and $t^{n}$.
For instance, the local error of variable $q_k$ can be written as
\begin{equation}
	E^{\text{rel}}_{\bm{x},k} = \frac{\lvert q^{n-1}_k - q^{n}_k \rvert}{\lvert q^{n-1}_k \rvert}.
\end{equation}
}

{
It is worth to point out that although these kinds of historical estimators work successfully in variety of test problems, relying on previous solution only may be insufficient for rapidly moving, strong shock cases.
For more details, see discussions in \citet{2012JCoPh.231..718K}.
}

\paragraph{{L{\"o}hner type estimator}}
{
The L{\"o}hner's error estimator \citet{1987CMAME..61..323L} was originally developed for finite-element simulations.
This is computationally efficient since it requests the ``current'' ($t^{n}$) solution only.
The multi-dimensional generalisation is given by
\begin{equation}
	E^{\text{rel}}_{\bm{x},k} = \sqrt{\frac{\sum\limits_p \sum\limits_q \left( \frac{\partial^2 q_k}{\partial x_p \partial x_q} \Delta x_p \Delta x_q \right)^2}{\sum\limits_p \sum\limits_q \left[ \left( \lvert \frac{\partial q_k}{\partial x_p} \rvert_{\bm{x}+\Delta x_p} + \lvert \frac{\partial q_k}{\partial x_p} \rvert_{\bm{x}-\Delta x_p} \right)\Delta x_p + f_{\text{wave},l} \lvert \bar{q_k} \rvert_{p,q} \right]^2} },
\end{equation}
where the indices $p$ and $q$ run over all dimensions $p,q = 1, \cdots, N_D$. 
The last term in the denominator $f_{\text{wave},l} \lvert \bar{q_k} \rvert_{p,q}$ prevent refinement of small ripples, where $f_{\text{wave},l}$ is the level dependent ``wave-filter'' parameter which is typically chosen of the order $10^{-2}$.
$\lvert \bar{q_k} \rvert_{p,q}$ here is an average value computed from surrounding corners,
\begin{equation}
	\lvert \bar{q_k} \rvert_{p,q} := \lvert {q_k}\rvert_{\bm{x}+\Delta x_p+\Delta x_q} + 2 \lvert {q_k}\rvert_{\bm{x}} + \lvert {q_k}\rvert_{\bm{x}+\Delta x_p+\Delta x_q}.
\end{equation}
Unless explicitly stated, the error is estimated based on the density $\rho$ only by using the L{\"o}hner's error estimator when adaptive mesh refinement module is activated.
}

\subsection{\label{sec:cfl_condition}{Time stepping}}
{
To evolve the hyperbolic partial differential equations stably, the size of the time step must satisfy the Courant-Friedrichs-Lewy (CFL) conditions \citet{1928MatAn.100...32C}.
Physically, this conditions ensure that the propagation speed of any travelling wave is always smaller than the ``numerical speed'' $\sim \Delta x / \Delta t$.
}

{
Before we present how we determine the time step $\Delta t$, we briefly outline the block-tree adaptive mesh refinement data structure as well as the notations we use here.
As discussed in section~\ref{sec:block-tree_amr}, in \texttt{Gmunu}, the whole computational domain is decomposed into $N_\text{leaf}$ active grid leafs, each of them contains $N_\text{block}$ blocks, each block contains $\left(N_\text{grid}\right)^{N_\text{dim}}$ grid cells, where $N_\text{grid}$ is the number of grids per dimension and $N_\text{dim}$ is the number of dimensions.
In addition, to simplify the implementation significantly, the same time step $\Delta t$ is assigned for all levels.
In the following, $1 \leq i \leq N_\text{dim}$ denotes the dimension, $1 \leq j \leq N_\text{block}$ denotes the block index and finally $1 \leq k \leq N_\text{leaf}$ denotes the leaf index.
}

{
In \texttt{Gmunu}, the time step $\Delta t$ is given by:
\begin{equation}
	\Delta t = \min_{1\leq k \leq N_\text{leaf}} \left( c_\text{CFL} \tau_{k} \right),
\end{equation}
where $c_\text{CFL}$ is the Courant-Friedrichs-Lewy factor with range $\left(0,1\right]$ and typically below 0.9; $\tau_{k}$ is the unrestricted time step at the $k$-th leaf.
In \texttt{Gmunu}, there are three possible ways to evaluate the unrestricted time step $\tau_{k}$, they are ``minimun'', ``summax'' and ``maxsum'':
\begin{align}
	\tau_{k}^{-1} = \begin{cases}
		\max\limits_{1 \leq i \leq N_\text{dim}} \left( \max\limits_{1 \leq j \leq N_\text{block}} \left( \frac{c^{\max}_i}{{\Delta_i}} \right)\right) &, \text{minimum} ; \\
		\sum\limits^{N_\text{dim}}_{i=1} \left( \max\limits_{1 \leq j \leq N_\text{block}} \left( \frac{c^{\max}_i}{{\Delta_i}} \right)\right)  &, \text{summax} ;\\
		\max\limits_{1 \leq j \leq N_\text{block}}\left( \sum\limits^{N_\text{dim}}_{i=1}\frac{c^{\max}_i}{{\Delta_i}} \right) &, \text{maxsum (default case)} ,
	\end{cases}
\end{align}
where $c^{\max}_i$ is the maximal signal propagation speed (at cell-centres) in the $i$ direction, usually the maximum of the absolute value of eigenvalue $ \max\limits_i \left( \lvert \lambda_i \rvert \right)$ (see section~\ref{sec:characteristic_speed}) is used.
In addition, $\Delta_i$ here is the spatial step size in direction $i$.
For example, in Cartesian coordinates $\left(x,y,z\right)$, $\Delta_i$ are simply $\left( \Delta x, \Delta y, \Delta z \right)$.
It is worth to point out that, in spherical-polar coordinates $\left( r, \theta, \phi \right)$, the corresponding $\Delta_i$ are $\left( \Delta r, r \Delta \theta, r \sin\theta \Delta \phi \right)$, the Courant-Friedrichs-Lewy become rigorous in multi-dimensional cases at the centre $(r \rightarrow 0$) or at the pole ($\theta\rightarrow 0, \pi$) in three-dimensional cases.
Many approaches proposed to deal with the rigorous time step constraint in spherical-polar coordinates (see \citet{2020LRCA....6....3M} and references therein).
In \texttt{Gmunu}, we made use of adaptive mesh refinement (see section~\ref{sec:amr}), the grids are enforced to be coarsened to keep $r \Delta \theta \sim \Delta r$ when $r$ is small and similarly, we require $r\sin\theta \Delta \sim \Delta r$ as $\theta \rightarrow 0$ or $\theta \rightarrow \pi$.
}

\section{Numerical tests}\label{sec:numerical_tests}
In the remainder of this paper, we present a selection of representative test problems with our code.
The tests {range} from special relativistic (magneto-)hydrodynamics to general relativistic (magneto-)hydrodynamics, from {one to multiple dimensions} and {in Cartesian}, cylindrical and spherical coordinates.
Unless otherwise specified, all simulations reported in this paper were performed with TVDLF approximate Riemann solver, 5-th order reconstruction method MP5 and SSPRK3 for the time integration. 

\subsection{Special Relativistic Hydrodynamics}
\subsubsection{{Two-dimensional smooth problem}}
{
A smooth test for relativistic hydrodynamics code proposed in \citet{2012CCoPh..11..114H}, which describes a wave propagating in a two-dimensional space, is well suited for checking the order of accuracy of a numerical code at the smooth part.
In particular, we perform the simulations with {Cartesian} coordinates $(x,y)$ on {a} flat spacetime, the computational domain covers the region $0 \leq x \leq 3 / \sqrt{2}$ and $0 \leq y \leq 2$. 
The initial condition is given as
\begin{align}
	&\rho = 1 + A \sin \left[ 2\pi \left( x \cos\theta + y \sin\theta \right) \right], \\
	&p = 1, \\
	&v_x = v_0, \qquad v_y = 0, 
\end{align}
where the wave is propagating at an angle $\theta = \pi/6$ relative to the horizontal axis, $A=0.2$ and $v_0=0.2$.
We consider an ideal-gas equation of state $ p = (\Gamma-1)\rho \epsilon$ with $\Gamma = 5/3$.
This problem has the exact solution
\begin{align}
	&\rho = 1 + A \sin \left[ 2\pi \left( \left( x \cos\theta + y \sin\theta \right) - \left( v_x \cos\theta + v_y\sin\theta\right)t \right) \right], \\
	&p = 1, \\
	&v_x = v_0, \qquad v_y = 0.
\end{align}
The discretization of the computational domain is set to be $\left[N, 2N \right]$ with an integer $N$ which controls the resolution.
In this test, we use 2-nd order accurate strong-stability preserving Runge-Kutta (SSPRK2) time integrator, Harten, Lax and van Leer (HLL) Riemann solver \citet{harten1983upstream} with 2-nd order Montonized central (MC) limiter \citet{1974JCoPh..14..361V}. 
}

{
To quantify the convergence rate at $t=2$, we define the relative numerical errors $\delta_N$ as
\begin{equation}
	\delta_N = \frac{ \sum\limits_{i}^N \sum\limits_{j}^N \lvert \rho^N_{i,j} - \rho^{\text{exact}}_{i,j} \rvert}{\sum\limits_{i}^N \sum\limits_{j}^N \lvert \rho^{\text{exact}}_{i,j} \rvert},
\end{equation}
and the convergence rate $R_N$ can be obtained by
\begin{equation} \label{eq:convergence_rate}
	R_N = \log_2 \left( \frac{\delta_{N/2}}{\delta_N} \right).
\end{equation}
Table~\ref{tab:hydro_convergence_rate} shows the numerical errors and convergence rates of this problem at $t=2$ at different resolution $N$.
The convergence rate can be virtually present with a numerical-errors-versus-resolution plot, as shown in the figure~\ref{fig:HD_convergence_rate}.
As expected, the second-order convergence is achieved with this setting.
\begin{table}
	\centering
	\begin{tabular}{c|c|c}
		$N$  & $\delta_N$ & $R_N$ \\ \hline
		32   & 1.859E-3   & --    \\ 
		64   & 4.839E-4   & 1.94  \\ 
		128  & 1.246E-4   & 1.96  \\ 
		256  & 3.120E-5   & 2.00  \\ 
		512  & 7.692E-6   & 2.02  \\ 
		1024 & 1.896E-6   & 2.02  \\ 
		2048 & 4.690E-7   & 2.02
	\end{tabular}
	\caption{\label{tab:hydro_convergence_rate}
		{
		Numerical errors and convergence rates of the two-dimensional relativistic hydrodynamics smooth problem at $t=2$.
		In this test, 2-nd order accurate strong-stability preserving Runge-Kutta (SSPRK2) time integrator, Harten, Lax and van Leer (HLL) Riemann solver \citet{harten1983upstream} with 2-nd order Montonized central (MC) limiter \citet{1974JCoPh..14..361V} are used. 
		As expected, the second-order convergence is achieved with this setting.
		}
		}
\end{table}
\begin{figure}
	\centering
	\includegraphics[width=\columnwidth, angle=0]{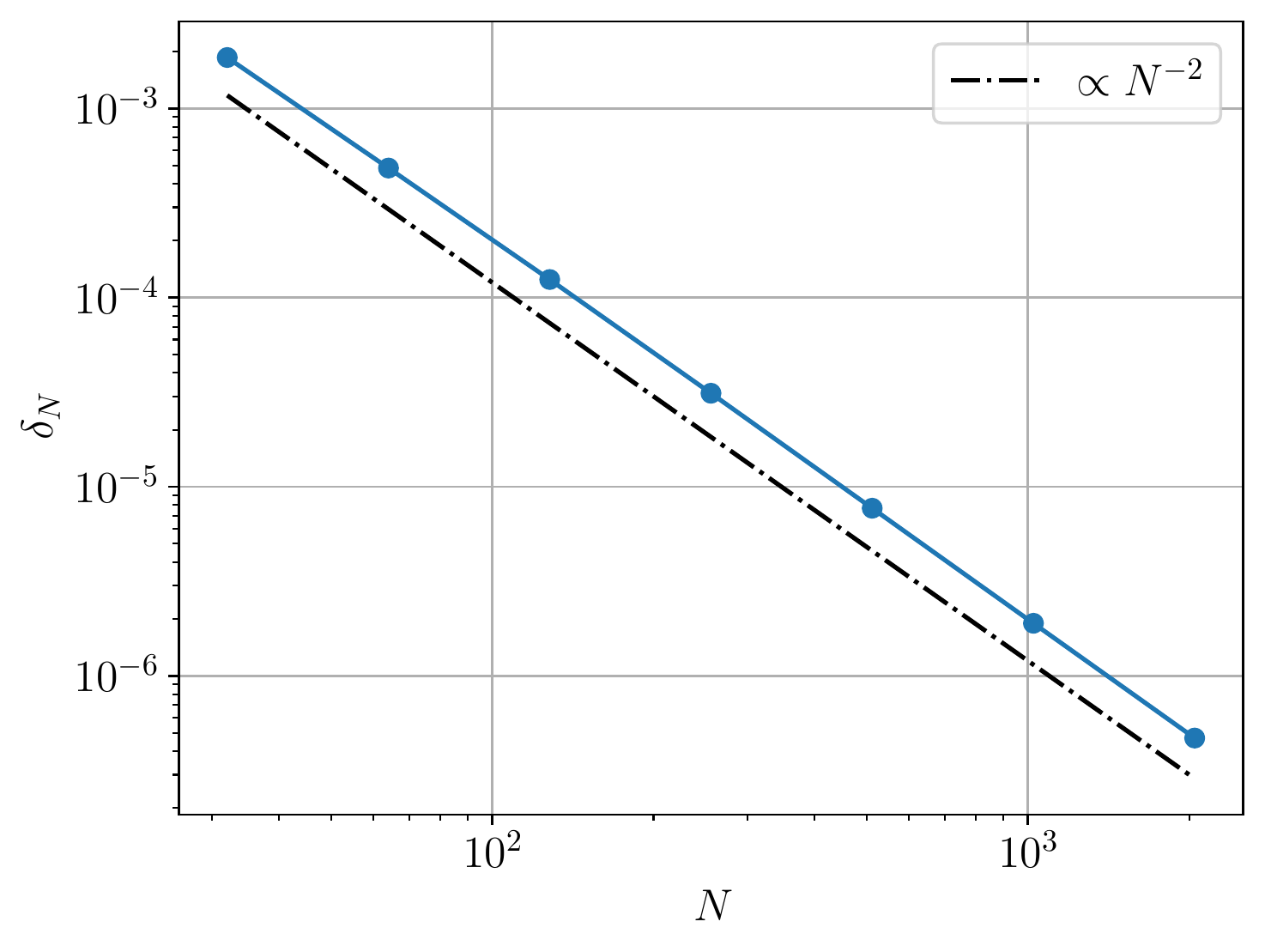}
	\caption{
		{
		Numerical errors versus resolution (blue line) for the two-dimensional relativistic hydrodynamics smooth problem.
		Second-order ideal scaling is given by the dashed black line.
		In this test, 2-nd order accurate strong-stability preserving Runge-Kutta (SSPRK2) time integrator, Harten, Lax and van Leer (HLL) Riemann solver \citet{harten1983upstream} with 2-nd order Montonized central (MC) limiter \citet{1974JCoPh..14..361V} are used. 
		As expected, the second-order convergence is achieved with this setting.
		}
		}
	\label{fig:HD_convergence_rate}	
\end{figure}
}

\subsubsection{Relativistic Shock Tubes}
\begin{figure}
	\centering
	\includegraphics[width=1.0\columnwidth, angle=0]{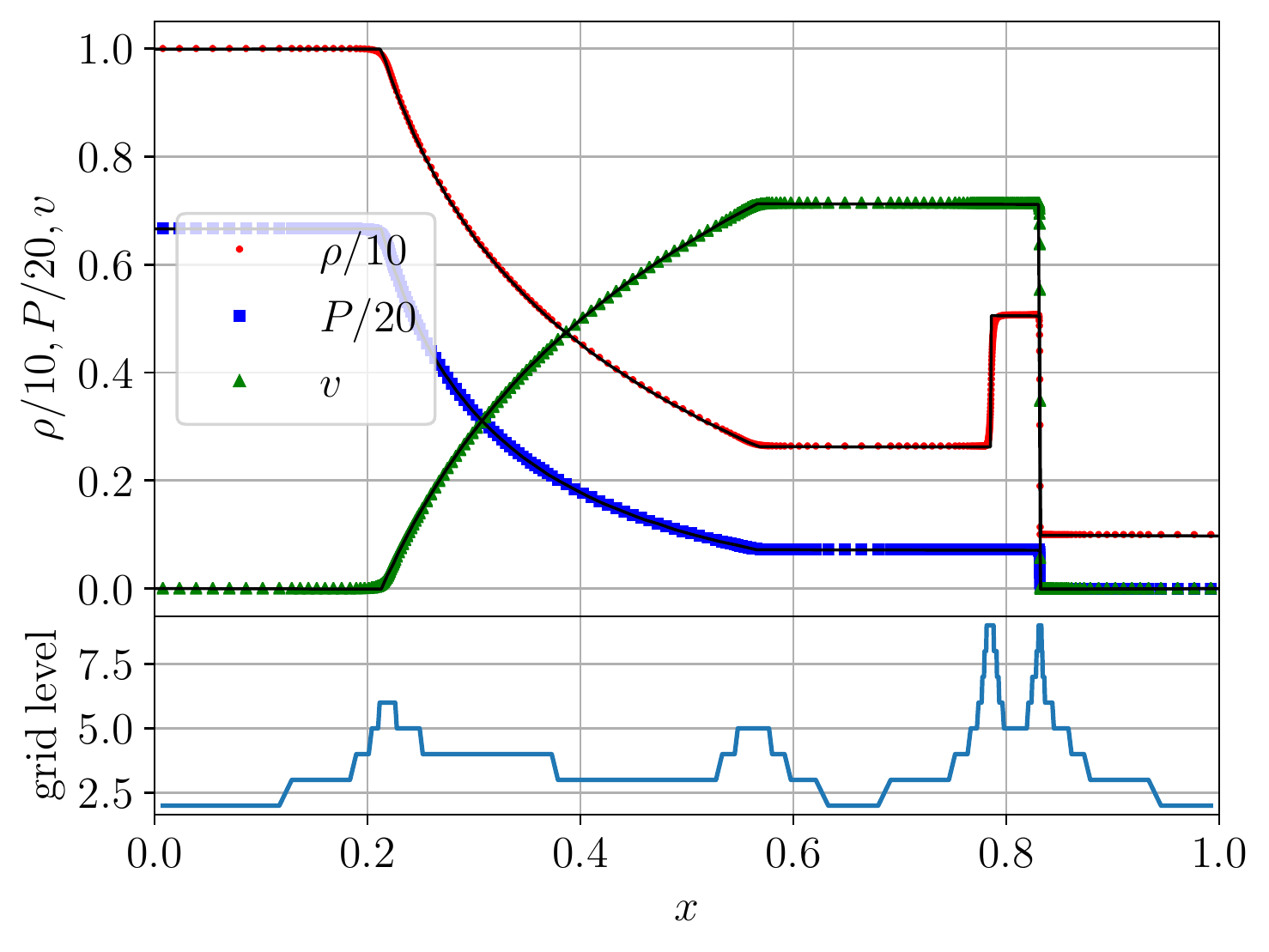}
	\caption{
		The upper panel shows the density (red dots), pressure (blue squares) and velocity (green triangles) profile at $t=0.4$ for the relativistic shocktube test problem.
		The solid lines are the analytic solutions.
		The numerical results obtained by \texttt{Gmunu} agree with the analytic solutions.
		The lower panel shows the {grid-level} at different location of the computational domain.
		{The grid-level} is higher to provide finer resolution when the density is sharper.
		}
	\label{fig:HD_shocktube_1d}	
\end{figure}

We follow \citet{SRHD} in this one-dimensional shock tube problem. 
In particular, we perform the simulation with {Cartesian} coordinates on {a} flat spacetime.
Instead of simulating this problem with {a} uniform grid, we activate the block-based AMR module in this case.
For instance, the computational domain covers the region $0 \leq x \leq 1$ with 16 base grid points and {allows for} 10 AMR levels (i.e. an effective resolution of 8192). 
The initial condition is given as
\begin{align}
	\left( \rho, p, v^x \right) = 
	\begin{cases}
		\left( 10 ,40/3, 0 \right)	&	\text{if } x < 0.5 ,\\
		\left( 1, 0, 0 \right)	&	\text{if } x > 0.5 .
	\end{cases}
\end{align}
We consider an ideal-gas equation of state $ p = (\Gamma-1)\rho \epsilon$ with $\Gamma = 5/3$.
The upper panel of the figure \ref{fig:HD_shocktube_1d} shows the comparison between the numerical results and the analytic solutions for the density, pressure and velocity profiles at $t = 0.4$.
The figure shows that our numerical results agree with the analytic solutions.
The lower panel shows the {grid-level} at different location of the computational domain.
{The grid-level} is higher to provide finer resolution when the density is sharper.

\subsubsection{Two-dimensional Riemann Problem}
\begin{figure}
	\centering
	\includegraphics[width=1.0\columnwidth, angle=0]{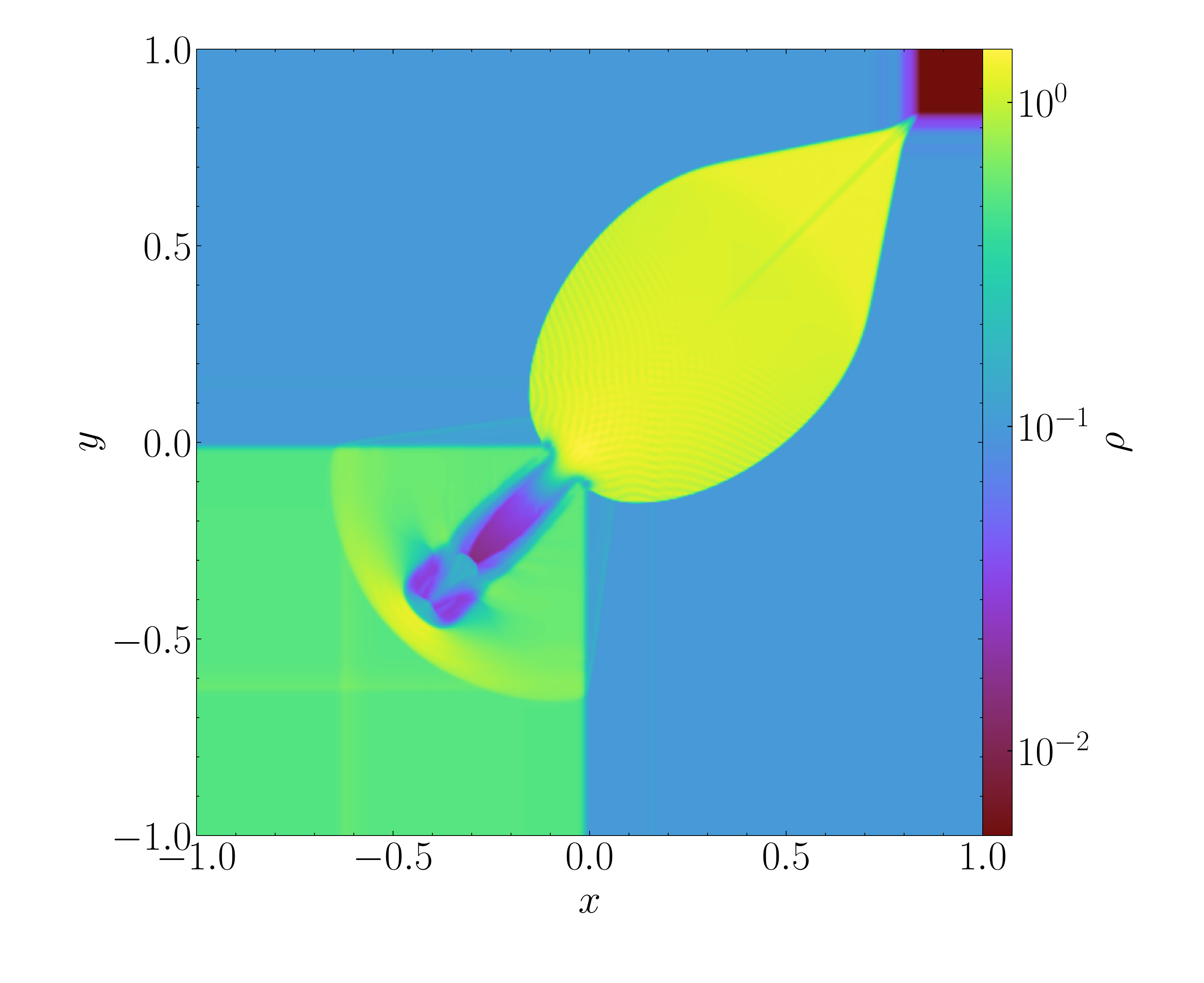}
	\caption{The density profile of {the} two-dimensional relativistic Riemann problem \citet{2005ApJS..160..199M} at time $ t = 0.8$.
	{The result agrees qualitatively with \citet{2005ApJS..160..199M}.}}
	\label{fig:HD_riemann_2d}	
\end{figure}

To test how \texttt{Gmunu} works {in two-dimensional Cartesian coordinates}, we picked a demanding highly relativistic two-dimensional Riemann problem \citet{2002A&A...390.1177D}.
Here, we follow the modified version of this test presented in \citet{2005ApJS..160..199M}, {in} which elementary waves are introduced at every interface.
The initial condition is given as
\begin{align}
	\left( \rho, p, v^x , v^y \right) = 
	\begin{cases}
		\left( \rho_1, p_1 , 0 , 0  \right)	&	\text{if } x > 0, y > 0 ,\\
		\left( 0.1 , 1 , 0.99 , 0  \right)	&	\text{if } x < 0, y > 0 ,\\
		\left( 0.5 , 1 , 0 , 0  \right)	&	\text{if } x < 0, y < 0 ,\\
		\left( 0.1 , 1 , 0 , 0.99  \right)	&	\text{if } x > 0, y < 0 ,\\
	\end{cases}
\end{align}
where $\rho_1 = 5.477875 \times 10^{-3} $, $p_1 = 2.762987 \times 10^{-3}$.
Here we consider an ideal-gas equation of state $ p = (\Gamma-1)\rho \epsilon$ with $\Gamma = 5/3$.
This test is run with a uniform grid $512 \times 512$ which covers the region $\left[-1,1\right]$ for both $x$ and $y$.
Figure~\ref{fig:HD_riemann_2d} shows the density profile at $t = 0.8$.
\texttt{Gmunu} is able to evolve this demanding test without {crashing the code}. 

\subsubsection{Two-dimensional axisymmetric jet in cylindrical geometry}
We study the propagation of a two-dimensional axisymmetric relativistic jet in cylindrical coordinates.
Not only {would we} like to test if \texttt{Gmunu} works properly in cylindrical geometry, to test the code's robustness, we simulated the model C2 in \citet{1997ApJ...479..151M}, which contains strong relativistic shocks, instabilities and shear flows and is highly supersonic.
The computational domain covers $0 \leq r \leq 15$ and $0 \leq z \leq 45$ with resolution $512 \times 1536$.
Initially, the jet is configured in the region $r \leq 1$ and $z \leq 1$ with density $\rho_b = 1 \times 10^{-2} $, pressure $p_b = 1.70305 \times 10^{-4}$, the velocity along z-axis $v_z = v_b = 0.99 c$ (which corresponds to a Lorentz factor $\sim$ 7).
Here we consider the ideal-gas equation of state with $\Gamma = 5/3$.
The rest of the computational domain is filled with an ambient medium with density $\rho_m = 1$, pressure $p_m = p_b$, and zero velocity.
We apply reflecting boundary conditions at the symmetric axis while the out-going boundary conditions were applied at all outer boundaries except that we keep the value unchanged inside the jet inlet $z=0, r<1$.
In this test, we use 3-rd order reconstruction method PPM.  

\begin{figure}
	\centering
	\includegraphics[width=\columnwidth, angle=0]{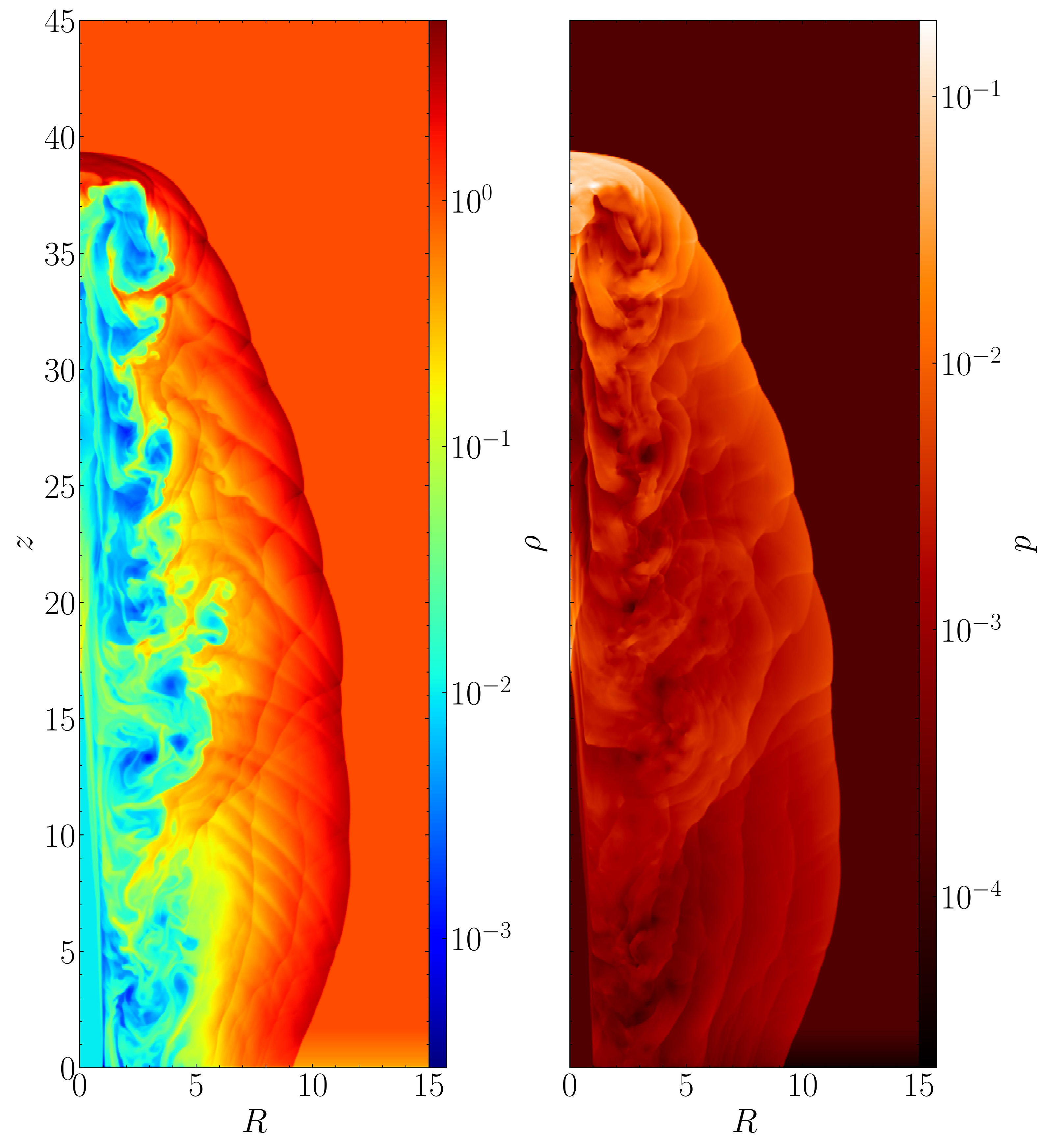}
        \caption{Density distribution (\emph{left panel}) and the pressure (\emph{right panel}) of the axisymmetric jet model C2 in \citet{1997ApJ...479..151M} at $t = 100$.
		The jet material interacts with the ambient medium and forms an expanding bow shock and {develops} the Kelvin-Helmholtz instability.
		The key structures of the jet, e.g.~, the head location, the shape of the bow shock and the development of the Kelvin-Helmholtz instability agrees with \citet{1997ApJ...479..151M}.
		}
	\label{fig:HD_jet_2d}	
\end{figure}

Figure~\ref{fig:HD_jet_2d} shows the density distribution of the axisymmetric jet at $t = 100$.
As shown in figure \ref{fig:HD_jet_2d}, an expanding bow shock is formed and the Kelvin-Helmholtz instability is developed.
The key structures of the jet, e.g. the head location, the shape of the bow shock and the development of the Kelvin-Helmholtz instability {all agree} with \citet{1997ApJ...479..151M}.

\subsection{Special Relativistic Magneto-Hydrodynamics}
\subsubsection{{Large-amplitude circularly polarized Alfv{\'e}n waves}}
{
Large-amplitude circularly polarized Alfv{\'e}n waves test was first proposed in \citet{2007A&A...473...11D}.
This test describes the propagation of circularly polarized Alfv{\'e}n waves with large-amplitude along a uniform background magnetic field $\vec{B}_0$.
Here, by following \citet{2014CQGra..31a5005M}, we perform the simulations with one-dimensional Cartesian coordinates $x$ on a flat spacetime, the computational domain covers the region $0 \leq x \leq 1$ with periodic boundary condition. 
The initial condition is given as
\begin{align}
	&\rho = 1.0, && p = 0.5 \\
	&v^x = 0, && v^y = -v_A A_0 \cos \left( k x\right), && v^z = -v_A A_0 \sin \left( k x\right) , \\
	&B^x = B_0, && B^y = A_0 B_0 \cos \left( k x\right), && B^z = A_0 B_0 \sin \left( k x\right) , 
\end{align}
where $k = 2\pi / L_x$ is the wave vector, $B_0 = 1$ is the constant magnetic field for $B^x$, $A_0 = 1$ is the amplitude parameter and finally the square of the Alfv{\'e}n speed $v_A^2$ can be expressed as
\begin{equation}
	v^2_A = \frac{2 B_0^2}{\rho h + B_0^2 \left( 1 + A_0^2 \right) } \left[ 1 + \sqrt{1 - \left( A_0^2 \frac{2 B_0^2}{\rho h + B_0^2 \left( 1 + A_0^2 \right) } \right)^2 }\right]^{-1}.
\end{equation}
We consider an ideal-gas equation of state $ p = (\Gamma-1)\rho \epsilon$ with $\Gamma = 5/3$.
The simulation is allowed up to $t = T = 2$ (one period).
The discretization of the computational domain is set to be $N$ with an integer $N$ which controls the resolution.
In this test, we use 2-nd order accurate strong-stability preserving Runge-Kutta (SSPRK2) time integrator, Harten, Lax and van Leer (HLL) Riemann solver \citet{harten1983upstream} with 2-nd order Montonized central (MC) limiter \citet{1974JCoPh..14..361V}. 
}

{
To quantify the convergence rate at $t=2$, we compute the $L_1$-norm of the difference of the difference between the initial and final values of the $z$-component of the magnetic field $B^z$ as
\begin{equation} \label{eq:L1_norm_Bz}
	\lvert\lvert B^z(t=2) - B^z(t=0) \rvert\rvert_1 := \frac{\sum\limits_i \lvert B^z(t=2) - B^z(t=0) \rvert \Delta V_i}{\sum\limits_i \Delta V_i},
\end{equation}
and the convergence rate $R_N$ follows equation~\ref{eq:convergence_rate}.
}

\begin{figure}
	\centering
	\includegraphics[width=\columnwidth, angle=0]{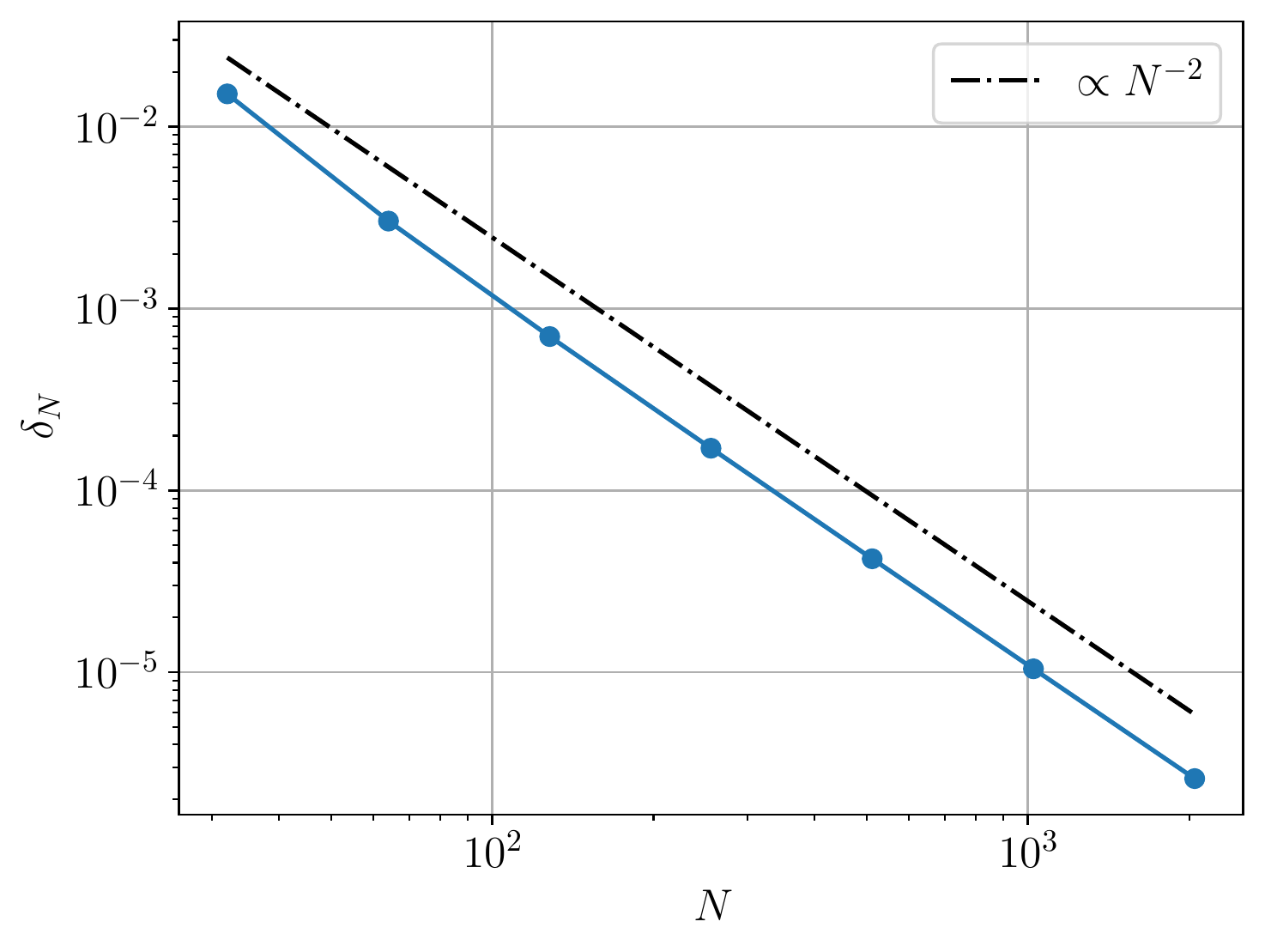}
	\caption{
		{
		Numerical errors versus resolution (blue line) for the one-dimensional ideal relativistic-magnetohydrodynamics Alfv{\'e}n wave problem.
		Second-order ideal scaling is given by the dashed black line.
		In this test, 2-nd order accurate strong-stability preserving Runge-Kutta (SSPRK2) time integrator, Harten, Lax and van Leer (HLL) Riemann solver \citet{harten1983upstream} with 2-nd order Montonized central (MC) limiter \citet{1974JCoPh..14..361V} are used. 
		As expected, the second-order convergence is achieved with this setting.
		}
		}
	\label{fig:MHD_convergence_rate}	
\end{figure}

{
Table~\ref{tab:mhd_convergence_rate} shows the $L_1$-norm of the difference of the difference between the initial and final values of the $z$-component of the magnetic field $B^z$ $\lvert\lvert B^z(t=2) - B^z(t=0) \rvert\rvert_1$ and convergence rates of this problem at $t=2$ at different resolution $N$.
The convergence rate can be virtually present with a numerical-errors-versus-resolution plot, as shown in the figure~\ref{fig:MHD_convergence_rate}.
As expected, the second-order convergence is achieved with this setting.
}
\begin{table}
	\centering
	\begin{tabular}{c|c|c}
		$N$  & $\lvert\lvert B^z(t=2) - B^z(t=0) \rvert\rvert_1$ & $R_N$ \\ \hline
		32   & 1.859E-3   & --    \\ 
		64   & 4.839E-4   & 2.32  \\ 
		128  & 1.246E-4   & 2.11  \\ 
		256  & 3.120E-5   & 2.04  \\ 
		512  & 7.692E-6   & 2.02  \\ 
		1024 & 1.896E-6   & 2.01  \\ 
		2048 & 4.690E-7   & 2.01
	\end{tabular}
	\caption{\label{tab:mhd_convergence_rate}
		{
		Numerical errors and convergence rates of the large-amplitude circularly polarized Alfv{\'e}n waves test at $t=2$.
		In this test, 2-nd order accurate strong-stability preserving Runge-Kutta (SSPRK2) time integrator, Harten, Lax and van Leer (HLL) Riemann solver \citet{harten1983upstream} with 2-nd order Montonized central (MC) limiter \citet{1974JCoPh..14..361V} are used. 
		As expected, the second-order convergence is achieved with this setting.
		}
		}
\end{table}

\subsubsection{Relativistic Shock Tubes}
Similar {to} relativistic hydrodynamics, there are shock tube tests in MHD.
We follow \citet{2001ApJS..132...83B} in this one-dimensional shock tube problem. 
In particular, we perform the simulation with {Cartesian} coordinates on {a} flat spacetime.
The initial condition is given as
\begin{align}
	\left( \rho, p, B^x, B^y \right) = 
	\begin{cases}
		\left( 1,1,0.5,1 \right)	&	\text{if } x < 0 ,\\
		\left( 0.125,0.1,0.5,-1 \right)	&	\text{if } x > 0.
	\end{cases}
\end{align}
We consider an ideal-gas equation of state $ p = (\Gamma-1)\rho \epsilon$ with $\Gamma = 2$.

\begin{figure}
	\centering
	\includegraphics[width=1.0\columnwidth, angle=0]{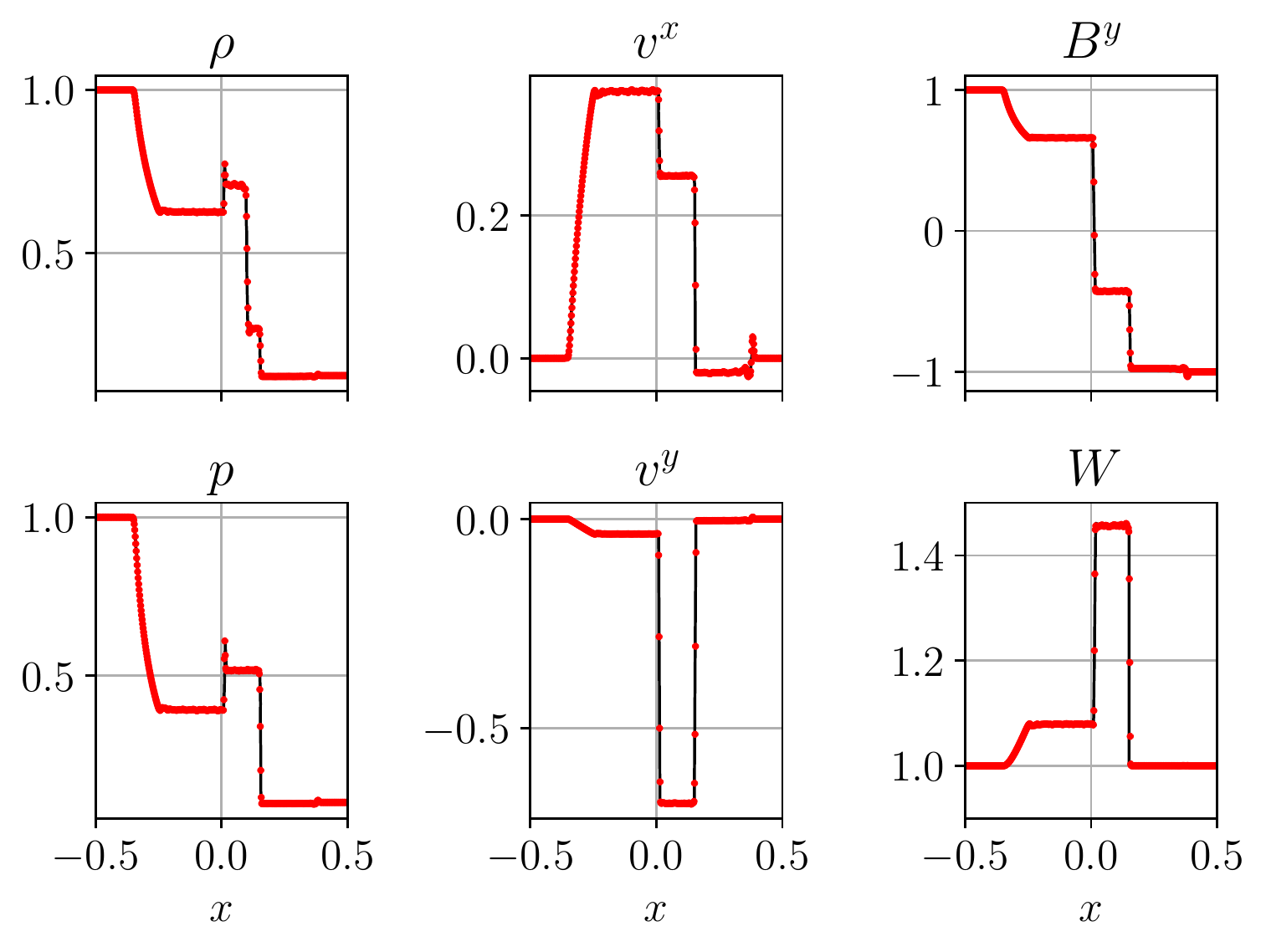}
	\caption{The density $\rho$ (\emph{upper left}), {pressure} $p$ (\emph{lower left}), velocity components $v^x$ (\emph{upper middle}) and $v^y$ (\emph{lower middle}), the y-component of the magnetic field $B^y$ (\emph{upper right}) and the Lorentz factor $W$ (\emph{lower right}) for the shock tube test at $t = 0.4$.
The red dots show the numerical results obtained by \texttt{Gmunu}, which agree with the reference solutions (black solid lines) \citet{2001ApJS..132...83B}.
		}
	\label{fig:MHD_shocktube_1d}	
\end{figure}

Figure~\ref{fig:MHD_shocktube_1d} compares the numerical results obtained by \texttt{Gmunu} (red dots) with the reference solutions (black solid lines) \citet{2001ApJS..132...83B} at $t=0.4$.
It illustrates the shock-capturing ability of \texttt{Gmunu} and the results agree with the reference results.

\subsubsection{Cylindrical blast wave}
The cylindrical blast wave is a well-known difficult multi-dimensional SRMHD test problem.
This problem describes an expanding blast wave in a plasma with an initially uniform magnetic field.
Here, we follow the parameters presented in \citet{1999MNRAS.303..343K}.
The initial condition of this test problem is determined with radial parameters $r_\text{in}$ and $r_\text{out}$.
The density (and also the pressure, in the same form) profile is given by:
\begin{align}
	\rho(r) = 
	\begin{cases}
		\rho_\text{in}	&	\text{if } r \leq r_\text{in} ,\\
		\exp \left[ \frac{ \left(r_\text{out} - r\right) \ln \rho_\text{in} +  \left(r - r_\text{in}\right) \ln \rho_\text{out} }{r_\text{out} - r_\text{in}} \right]	&	\text{if }  r_\text{in} \leq r \leq r_\text{out} ,\\
		\rho_\text{out}	&	\text{if } r \geq r_\text{out} ,
	\end{cases}
\end{align}
where the parameters are: 
\begin{align}
	& r_\text{in} = 0.8, && r_\text{out} = 1.0; \\
	& \rho_\text{in} = 10^{-2}, && \rho_\text{out} = 10^{-4} ;  \\
	& p_\text{in} = 1.0, && p_\text{out} = 3 \times 10^{-5};  \\
	& B^i = (0.1, 0, 0), && v^i = (0,0,0) .
\end{align}
Here we consider the ideal-gas equation of state with $\Gamma = 4/3$.
The computational domain covers $[-6,6]$ for both $x$ and $y$ directions with the resolution {$128 \times 128$}.

\begin{figure}
	\centering
	\includegraphics[width=1.0\columnwidth, angle=0]{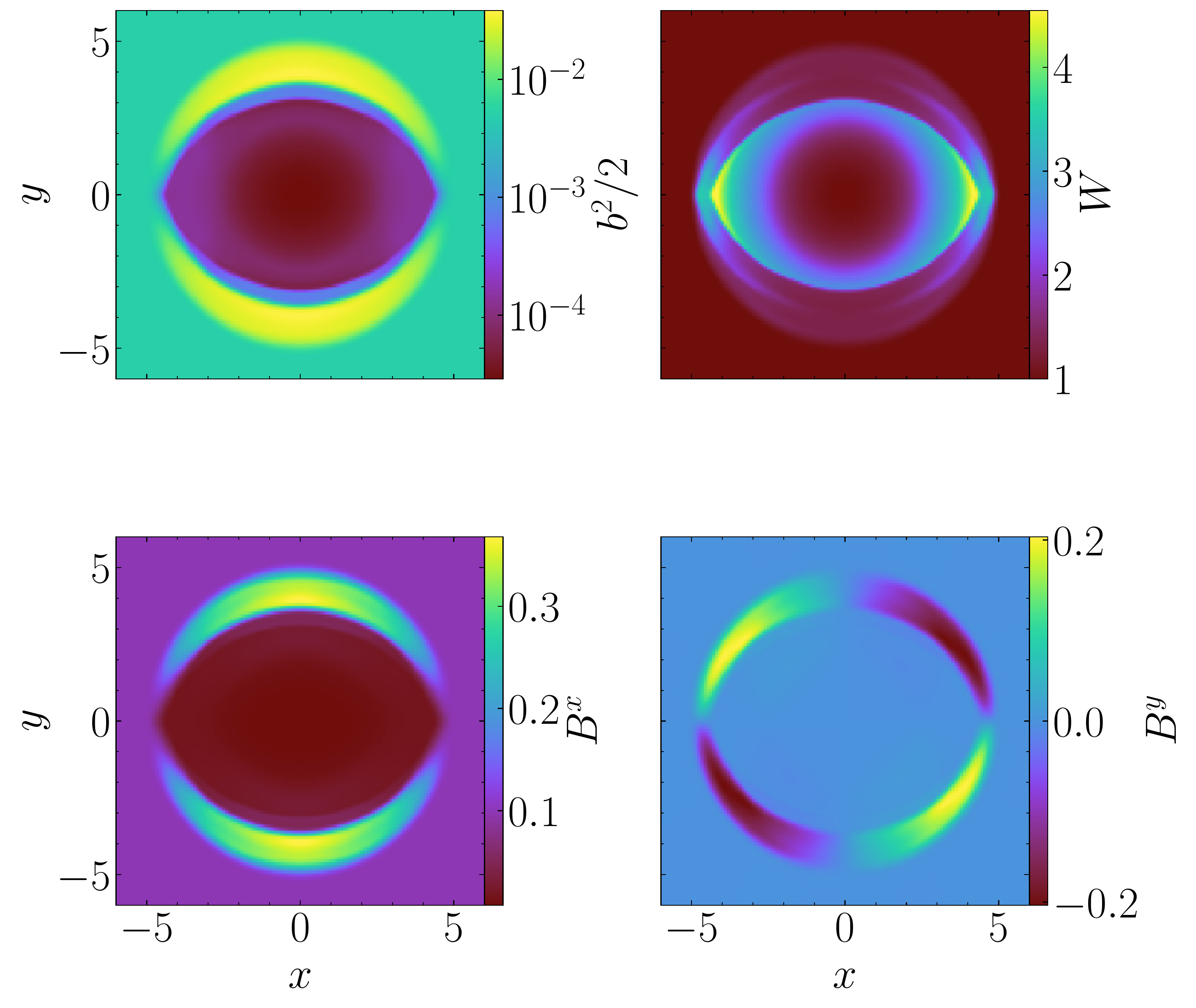}
	\caption{The two-dimensional profile for the cylindrical blast wave of the magnetic field strength $B^i B_i$ (\emph{upper left}), Lorentz factor $W$(\emph{upper right}), $B^x$(\emph{lower left}), $B^y$(\emph{lower right}) at $t = 4.0$.
		}
	\label{fig:MHD_cylindrical_blast_wave_2d}	
\end{figure}

\begin{figure}
	\centering
	\includegraphics[width=1.0\columnwidth, angle=0]{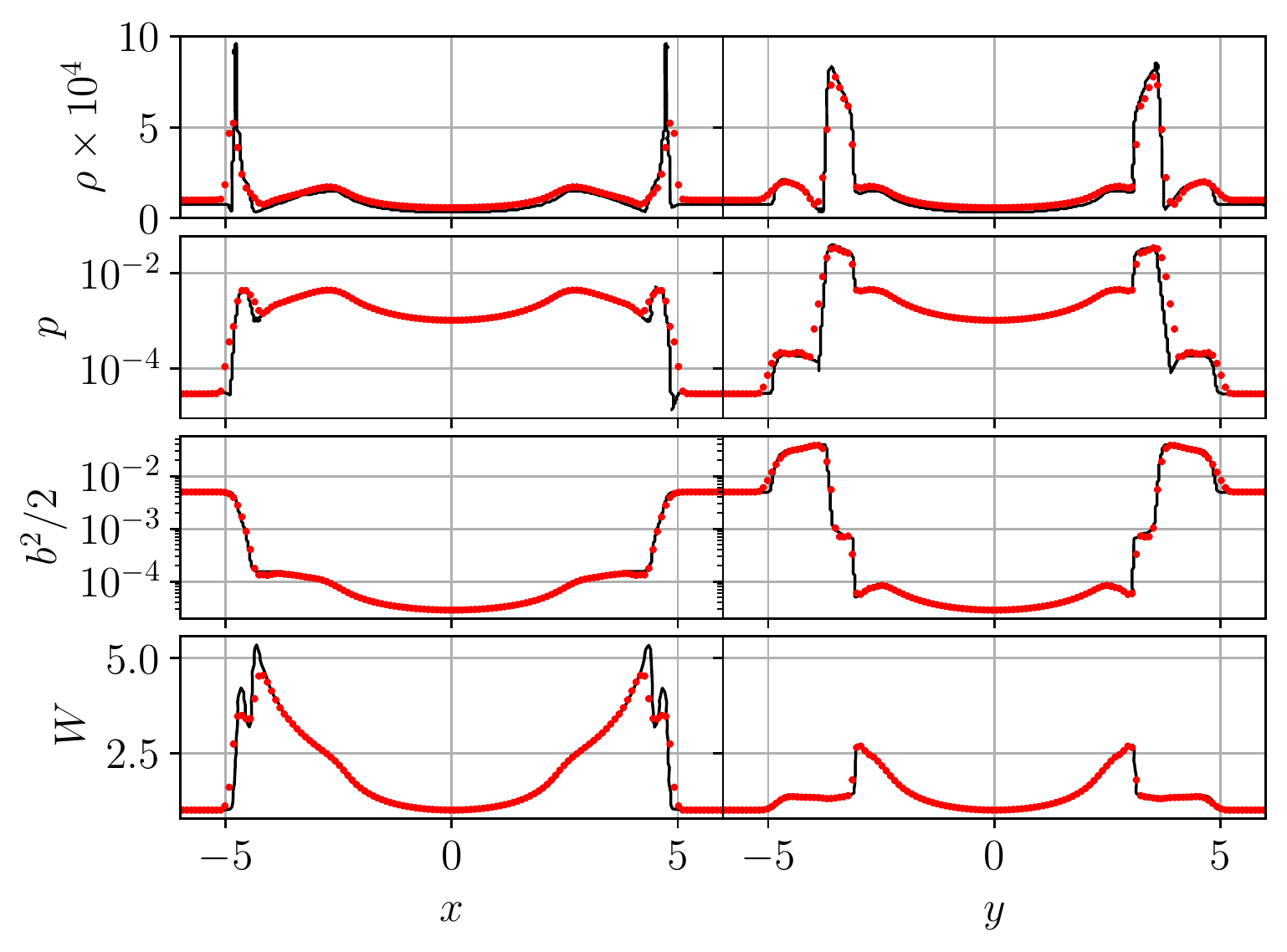}
	\caption{One-dimensional slices along the $x-$ axis (\emph{left {column}}) and $y-$ axis (\emph{right {column}}) for the density $\rho$ (\emph{{top row}}), {pressure} $p$ (\emph{{second row}}), magnetic pressure $b^2 / 2$ (\emph{{third row}}) and Lorentz factor $W$ (\emph{{fourth row}}) for the MHD cylindrical blast wave test at $t = 4.0$.
	The red dots show the numerical results obtained by \texttt{Gmunu}{, which agree with the reference solutions (black solid lines) \citet{2014CQGra..31a5005M}}.
		}
	\label{fig:MHD_cylindrical_blast_wave_2d_slices}	
\end{figure}

Figure~\ref{fig:MHD_cylindrical_blast_wave_2d} shows the two-dimensional profile of the magnetic field strength $B^i B_i $ ,$B^x$, $B^y$ and the Lorentz factor $W$ at $t = 4.0$. 
To compare our results with other groups {(e.g. \citet{2014CQGra..31a5005M})} {in} more detail, we also plot one-dimensional slices along the $x-$ and $y-$ axes for the rest mass density $\rho$, pressure $p$, magnetic pressure $b^2/2$ and the Lorentz factor $W$ at $t=4$, as shown in fig. \ref{fig:MHD_cylindrical_blast_wave_2d_slices}.
In this test, the numerical results obtained by \texttt{Gmunu}, which agree with the reference solutions {\citet{2014CQGra..31a5005M}}.

\subsubsection{Loop advection}
The advection of a weakly magnetized loop is a well known test to examine divergence-control technique in a MHD code.
This test is performed on an uniform background with $\rho = 1$, $p = 1$, $v^x=0.2$ and $v^y = 0.1$.
The initial condition of the magnetic field $B^i$ is given as
\begin{align}
	B^i = 
	\begin{cases}
		\left( -A_0 y / r, A_0 x / r, 0 \right)	&	\text{if } r < R ,\\
		\left( 0, 0, 0 \right)	&	\text{if } r > R ,
	\end{cases}
\end{align}
where $R=3$ is the radius of the advecting magnetic loop, $r \equiv \sqrt{x^2 + y^2}$ and $A_0$ is chosen to be $10^{-3}$.
We consider an ideal-gas equation of state $ p = (\Gamma-1)\rho \epsilon$ with $\Gamma = 4/3$.
The computational domain is set to be \emph{periodic} at all boundaries and covers the region $-1 \leq x \leq 1$ and $-0.5 \leq y \leq 0.5$ with the base grid points $n_x \times n_y = 32 \times 16$ and allowing 5 AMR {levels} (i.e., an effective resolution of $512 \times 256 $).
{
Note that in this test, the refinement is determined based on the strength of the magnetic field.
In particular, the grid is refined if the square of the magnetic field $B^iB_i$ is larger than $10^{-10}$ while it is coarsen otherwise.
}

\begin{figure}
	\centering
	\includegraphics[width=1.0\columnwidth, angle=0]{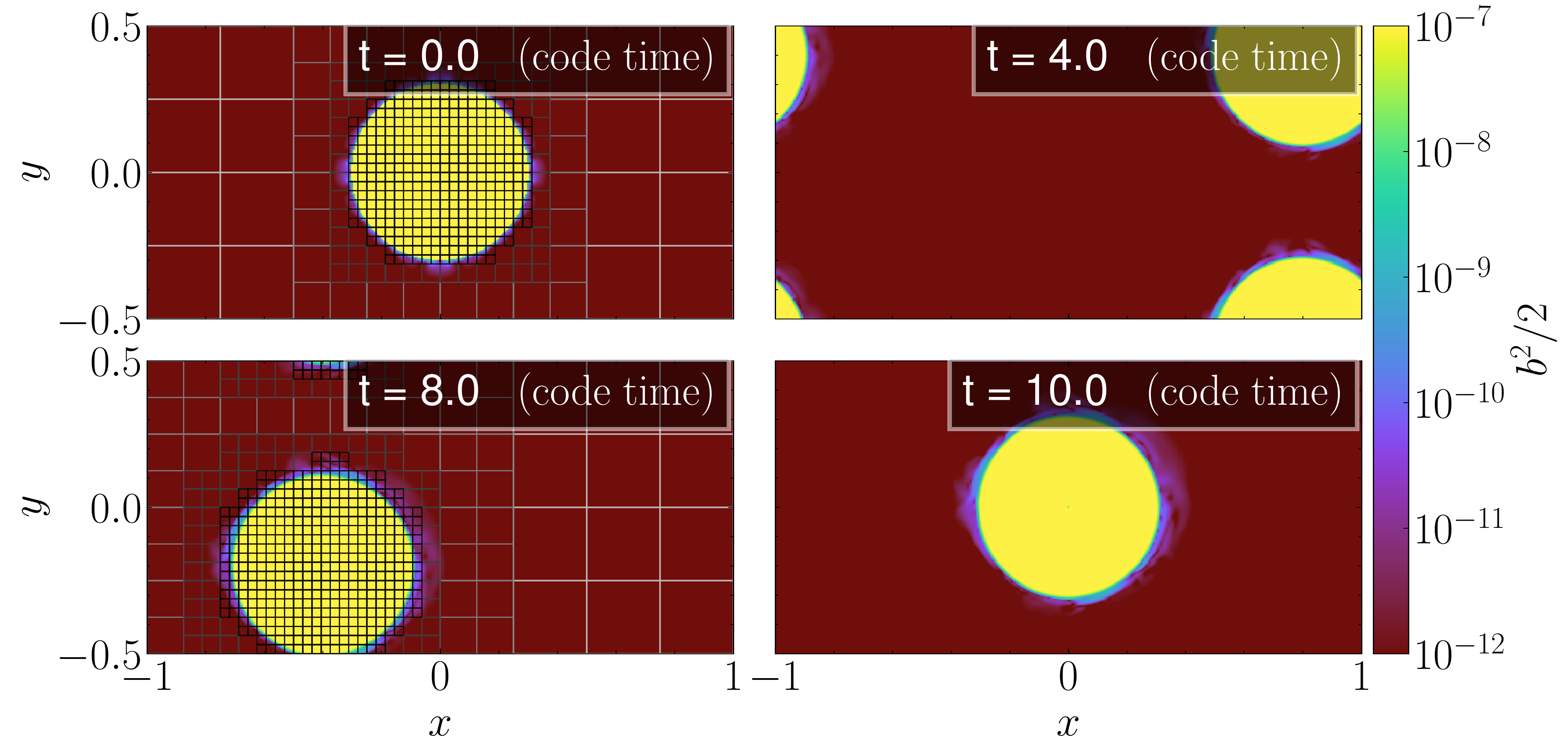}
	\caption{The evolution of the magnetic pressure $b^2/2$ for the loop advection test at various time slices.
	         The AMR blocks of $8 \times 8$ cells are shown on the left panels.
		 The shape of the loop is preserved well at $t=10$, where the magnetic field has translated with 1 cycle.
		}
	\label{fig:MHD_loop_advection_2d_time_series}	
\end{figure}

Figure~\ref{fig:MHD_loop_advection_2d_time_series} gives an example of the evolution of the magnetic pressure $b^2/2$ for the loop advection test at different {times}.
The shape of the loop is preserved well at $t=10$, where the magnetic field has translated with 1 cycle.

\begin{figure}
	\centering
	\includegraphics[width=1.0\columnwidth, angle=0]{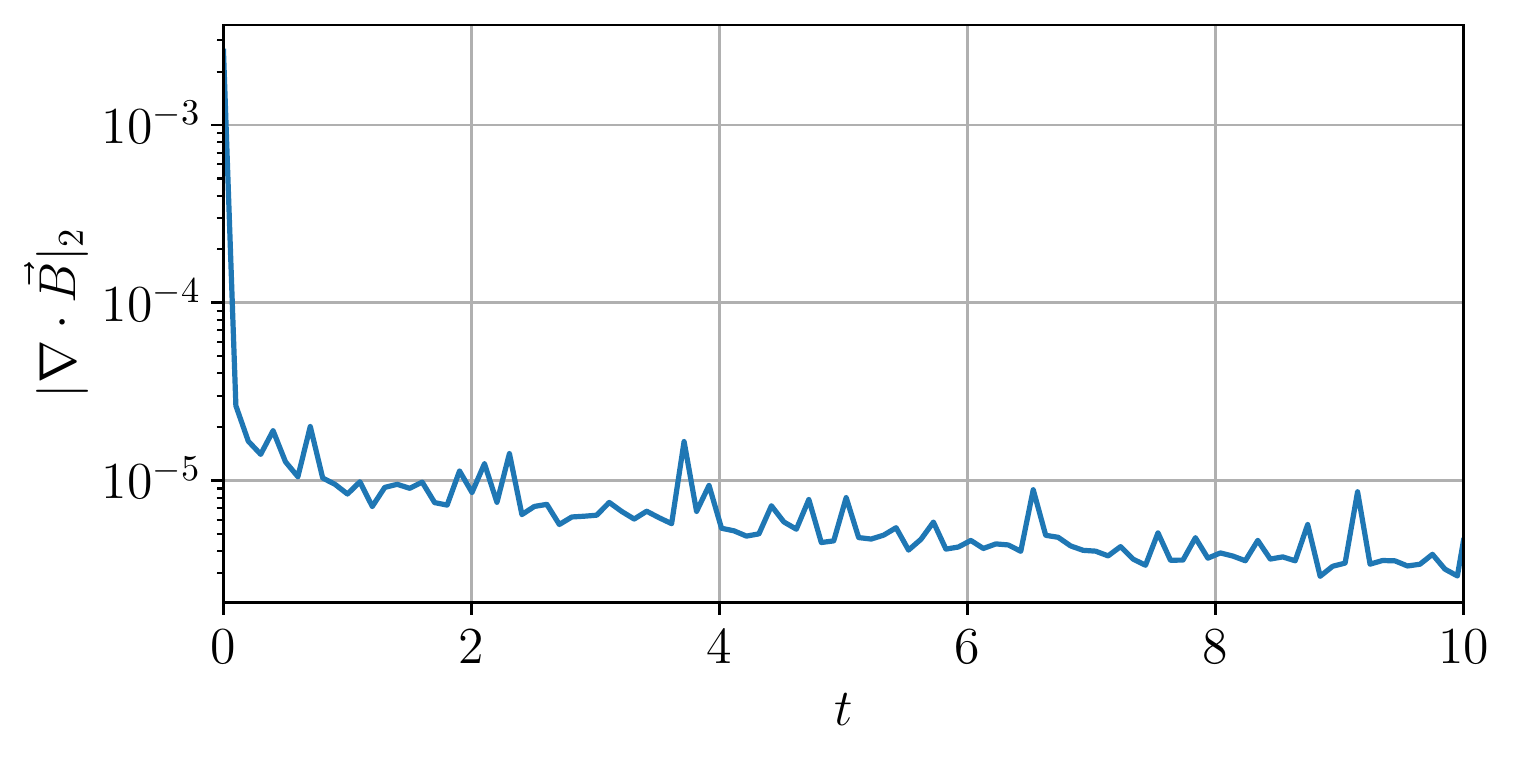}
	\caption{ The $L_2$-norm of $\nabla \cdot \vec{B}$ versus time for the advected field loop test.
		The $|\nabla \cdot \vec{B}|_2$ is suppressed to lower than $10^{-5}$ immediately when the evolution started and is well controlled for the rest of the evolution.
		}
	\label{fig:MHD_loop_advection_2d_L2_divB}	
\end{figure}

Figure~\ref{fig:MHD_loop_advection_2d_L2_divB} shows the {evolution of the} $L_2$-norm of $\nabla \cdot \vec{B}$, defined as 
\begin{equation}
	|\nabla \cdot \vec{B}|_2 \equiv \sqrt{\frac{1}{V} \int |\nabla \cdot \vec{B}|^2 dV},
\end{equation}
{which can be used to indicate the validity of the divergence-control.}
The $L_2$-norm of $\nabla \cdot \vec{B}$ is suppressed to lower than $10^{-5}$ immediately when the evolution started and is well controlled for the rest of the evolution.
Overall, the elliptic divergence cleaning {(see section \ref{sec:elliptic_cleaning})} works well to control monopole errors for this test case.

{
It is worth to point out that, although elliptic cleaning of the divergence of magnetic field is technically acausal, no artefacts from these tests are observed.
The main reason is that, the magnetic monopole of the initial setup of the runs are below the machine precision, some are even identically equal to zero.
Although the divergence of magnetic field is small, it is cleaned by using elliptic solver at each time step.
While this act is technically acausal, removing negligible and non-physical parts of the magnetic field at each time step will not significantly affect the evolution of the systems.
Artefacts may arises when the divergence of magnetic field is non-negligible before cleaning, however, this situation itself is non-physical.
As mentioned, more detailed and systematic studies of different divergence handling approaches is planned in the future.
}

\subsection{\label{sec:GRMHD_tests}General relativistic (magneto-)hydrodynamics in dynamical spacetime}
\subsubsection{\label{sec:user_refinement_criteria}{Refinement criteria}}
{
In addition to the error estimators mentioned above, user-defined additional conditions are available in \texttt{MPI-AMRVAC}.
Note that all the error estimators discussed are mostly local (block-based calculations), applying them only may not sufficient to have a optimized mesh refinement.
For example, numerical studies suggest that the ratio of the radius of the compact objects $R$ and the grid size $\Delta x$ should be larger than 50 for neutron stars \citet{doi:10.1142/9692}.
In our experience, it is hard to simultaneously resolve the interior of the star and the star surface properly by applying a local estimator only. 
This is because the density gradients at the star surface could be extremely large.
As a result, the interior of the star seems sufficiently smooth (refining is not necessary) compare with the star surface.
Moreover, low density matters often eject from the star surface into the vacuum, which will trigger local error estimator to refine the mesh in these regions.
The computational cost are wasted if those ejecta is not the main focus of the studies.
Thus, it is usually beneficial to include addition conditions on top of the local error estimators if prior knowledges of the system are available.
}

{
The lapse function $\alpha$ can be used as an indicator for the grid refinement \citet{2002ApJ...572L..39S, 2004PhRvD..69h4024S}.
We defined a relativistic gravitational potential $\Phi := 1 - \alpha$.
Since $\Phi$ is approximately proportional to $M/R$, $\Phi^{-1}$ can be used as a measure of the characteristic length scale.
The grid resolution can then be optimized by assigning the valid range of the potential for all available levels, e.g., $\Phi_l$.
This approach is adopted in all our general relativistic simulations to be discussed below.
The grid refinement used in this work is the following:
For any $\Phi$ larger than the maximum potential $\Phi_{\text{max}}$ (which is set as 0.2 in this work), the block is set to be finest.
While for the second finest level, the same check is performed with a new maximum potential which is half of the previous one, so on and so forth.
The grid is updated every 500 timesteps.
}
\subsubsection{\label{sec:BU8_2D}Stability of a rapidly rotating neutron star}
Here we study the evolution of a stable rapidly rotating neutron star with {a} dynamical background metric.
In this test, we consider a uniformly rotating model which is constructed with the polytropic equation of state with $\Gamma = 2$ and $K=100$ with central rest-mass density $\rho_c = 1.28 \times 10^{-3}$ and the angular velocity $\Omega = 2.633 \times 10^{-2}$ (in $c=G=M_\odot=1$ unit), which is also know as ``BU8'' in the literature \citet{2006MNRAS.368.1609D, XCFC}.
The initial neutron star model is generated with the open-source code \texttt{XNS} \citet{2011A&A...528A.101B,XNS1,XNS2,XNS3}.
The computational domain covers $0 \leq r \leq 30$, $0 \leq \theta \leq \pi/2$ with the resolution $n_r \times n_\theta = 640 \times 64$.
This test problem is simulated with the ideal-gas equation of state {$P = (\Gamma - 1)\rho\epsilon$ with $\Gamma = 2$}.
Long time evolution of this model is demanding since the rotational rate is close to the mass shedding limit.
{While maintaining this model stably is formidable, we challenge the robustness of our code with the use of the positivity preserving limiter by setting an extremely low ``atmosphere'' density $\rho_\text{atmo} = 10^{-20}$ (which is {below} \emph{machine precision}) and simulate the system with a 5-th order reconstruction method MP5.}
As in \citet{2020CQGra..37n5015C}, in order to increase the size of the time steps in our simulations, we treat $0 < r < 0.4$ as a spherically symmetric core (i.e., only radial motions are allowed).

\begin{figure}
	\centering
	\includegraphics[width=1.0\columnwidth, angle=0]{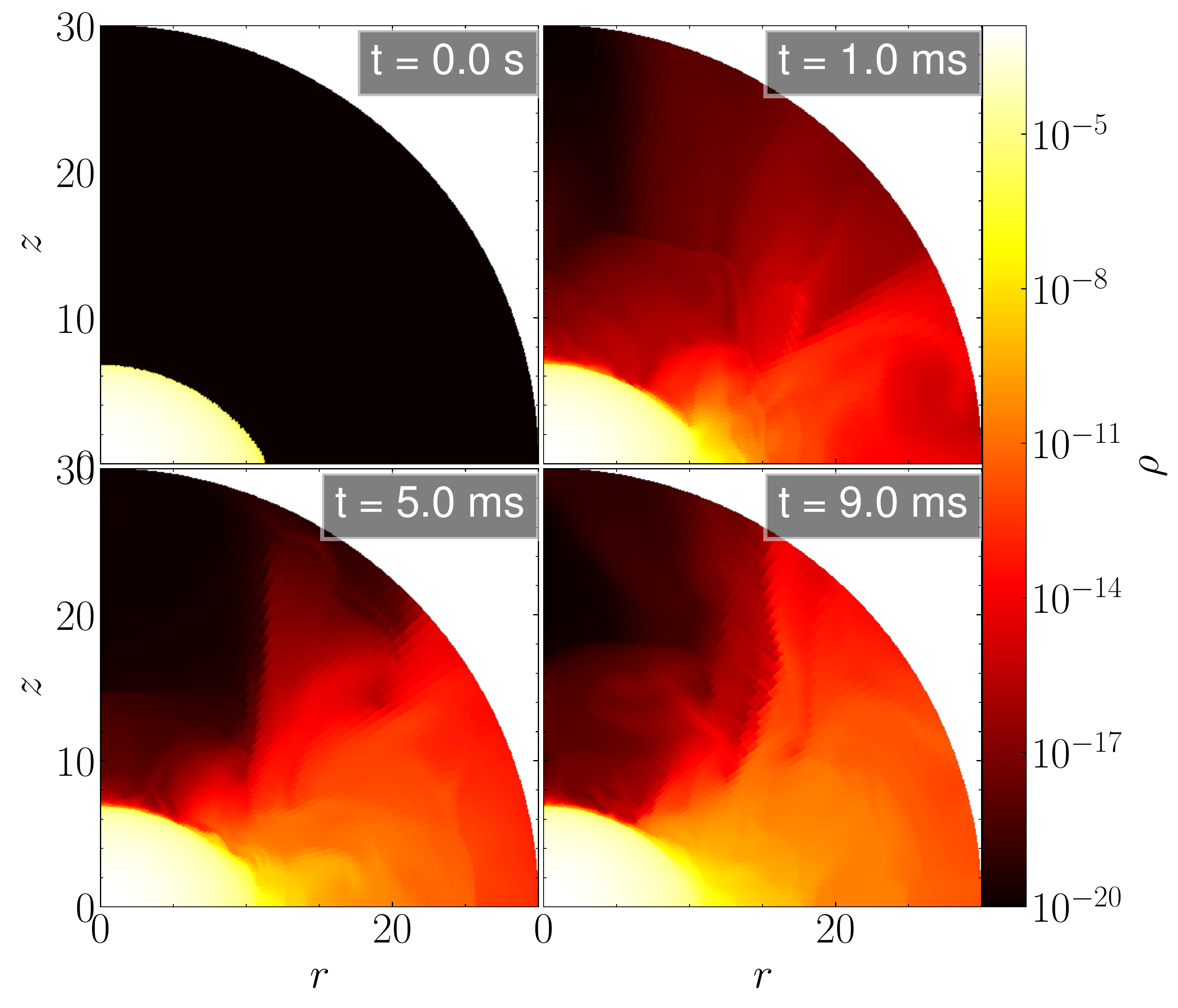}
	\caption{Example of the evolution of the density $\rho$ {of} the rapidly rotating neutron star BU8 at various time slices.
		As shown in the density map, since the ``atmosphere'' density $\rho_\text{atmo}$ is set to be $10^{-20}$, the low-density fluid (e.g. $ \rho \sim 10^{-9}$ to $ 10^{-13}$, which are the typical values of the ``atmosphere'' in the literature) is free to be evolved without crashing the code. 
		This can be achieved with the positivity preserving limiter (see section \ref{sec:PPlimiter}) and could significantly avoid violations of the conservation properties at the neutron star surface.
		}
	\label{fig:HD_BU8_time_series}	
\end{figure}

\begin{figure}
	\centering
	\includegraphics[width=1.0\columnwidth, angle=0]{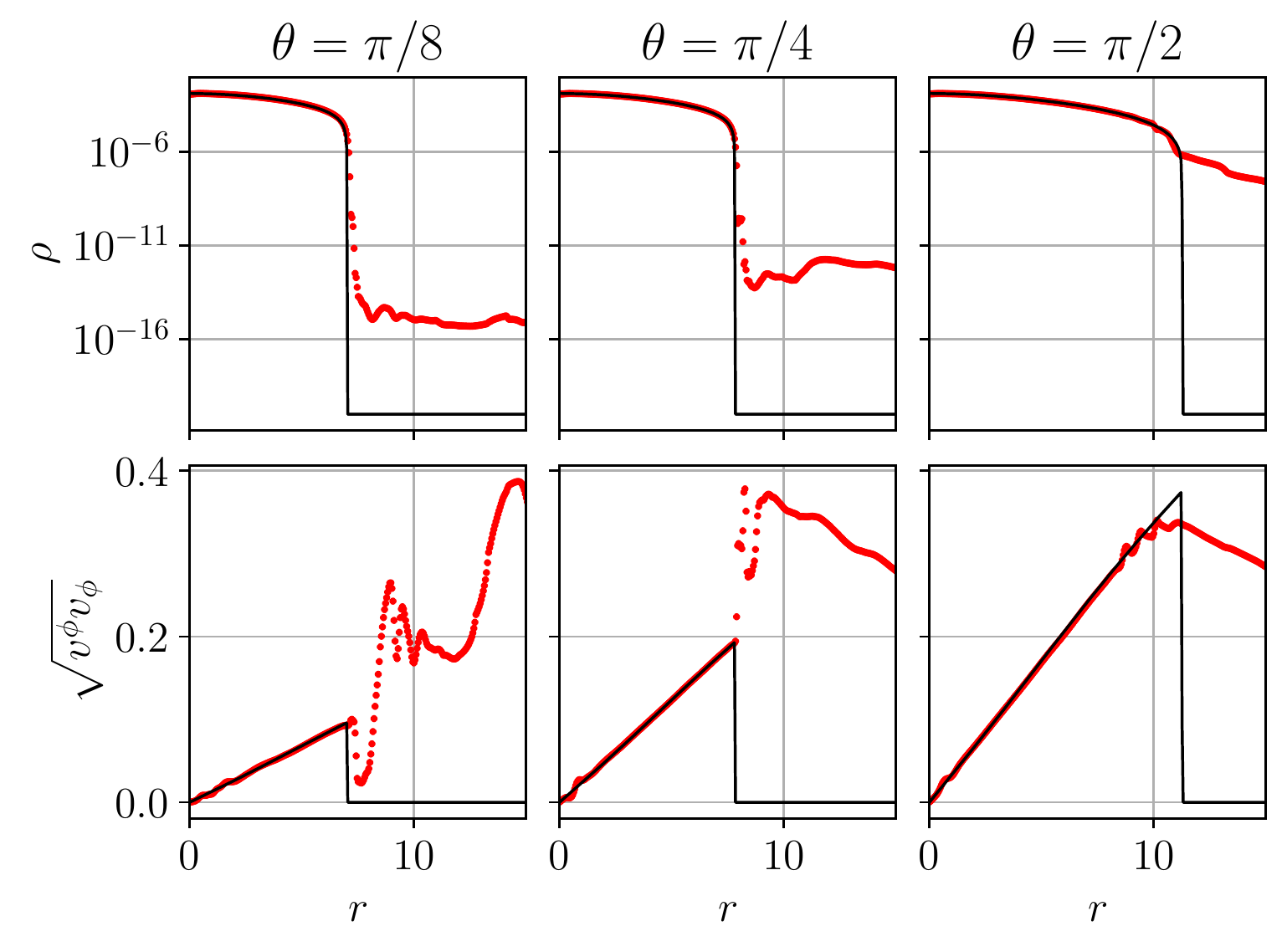}
	\caption{One-dimensional slices of the rapidly rotating neutron star BU8 along the $\theta = \pi / 8$ (\emph{left}), $\theta = \pi / 4$ (\emph{middle}) and $\theta = \pi / 2$ (\emph{right}) for the density $\rho$ (\emph{upper}) and the rotational velocity ${\sqrt{v_\phi v^\phi}}$ (\emph{lower}).
The black solid lines show the initial profiles while the red dots show the profiles $t = 9$ ms.
		}
	\label{fig:HD_BU8_slices}	
\end{figure}

With the positivity preserving limiter and the robust recovery of primitive variables scheme, \texttt{Gmunu} evolve such demanding systems stably even with an extremely low density of ``atmosphere'' up to at least $t=9$ ms without crashing the code.
Figure~\ref{fig:HD_BU8_time_series} gives an example of the evolution of this rapidly rotating neutron star model BU8 at different time.
Figure~\ref{fig:HD_BU8_slices} shows one-dimensional slices of the rapidly rotating neutron star BU8 along the $\theta = \pi / 8$, $\theta = \pi / 4$ and $\theta = \pi / 2$ for the density $\rho$ and the rotational velocity ${\sqrt{v_\phi v^\phi}} $ respectively.
The density and the velocity profiles are maintained well except that some low density ``atmosphere'' $ \rho \sim 10^{-9}$ to $ 10^{-17}$ is surrounding the neutron star.

\begin{figure}
	\centering
	\includegraphics[width=1.0\columnwidth, angle=0]{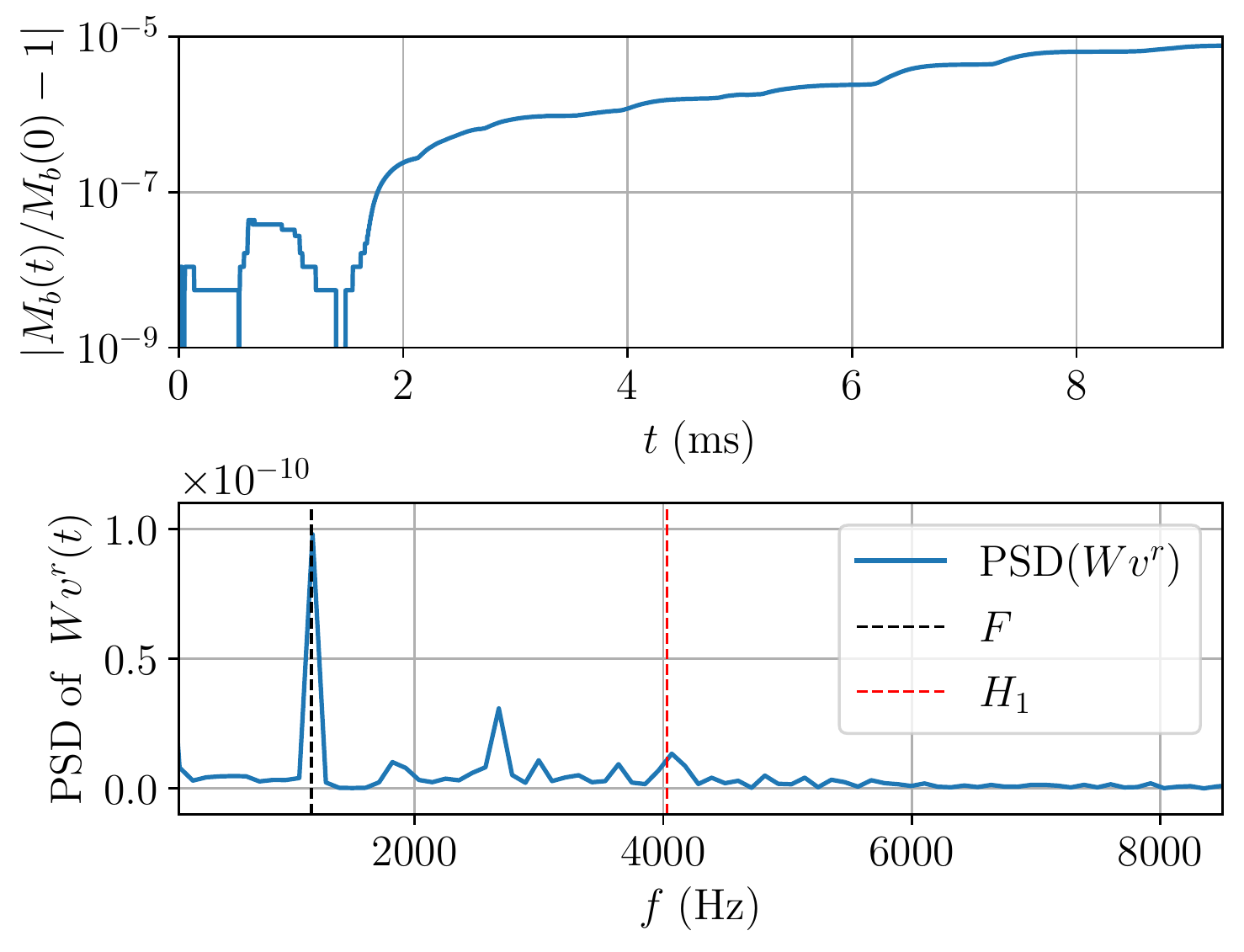}
	\caption{
                \emph{Upper panel}: The relative variation of the rest mass $M_b$ of the rapidly rotating neutron star BU8 in time.
                \emph{Lower panel}: The power spectral density of the radial velocity $Wv^r (t)$ at $r = 5$, $\theta = \pi/4$ (inside the neutron star).
                The vertical lines represent the known and well-tested eigenmode frequencies \citet{2006MNRAS.368.1609D}.
		Even for such rapidly rotating neutron star BU8 with extreme simulations settings (i.e., $\rho_\text{atmo} = 10^{-20}$ with MP5 reconstruction), \texttt{Gmunu} is able to maintain the profile up to 9 ms and the relative variation of the rest mass of the order {$10^{-5}$}.
		}
	\label{fig:HD_BU8_central}	
\end{figure}

To illustrate the conservation properties, we monitor the total rest mass $M_b$ of the whole system, where the rest mass $M_b$ is given by
\begin{equation}
	M_b = \int_{\Sigma_t} \psi^6 \rho W \sqrt{\hat{\gamma}} d^3 x.
\end{equation}
The upper panel of figure \ref{fig:HD_BU8_central} shows the relative variation of the rest mass $M_b$ in time.
Even for such rapidly rotating neutron star BU8 with extreme configurations, \texttt{Gmunu} is able to maintain the profile up to 9 ms and the relative variation of the rest mass of the order {$10^{-5}$}.
As an another indicator for {the validity of the} code, the lower panel of figure \ref{fig:HD_BU8_central} shows the power spectral density of the radial velocity $Wv^r (t)$ at $r = 5$, $\theta = \pi/4$ (inside the neutron star), which {agrees} with the well-tested eigenmode frequencies \citet{2006MNRAS.368.1609D}.

\begin{figure}
	\centering
	\includegraphics[align=c, width=\columnwidth, angle=0]{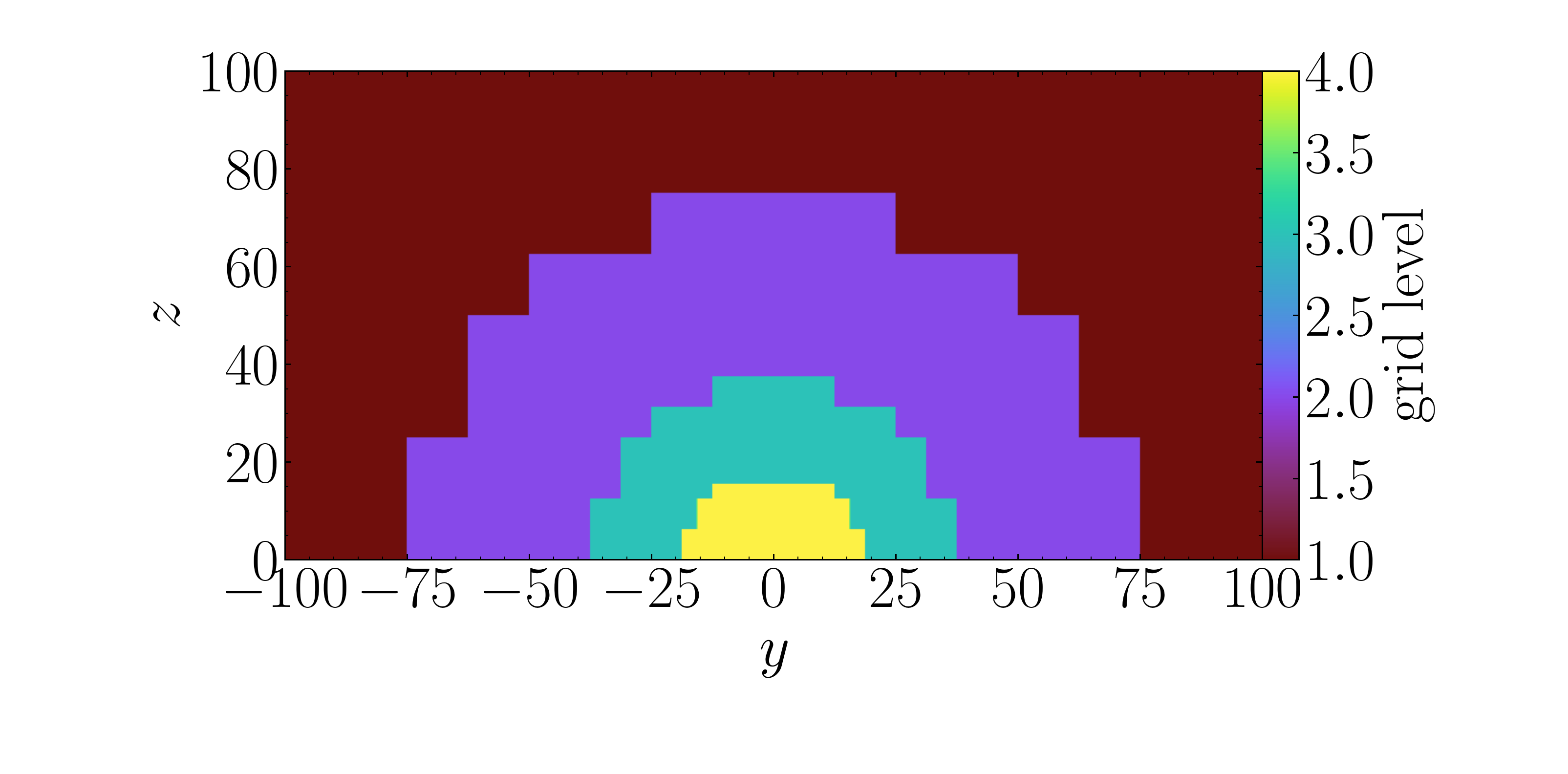}
	\includegraphics[align=c, width=\columnwidth, angle=0]{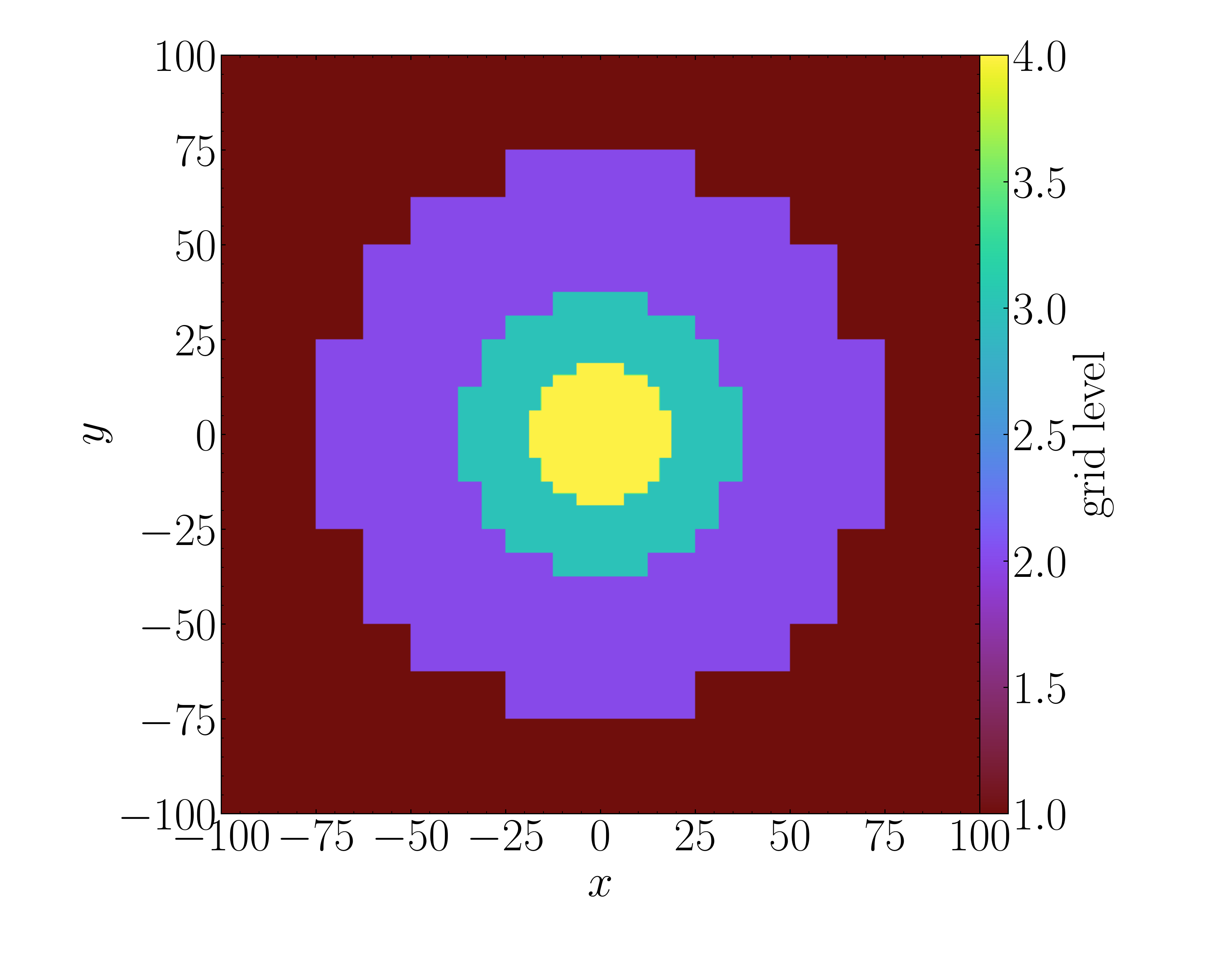}
	\caption{
		{
		The projection of grid levels along $x$-axis (\emph{upper panel}) and $z$-axis (\emph{lower panel}) in Cartesian coordinates.
		The computational domain covers $[-100,100]$ for both $x$,$y$ while $z \in [0,100]$, with the resolution {$N_x \times N_y \times N_z = 64 \times 64 \times 32$} and allowing 5 AMR level {(an effective resolution of $1024 \times 1024 \times 512$)}.
		}
		}
	\label{fig:HD_BU8_3D_grid_level}	
\end{figure}

\begin{figure}
	\centering
	\includegraphics[align=c, width=\columnwidth, angle=0]{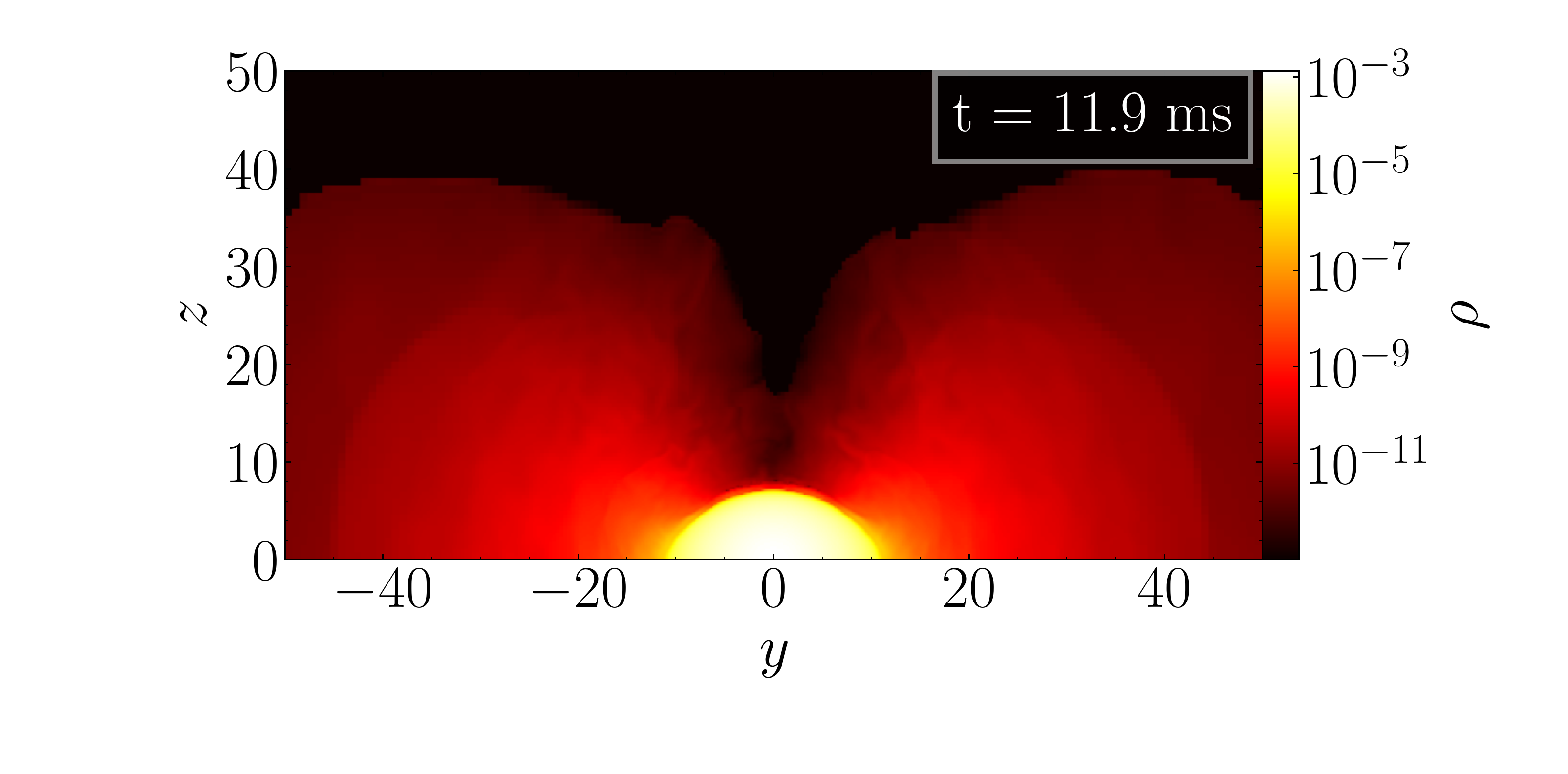}
	\includegraphics[align=c, width=\columnwidth, angle=0]{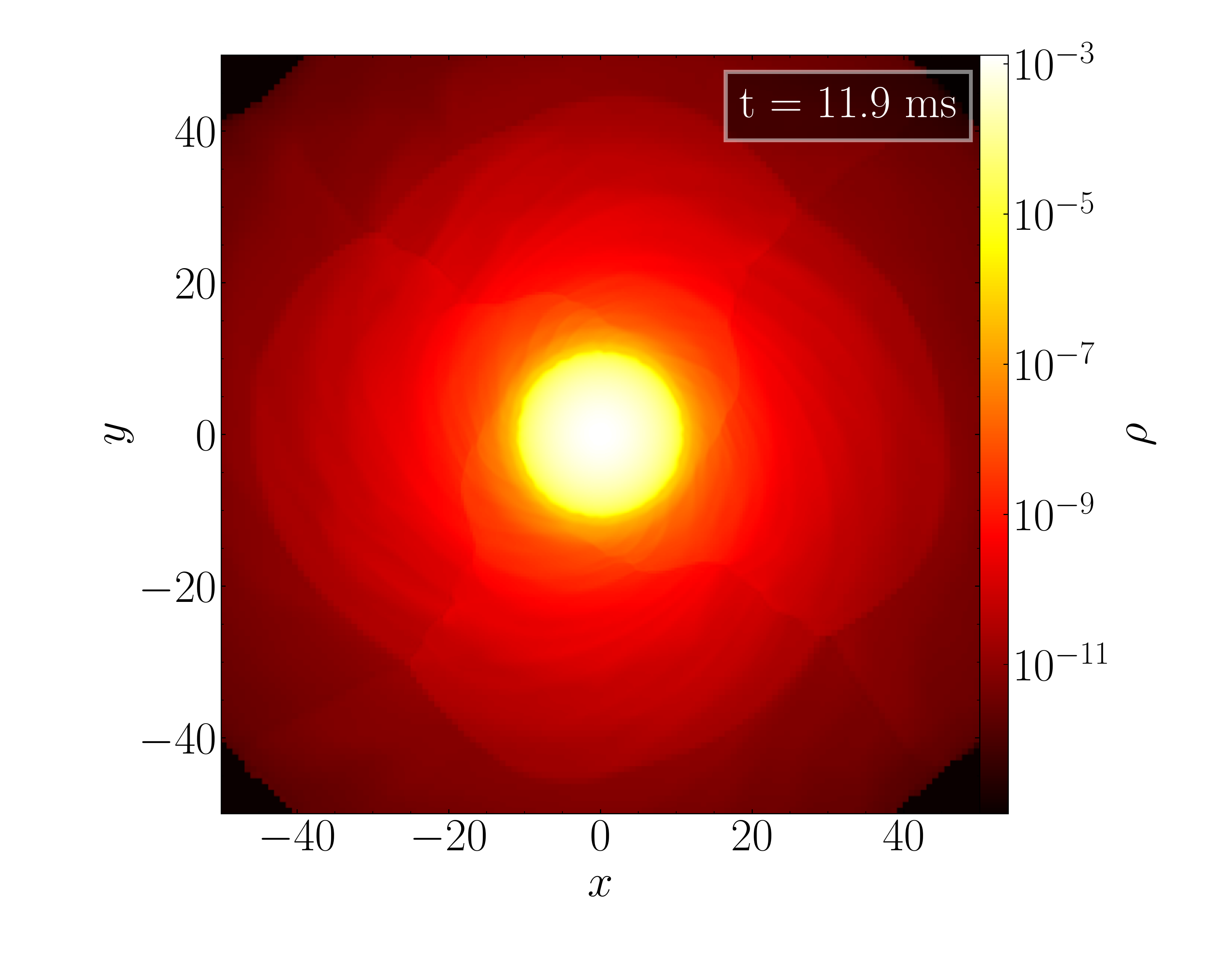}
	\caption{
		{
		The projection of density profile along $x$-axis (\emph{upper panel}) and $z$-axis (\emph{lower panel}) of a rapidly rotating neutron star in {Cartesian coordinates} with the annotated mesh lines at {$t=11.9$} ms.
		}
		}
	\label{fig:HD_BU8_3D_rho}	
\end{figure}

{
We simulated the same model in Cartesian coordinates $(x,y,z)$.
The computational domain covers $[-100,100]$ for both $x$,$y$ while $z \in [0,100]$, with the resolution {$N_x \times N_y \times N_z = 64 \times 64 \times 32$} and allowing 5 AMR level {(an effective resolution of $1024 \times 1024 \times 512$)}.
Figure~\ref{fig:HD_BU8_3D_grid_level} shows the grid level at different locations in the computational domain while figure~\ref{fig:HD_BU8_3D_rho} shows the projection of density profile of a rapidly rotating neutron star.
}

\begin{figure}
	\centering
	\includegraphics[width=\columnwidth, angle=0]{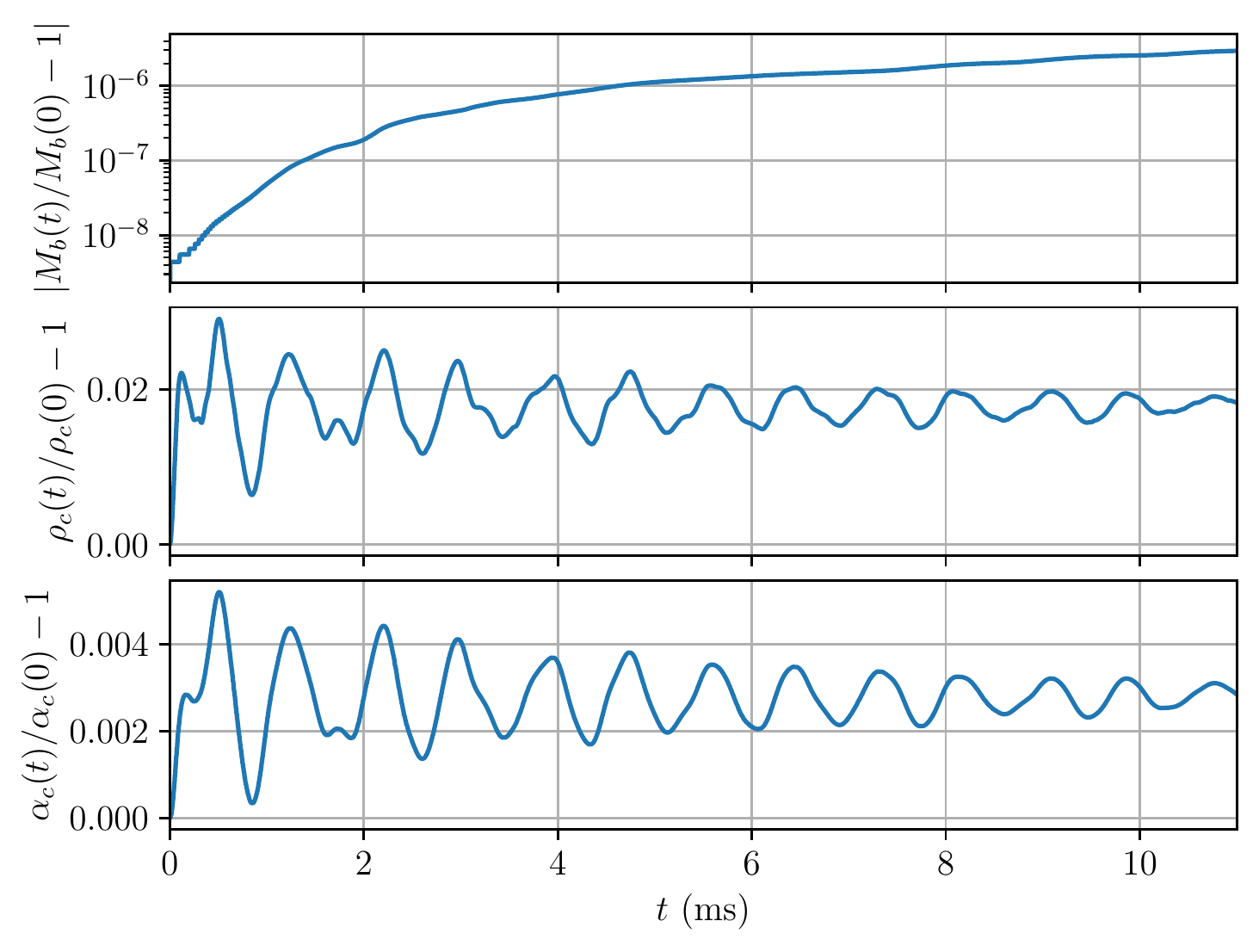}
	\caption{
		{
                \emph{Upper panel}: The relative variation of the rest mass $M_b$ in time.
		The conservation of the rest mass $M_b$ is preserved remarkably well from $t=0$ ms to $t=12$ ms where the relative variation is of the order {$10^{-5}$}.
                \emph{Middle panel}: The relative variation of the density $\rho_c$ in time.
		\emph{Lower panel}: The relative variation of the lapse function $\alpha_c$ in time.
		}
		}
	\label{fig:HD_BU8_3D_central}	
\end{figure}

\begin{figure}
	\centering
	\includegraphics[width=\columnwidth, angle=0]{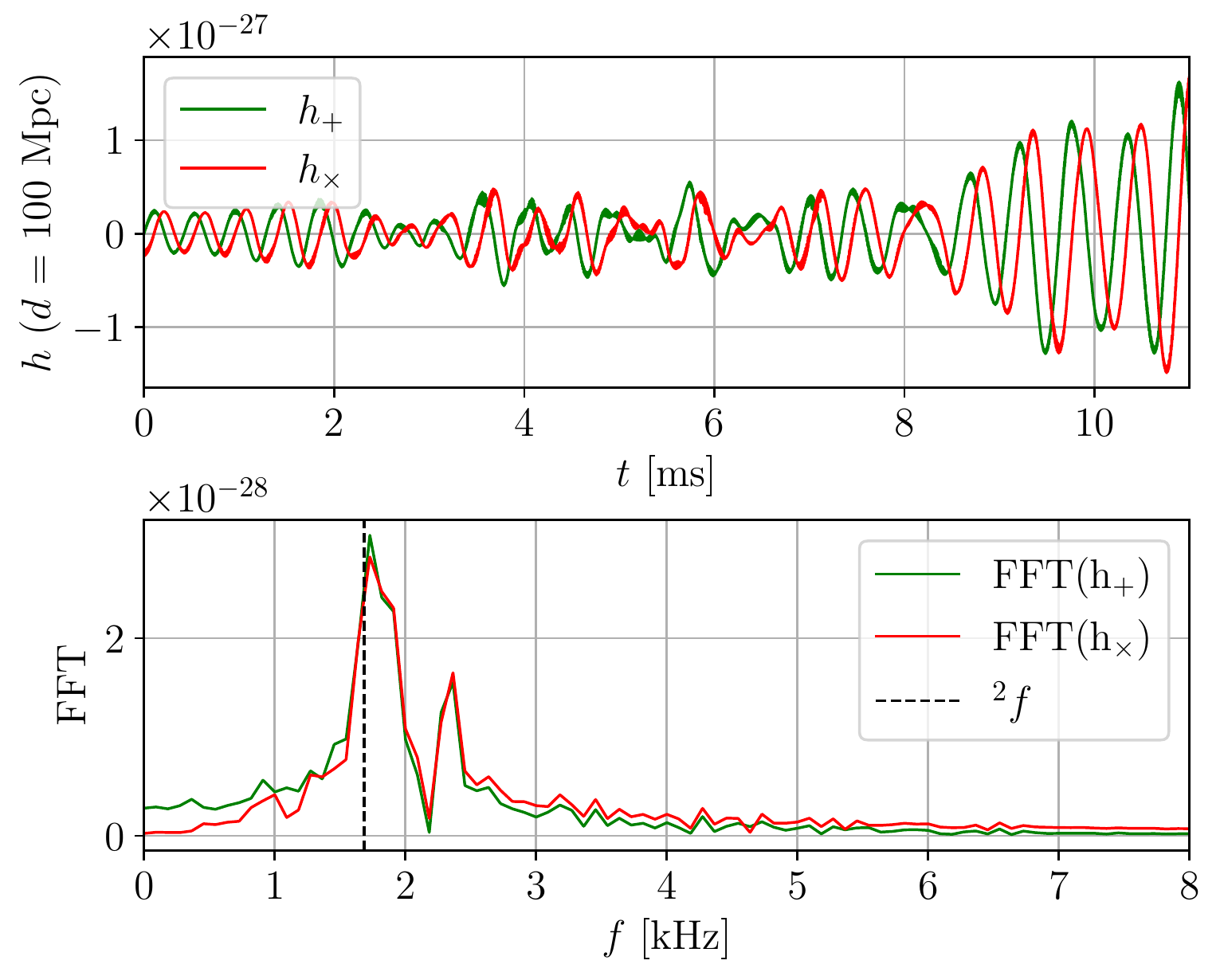}
	\caption{
		{
                \emph{Upper panel}: Gravitational waves extracted from the rapidly rotating neutron star BU8 in time at distance $d = 100 \rm{Mpc}$ at the equator.
                \emph{Lower panel}: The Fast Fourier Transform the gravitational waves.
		The vertical lines represent the known and well-tested eigenmode (${}^2f$ mode) frequencies \citet{2006MNRAS.368.1609D}.
		}
		}
	\label{fig:HD_BU8_3D_GW}	
\end{figure}

{
Figure~\ref{fig:HD_BU8_3D_central} shows the evolution of the Rapidly rotating neutron star BU8 in Cartesian coordinate.
\texttt{Gmunu} is able to maintain the profile up to 100 ms and the relative variation of the rest mass of the order {$10^{-5}$}.
}
{
Although conformally flat approximation is a gravitational-waveless approximation to general relativity, gravitational waves can still be extracted by using quadrupole formula.
Figure~\ref{fig:HD_BU8_3D_GW} shows the gravitational waves extracted at distance $d = 100 \rm{Mpc}$ at the equator in time domain and also in frequency domain, the dominating non-radial ${}^2f$ mode which agrees with the well-tested eigenmode frequencies \citet{2006MNRAS.368.1609D}.
}

\subsubsection{Differentially rotating strongly magnetized neutron star}\label{sec:magns}
Here we study the evolution of a differentially rotating strongly magnetized equilibrium neutron star.
As there are no similar studies in the literature except \citet{XECHO}, we use the same equilibrium model as in \citet{XECHO} here.
In this test, we construct an equilibrium model with {a} polytropic equation of state with $\Gamma = 2$ and $K=100$ with central rest-mass density $\rho_c = 1.28 \times 10^{-3}$.
The neutron star is differentially rotating with $\Omega_c = 2.575 \times 10^{-2}$, $A^2 = 70$ and is magnetized with magnetic polytropic index $m = 1$ and magnetic coefficient $K_m = 3$.
Here we note that this is a strong toroidal magnetic field, $\sim 5 \times 10^{17}$G inside the neutron star, which is roughly 10\% of the total internal energy.
This test problem is simulated with the polytrope equation of state with $\Gamma = 2$ and $K=100$.

\begin{figure}
	\centering
	\includegraphics[width=1.0\columnwidth, angle=0]{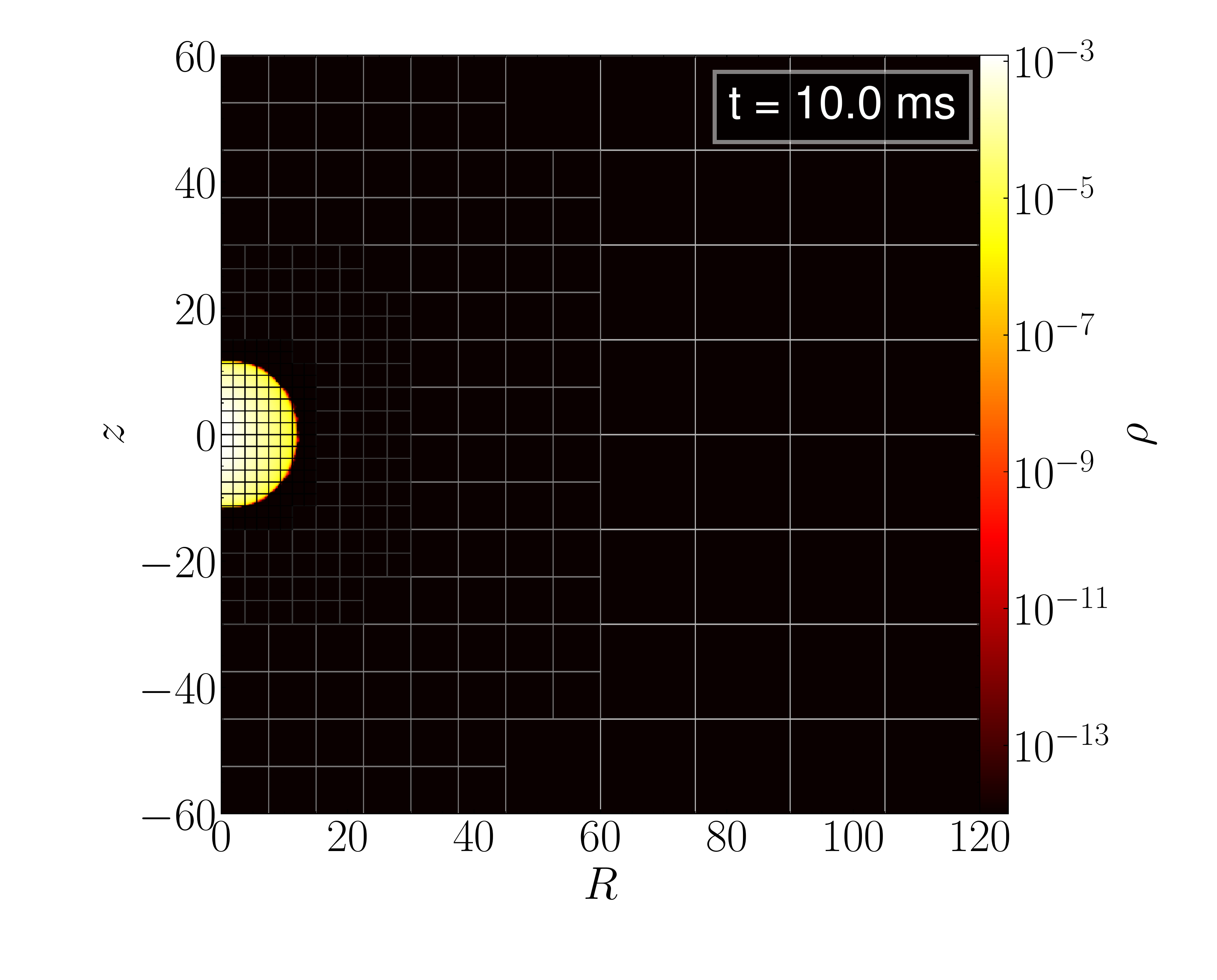}
	\caption{
                The density profile of a differentially rotating strongly magnetized equilibrium neutron star in cylindrical coordinates with the annotated mesh lines at $t=10$ ms.
		The computational domain covers $0 \leq R \leq 120$ and $-120 \leq z \leq 120$, with the resolution {$n_R \times n_z = 32 \times 64$} and allowing {5 AMR levels}.
                At the outer region ($R \sim 100$), the size of a block (containing $8 \times 8$ cells) is almost the size of the neutron star.
		}
	\label{fig:MHD_NS_cylindrical_2D_mesh}	
\end{figure}

We simulate this initial model in 2-dimensional cylindrical coordinates $(R, z, \varphi)$, where the computational domain covers $0 \leq R \leq 120$ and $-120 \leq z \leq 120$, with the resolution {$n_R \times n_z = 32 \times 64$} and allowing {5 AMR levels} (i.e., an effective resolution of {$512 \times 1024 $}).
As an example, figure \ref{fig:MHD_NS_cylindrical_2D_mesh} shows the density profile with the annotated mesh lines at $t=10$ ms.

\begin{figure}
	\centering
	\includegraphics[width=1.0\columnwidth, angle=0]{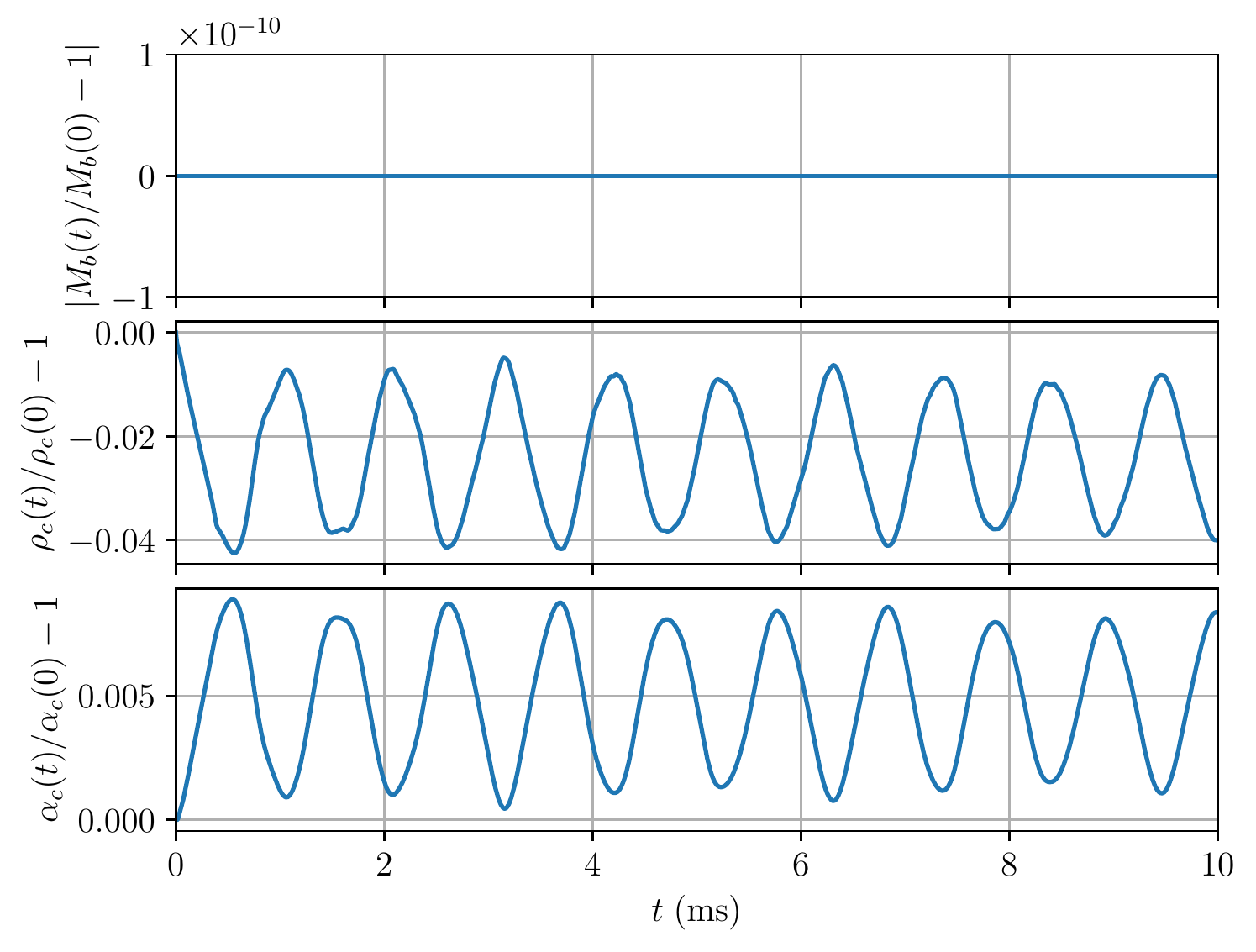}
	\caption{
                \emph{Upper panel}: The relative variation of the rest mass $M_b$ in time.
		The conservation of the rest mass $M_b$ is preserved remarkably well from $t=0$ ms to $t=10$ ms where the relative variation is {zero (so cannot be plotted in log scale)}.
                \emph{Middle panel}: The relative variation of the density $\rho_c$ in time.
                \emph{Lower panel}: The relative variation of the lapse function $\alpha_c$ in time.
		}
	\label{fig:MHD_NS_cylindrical_2D_central}	
\end{figure}

\begin{figure}
	\centering
	\includegraphics[width=1.0\columnwidth, angle=0]{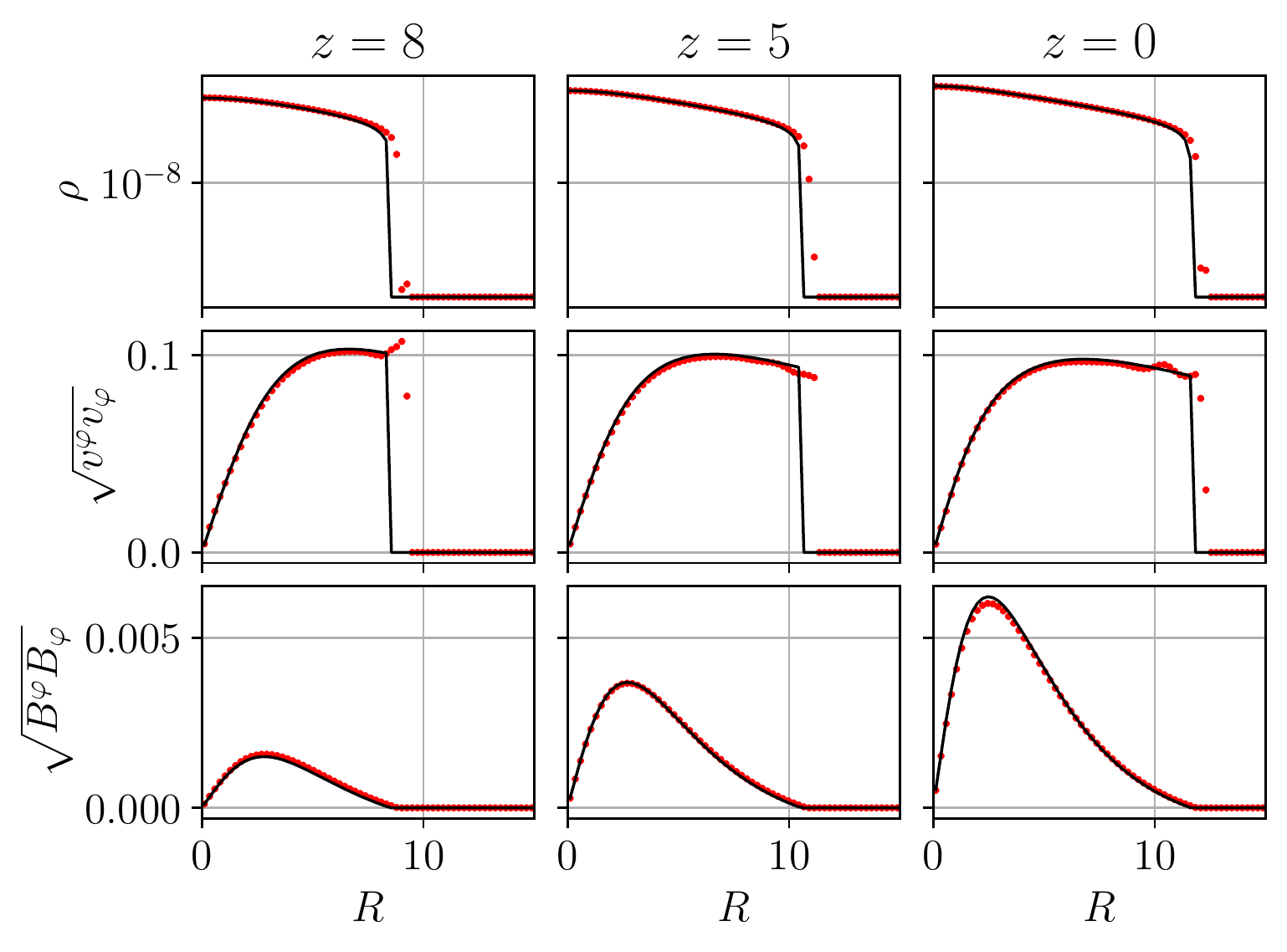}
	\caption{One-dimensional slices of a differentially rotating strongly magnetized equilibrium neutron star in spherical coordinates along the $z = 8$ (\emph{left column}), $ z = 5$ (\emph{middle column}) and $z = 0$ (\emph{right column}) for the density $\rho$ (\emph{upper row}), the rotational velocity $\sqrt{v_\varphi v^\varphi}$ (\emph{middle row}) and the magnetic field $\sqrt{B_\varphi B^\varphi}$.
The black solid lines show the initial profiles while the red dots show the profiles $t = 10$ ms.
		}
	\label{fig:MHD_NS_cylindrical_2D_slices}	
\end{figure}

Figure~\ref{fig:MHD_NS_cylindrical_2D_central} shows the evolution of this differentially rotating strongly magnetized equilibrium neutron star in cylindrical coordinate.
The rest mass $M_b$ is {unchanged} during the whole simulation ($t=0$ ms to $t=10$ ms).
Figure~\ref{fig:MHD_NS_cylindrical_2D_slices} compares the initial {$(t=0)$} density profile, rotational velocity and the magnetic field (black solid lines) with the same quantities (red dots) at $t = 10$ ms.
The profiles are maintained well except some slight distortions.

\subsubsection{Stability of a non-rotating neutron star}
\begin{figure}
	\centering
	\includegraphics[width=1.0\columnwidth, angle=0]{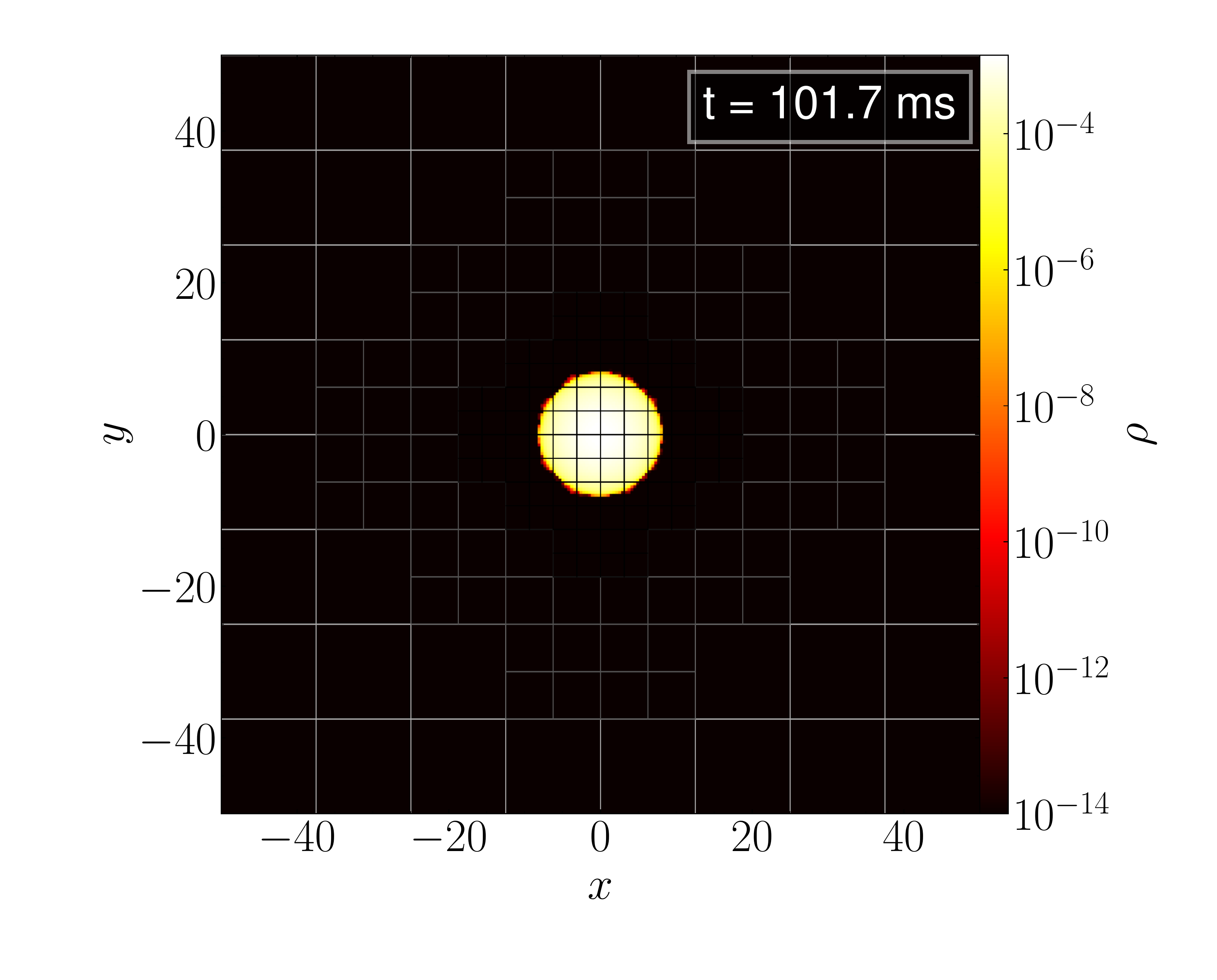}
	\caption{
		The projection of density profile along $z$-axis of a spherical neutron star in {Cartesian coordinates} with the annotated mesh lines at {$t=101.7$} ms.
		The computational domain {covers $[-100,100]$} for both $x$,$y$ and $z$, with the resolution {$n_x \times n_y \times n_z = 64 \times 64 \times 64$} and allowing {4} AMR level {(an effective resolution of $512^3$)}.
		}
	\label{fig:HD_BU0_3D_mesh}	
\end{figure}

We present a full 3-dimensional simulation of a spherically symmetric neutron star here. 
In this test, we consider a non-rotating model which is constructed with the polytropic equation of state with $\Gamma = 2$ and $K=100$ with central rest-mass density $\rho_c = 1.28 \times 10^{-3}$(in $c=G=M_\odot=1$ unit), which is also know as ``BU0'' in the literature \citet{2006MNRAS.368.1609D, XCFC}.
Actually, such a spherically symmetric model can be simulated in a one- or two-dimensional spherical coordinate.
Nevertheless, as a demonstration, we simulate this system in 3D {Cartesian} coordinates without imposing any symmetries, i.e.~ this problem is simulated in the full 3D configuration.
The computational domain {covers $[-100,100]$} for both $x$,$y$ and $z$, with the resolution {$n_x \times n_y \times n_z = 64 \times 64 \times 64$} and allowing {4} AMR level {(an effective resolution of $512^3$)}.
{The refinement setting is identical to section \ref{sec:magns}}.
As an example, figure \ref{fig:HD_BU0_3D_mesh} shows the density profile with the annotated mesh lines at {$t=101.7$ ms}.

\begin{figure}
	\centering
	\includegraphics[width=1.0\columnwidth, angle=0]{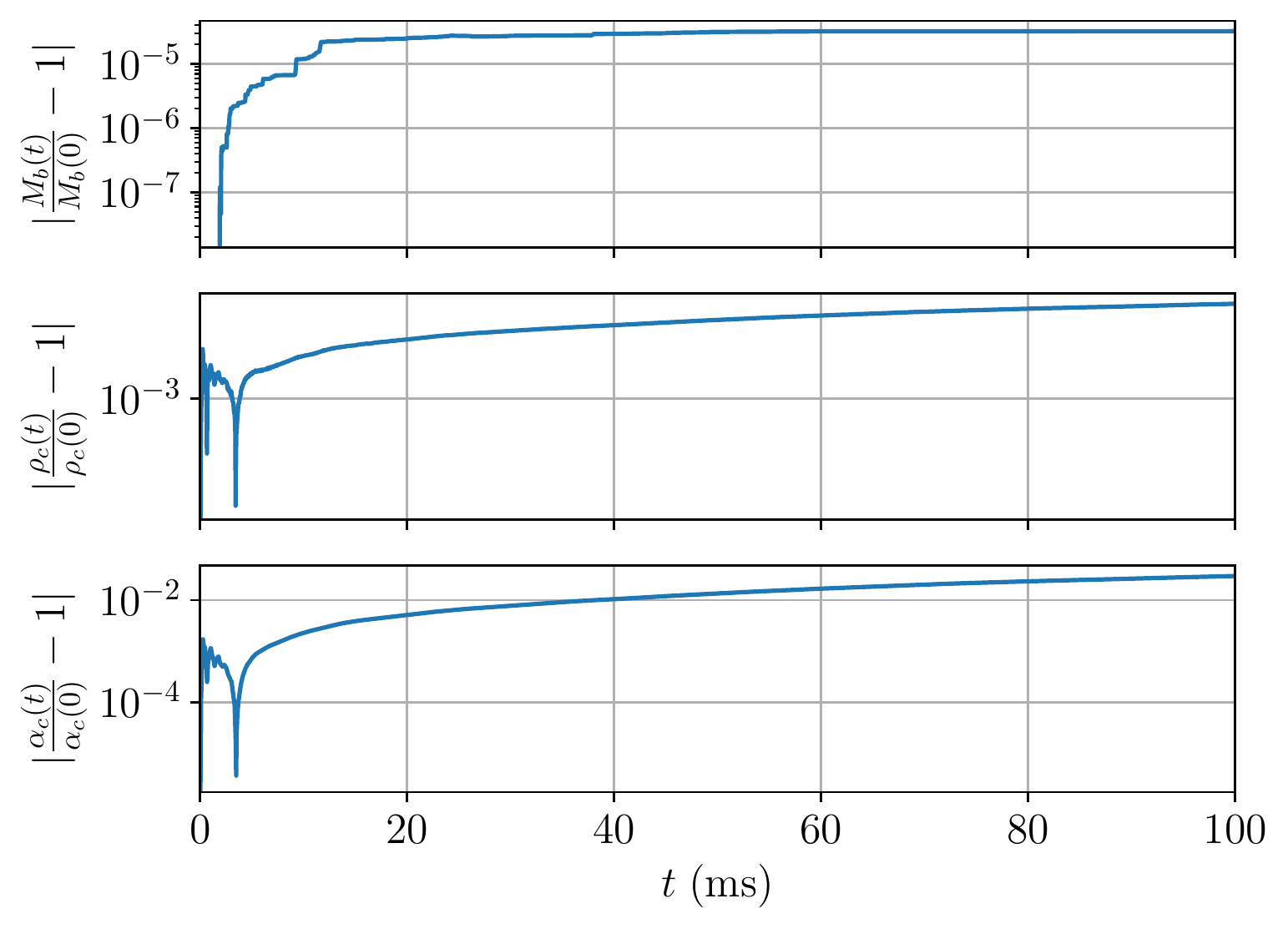}
	\caption{ {
                \emph{Upper panel}: The relative variation of the rest mass $M_b$ in time.
		The conservation of the rest mass $M_b$ is preserved remarkably well from $t=0$ ms to $t=100$ ms where the relative variation is of the order {$10^{-4}$}.
                \emph{Middle panel}: The relative variation of the density $\rho_c$ in time.
		\emph{Lower panel}: The relative variation of the lapse function $\alpha_c$ in time.}
		}
	\label{fig:HD_BU0_central}	
\end{figure}

\begin{figure}
	\centering
	\includegraphics[width=1.0\columnwidth, angle=0]{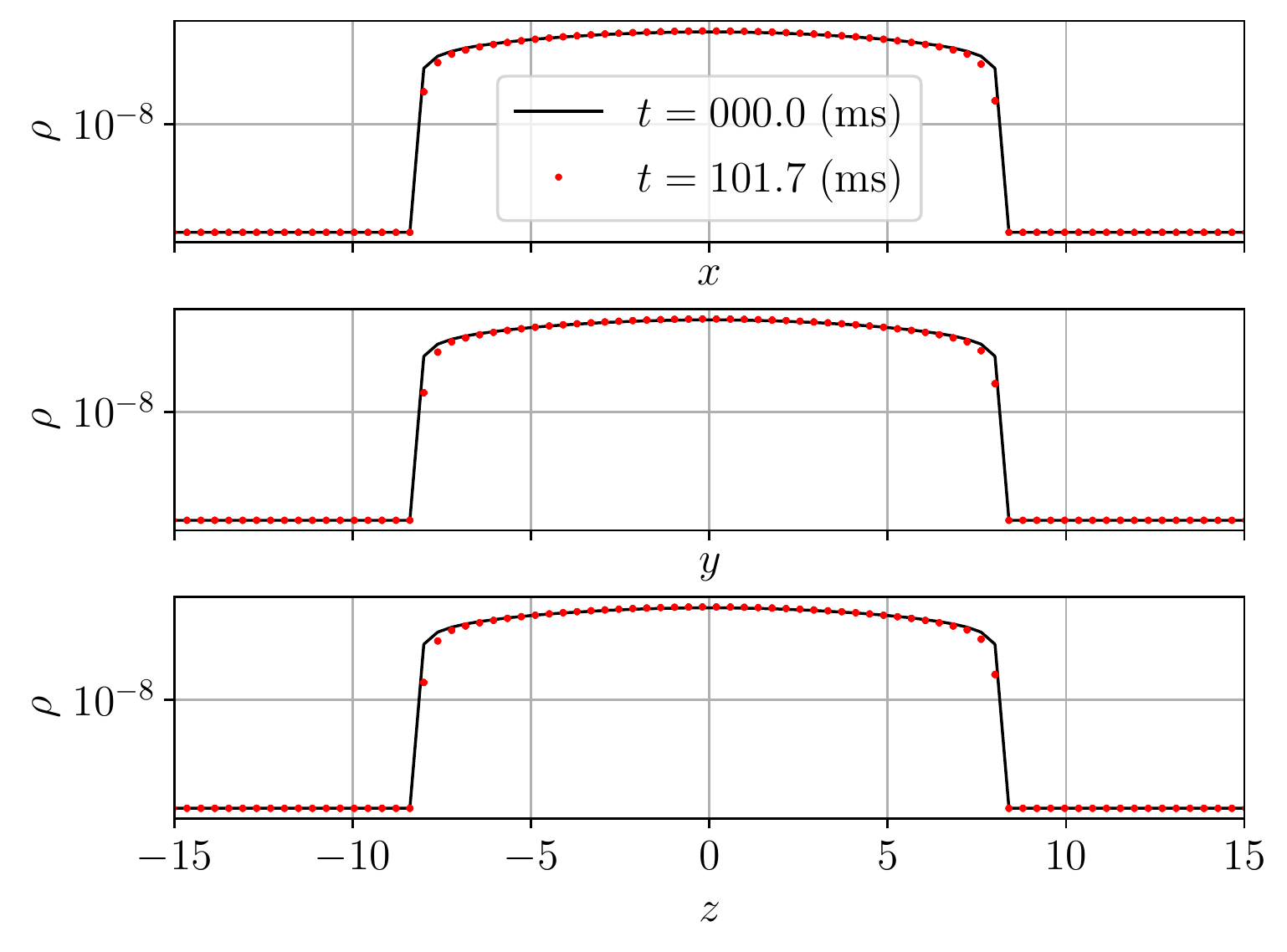}
	\caption{One-dimensional slices of the non-rotating equilibrium neutron star BU0 along the $x-$ axis (\emph{upper panel}), $y-$ axis (\emph{middle panel}) and $z-$ axis (\emph{lower panel}) for the density $\rho$.
	The black solid lines show the initial profiles while the red dots show the profiles {$t = 101.7$} ms.
		}
	\label{fig:HD_BU0_3d_slices}	
\end{figure}

{Figure~\ref{fig:HD_BU0_central} shows the evolution of the spherically symmetric neutron star BU0 in Cartesian coordinate.
\texttt{Gmunu} is able to maintain the profile up to 100 ms and the relative variation of the rest mass of the order {$10^{-4}$}.}
Figure~\ref{fig:HD_BU0_3d_slices} compares the initial density profile (black solid lines) with the same quantities (red dots) at $t = 101.7$ ms.
The density profile is maintained well.

\subsubsection{Migration of an unstable neutron star}
To see how \texttt{Gmunu} preform in the fully non-linear regime with significant changes and coupling in the metric and fluid variables, here we present a simulation of the migration of an unstable neutron star, which is one of the standard tests for hydrodynamical evolution coupled with dynamical spacetime in the fully non-linear regime \citet{2002PhRvD..65h4024F, 2010PhRvD..81h4003B, XCFC, XECHO}.
In this test, we consider an unstable neutron star, which lies on the unstable branch of the mass-radius curve.
{The neutron star} is constructed with the polytropic equation of state with $\Gamma = 2$ and $K=100$ with central rest-mass density $\rho_c = 8.00 \times 10^{-3}$(in $c=G=M_\odot=1$ unit), which is also {known} as ``SU'' in \citet{XCFC}.
We simulate this initial model in 2-dimensional cylindrical coordinates {$(R, z)$}, where the computational domain covers $0 \leq R \leq 60$ and $-60 \leq z \leq 60$, with the resolution $n_R \times n_z = 32 \times 64$ and allowing 4 AMR level (i.e., an effective resolution of $256 \times 512 $).
{The refinement setting is identical to section \ref{sec:magns}}.
We adopt the ideal-gas (gamma-law) equation of state $P = (\Gamma - 1)\rho\epsilon$ with $\Gamma = 2$ for the fluid so that we can also capture the shock heating effect.

\begin{figure}
	\centering
	\includegraphics[width=1.0\columnwidth, angle=0]{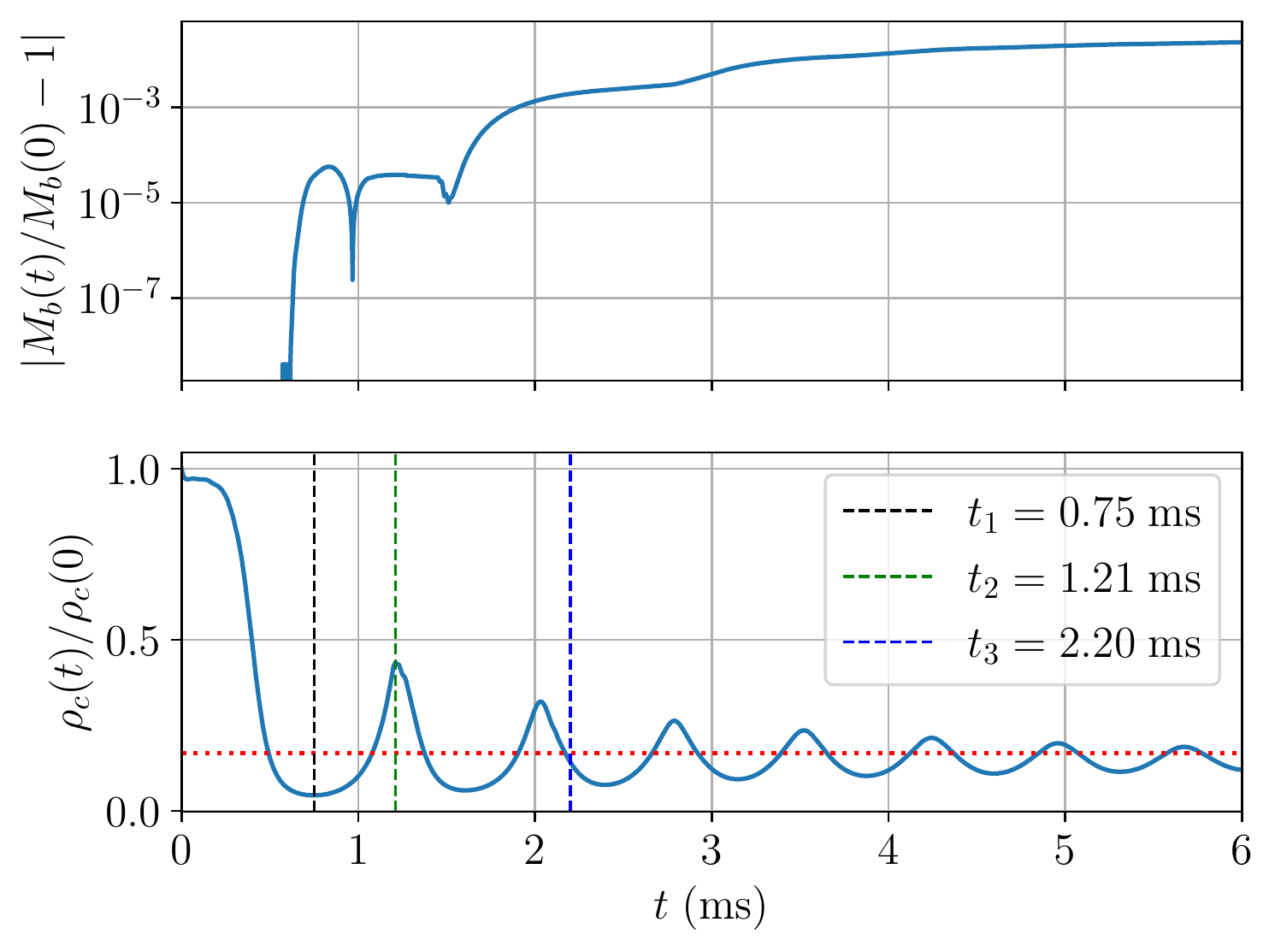}
	\caption{
		Evolution of an unstable spherically symmetric neutron star. 
                \emph{Upper panel}: The relative variation of the rest mass $M_b$ in time.
		\emph{Lower panel}: The central density $\rho_c/\rho_c(t=0)$ in time. 
		The dotted line represents the central density $\rho_c$ of the neutron star on the stable branch.
	}
	\label{fig:HD_mig_central}	
\end{figure}

\begin{figure}
	\centering
	\includegraphics[width=1.0\columnwidth, angle=0]{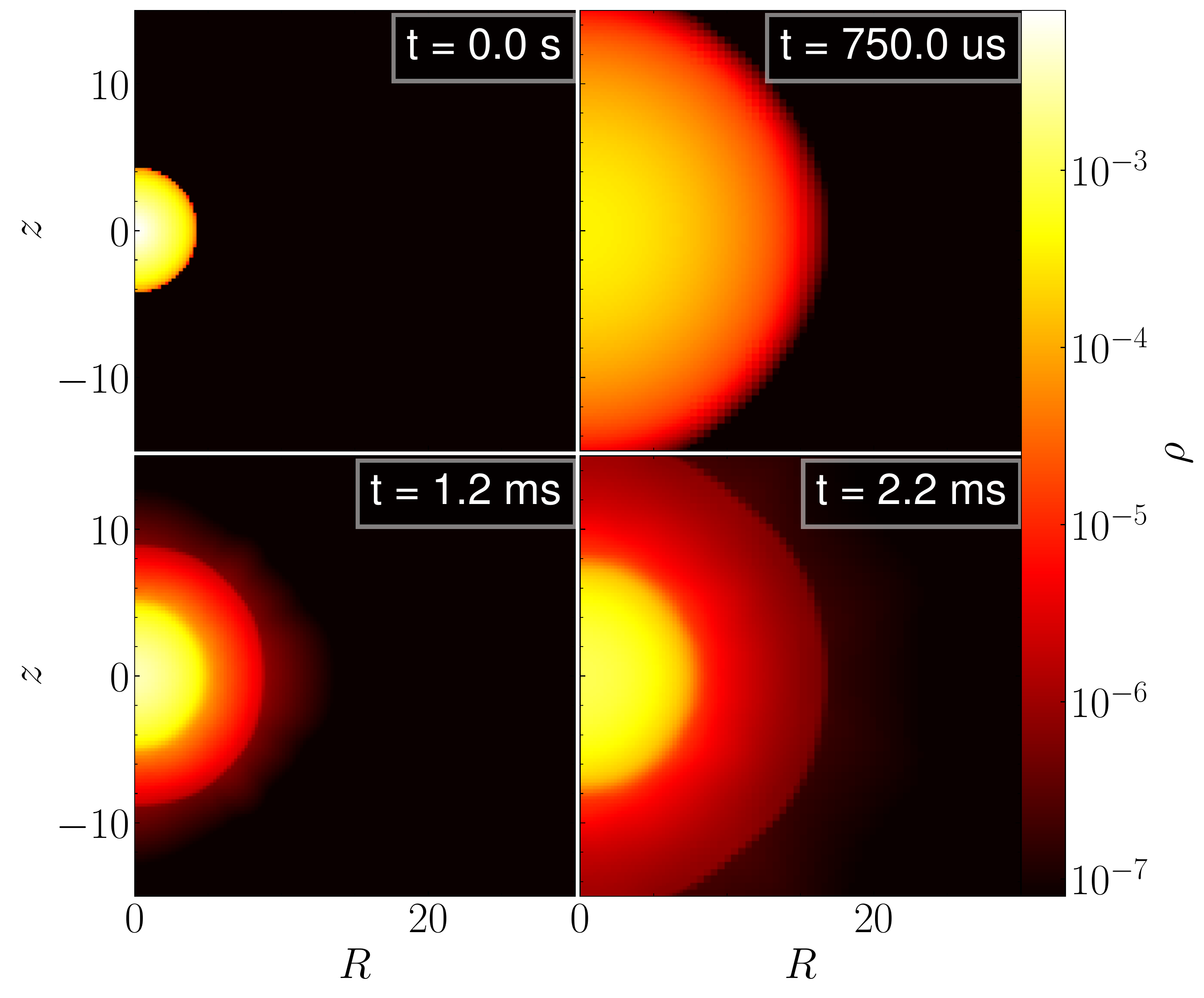}
	\caption{The evolution of the density $\rho$ an unstable neutron star SU at various time slices.
		Initially (\emph{upper left}), the star has the central density $\rho_c = 8.00 \times 10^{-3}$ with a radius $r = 4.267$.
		As the star evolves, the central density reduces and the radius of the star expands to a larger value.
		At {$t=0.75$} ms (\emph{upper right}), the central density of the system reach the lowest point and start to increase.
		At {$t=1.21$} ms (\emph{lower left}), the central density of the system hits the local maxima.
		A small amount of mass is ejected outside the surface of the star due to the shock heating effect.
                After some time, at $t=2.20$ ms (\emph{lower right}), the ejected mass surrounding the star covers most of the computational domain.
		}
	\label{fig:HD_mig_time_series}	
\end{figure}

As the star evolves and migrates to the corresponding stable configuration $\rho_c = 1.346 \times 10^{-3}$ with the same mass, the radius of the star expands to a large value.
Figure~\ref{fig:HD_mig_central} shows the evolution of the baryon mass $M_b$ and the central density $\rho_c$ as a function of time.
The oscillations of the central density $\rho_c$ are damped since shock waves are formed at every pulsation and some kinetic energy is dissipated into thermal energy.
A small amount of mass is ejected outwards from the surface of the star to the surrounding artificial low-density {($\rho_{\text{atmo}} = 10^{-14}$)} ``atmosphere'' whenever these shock waves hit the surface of the star, and thus the total baryon mass $M_b$ decays once the shock waves hit the outer numerical boundaries.
With weaker oscillation, the decay rate of the baryon mass is smaller.
This dissipation effect can also be seen in the density profile, as shown in figure \ref{fig:HD_mig_time_series}.
As a result, the baryon mass and the central density of the final equilibrium stable configuration is slightly lower than the expected value.

\section{\label{sec:}{Performance and Scaling}}
{
In this section, we present two tests to assess the strong scaling of \texttt{Gmunu}.
The scaling tests to be presented below were obtained on the Central Research Computing Cluster in The Chinese University of Hong Kong.
In particular, in all tests, we used computing nodes on the Central Cluster with dual Intel Xeon Gold 6130 processors, for a total of 32 cores per node, and one MPI process per core was used.
}

\subsection{{Scaling of special-relativistic (magneto-)hydrodynamics}}
\begin{figure}
	\centering
	\includegraphics[width=\columnwidth, angle=0]{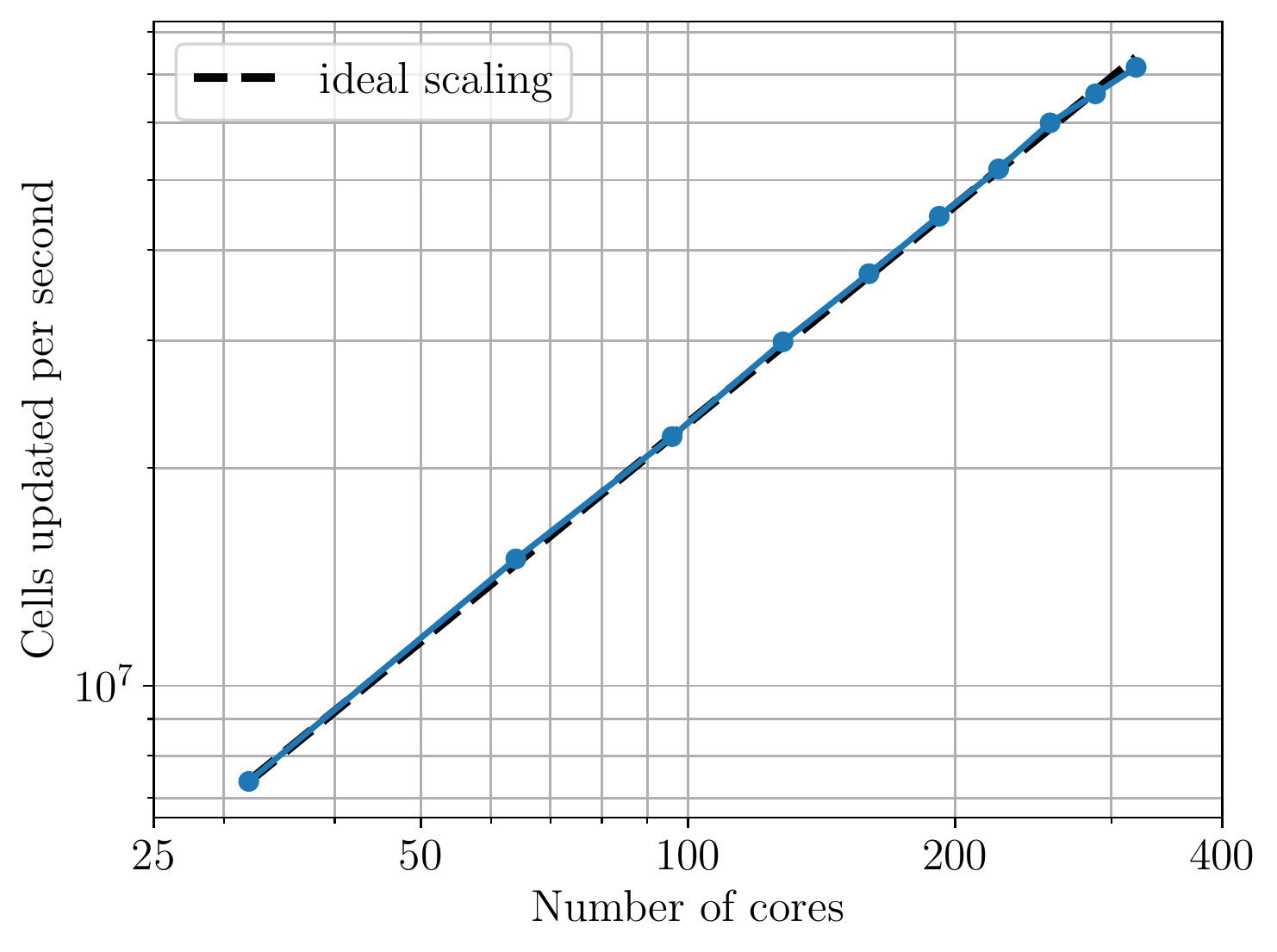}
	\caption{
		{Strong scaling of a relativistic hydrodynamics application with a set of uniform grid $1024^2$ decomposed into $128^2$ blocks with each block of $8^2$ cells.
		The figure shows the cells updated per second for increasing number of cores.
		The blue line shows the results obtained by \texttt{Gmunu} while the black dashed line is the ideal scaling.}
		}
	\label{fig:HD_riemann_2d_scaling}	
\end{figure}

{
In this subsection, we focus on the strong scaling of the hydrodynamics solver (hyperbolic sector) of \texttt{Gmunu}.
The test here is the two-dimensional Riemann problem, as discussed in section~\ref{sec:2D_riemann_test}, with a slightly different setting.
In particular, in this test, PPM reconstruction is used, and the simulation box consist of a set of uniform grid $1024^2$ decomposed into $128^2$ blocks with each block of $8^2$ cells.
Figure~\ref{fig:HD_riemann_2d_scaling} shows the cells updated per second at different number of cores.
}

\subsection{{Performance of the metric solver}}
{
In this section, we demonstrate the convergence properties and performance of our multigrid metric solver.
We solve the metric of model BU8 (see \ref{sec:BU8_2D}), which represents a rapidly rotating neutron star and far from spherically symmetric, in a \emph{full} three-dimensional setting.
For instance, the computational domain covers $[-200,200]$ for both $x$,$y$ and $z$, with the resolution {$N_x \times N_y \times N_z = 256 \times 256 \times 256$}, is decomposed into $16 \times 16 \times 16$ blocks with each block of $16 \times 16 \times 16$ cells.
In addition to this uniform grid setting, we also consider the same case with allowing 4 AMR level {(an effective resolution of $2048 \times 2048 \times 2048$)}.
To make it a fair and general comparision, instead of using the pre-solved initial data as the initial guess, we focus on solving the lapse function $\alpha$ with the flat space initial guess $\alpha = 1$.
In the following test, two upward and downward red-black Gauss-Seidel smoothing steps are used.
}

\subsubsection{{Convergence properties}}
\begin{figure}
	\centering
	\includegraphics[width=\columnwidth, angle=0]{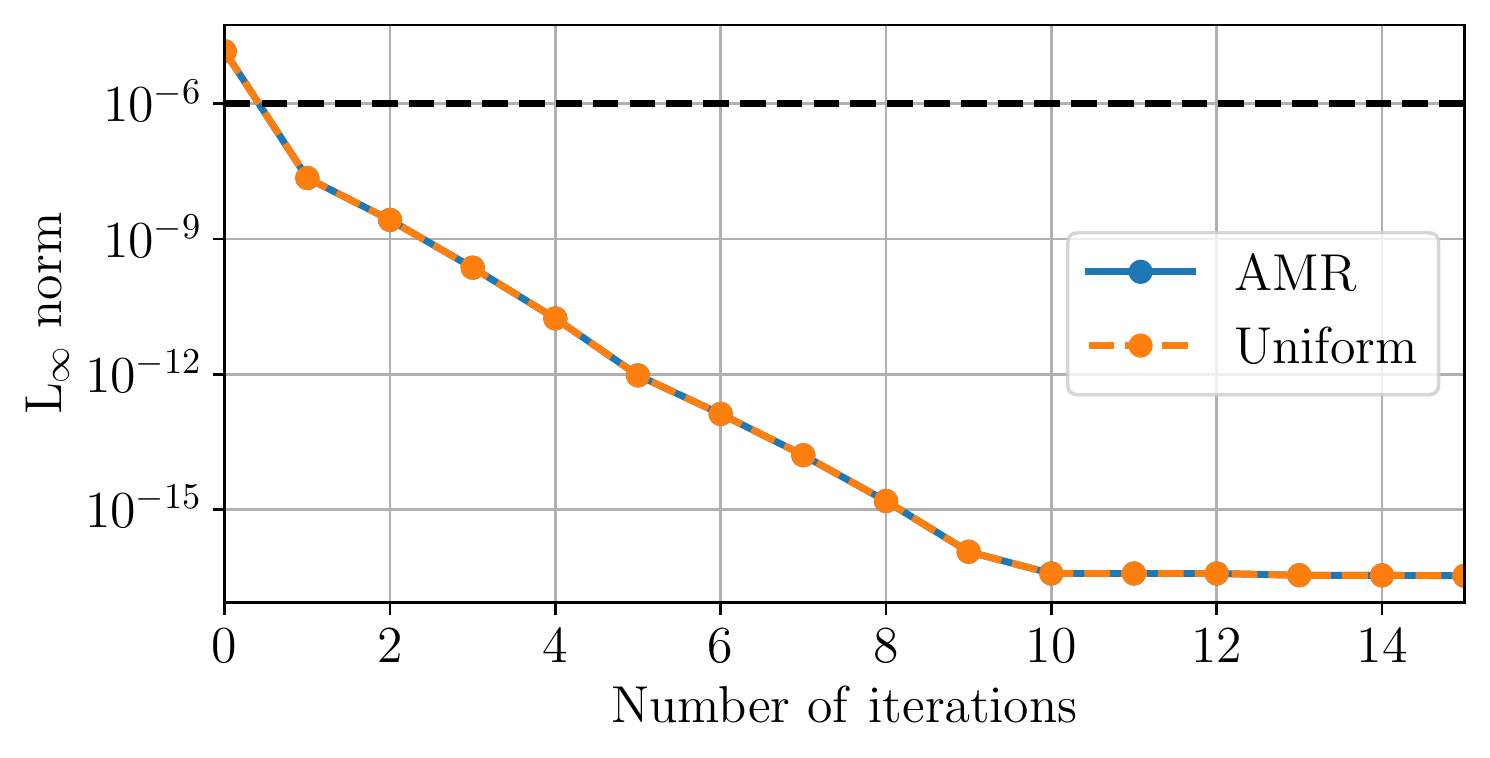}
	\caption{
		{$L_\infty$ norm of residual of equation~\eqref{eq:alpha} of an highly non-spherically symmetric model BU8 as a function of the number of full multigrid (FMG) iterations.
		The convergence properties with or without AMR activated are almost identical in this test case.
		Even if the multigrid solver starts from the flat space initial guess, \emph{one} iteration is sufficient to converge to the prescribed tolerance (horizational black dashed line), and the residual is reduced up to machine precision (i.e. $L_\infty \lesssim 10^{-16}$) after about 10 iterations.}
		}
	\label{fig:MG_BU8_alp_res}	
\end{figure}

{
Figure~\ref{fig:MG_BU8_alp_res} shows the $L_\infty$ norm of the residual of equation~\eqref{eq:alpha} as a function of the number of full multigrid (FMG) iterations.
The convergence properties with or without AMR activated are almost identical in this test case.
Even if the multigrid solver starts from the flat space initial guess, \emph{one} iteration is sufficient to converge to the prescribed tolerance (horizational black dashed line), and the residual is reduced up to machine precision (i.e. $L_\infty \lesssim 10^{-16}$) after about 10 iterations.
The $L_\infty$ norm of residual of equation~\eqref{eq:alpha} in both cases are identical, which implies the maximum residual in both cases are the same.
Indeed, in this test, the maximum residual is located at the outer boundary of the computational domain, where the resolution with AMR is set to be the lowest in purpose (see the discussion of section~\ref{sec:user_refinement_criteria}), which is identical to the uniform case.
}

{
In practice, at the beginning of the simulation, we use the initial data provided by \texttt{XNS} as initial guess. 
During the evolution, we use the previous solution as initial guess for the next iteration.
This makes the solver converge much faster as the solutions on previous time step are usually good approximation to the solution.
}

\subsubsection{{Strong scaling}}
\begin{figure}
	\centering
	\includegraphics[width=\columnwidth, angle=0]{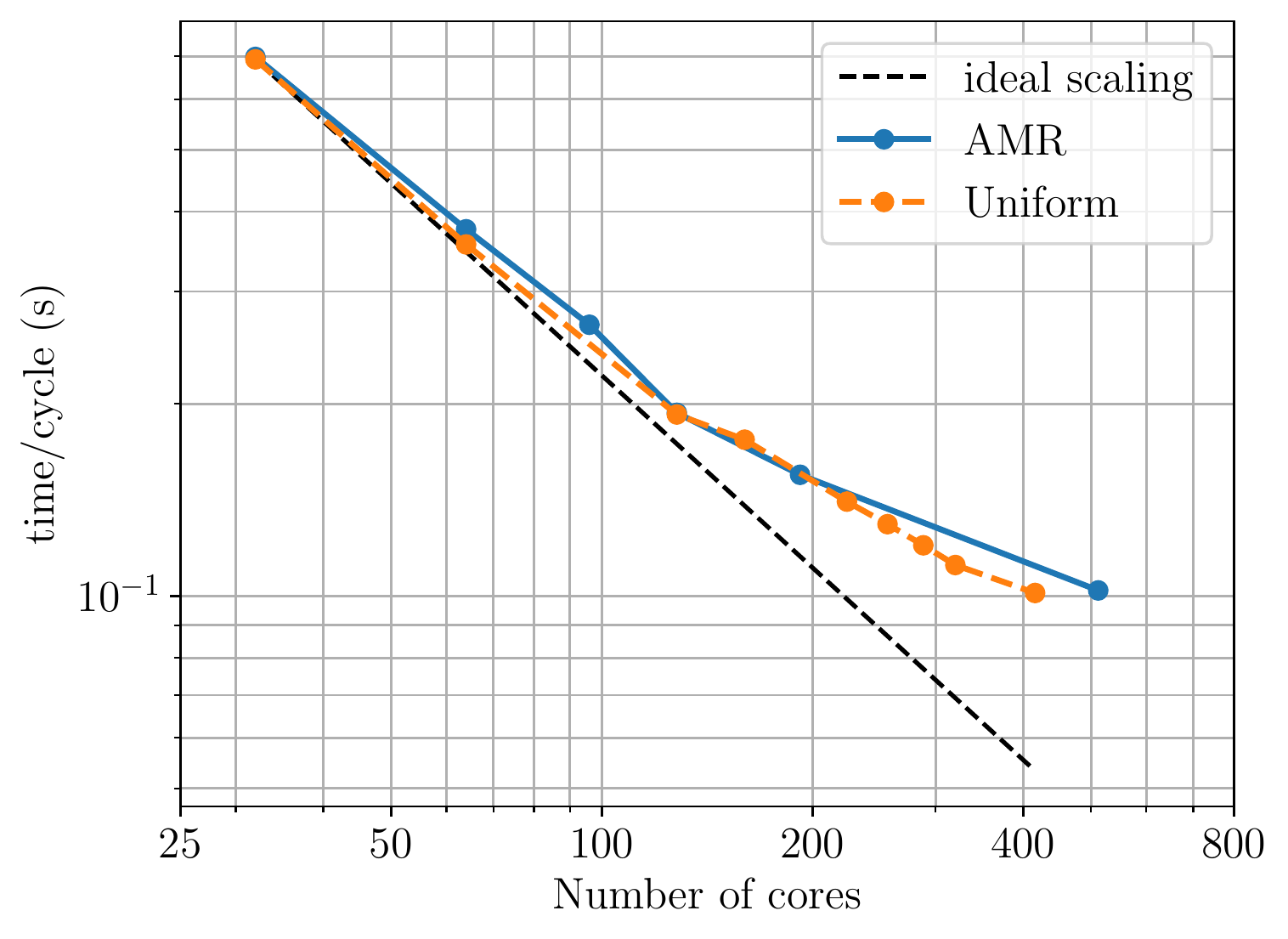}
	\caption{
		{Computational time per full multigrid (FMG) cycles when solving equation~\eqref{eq:alpha} of an highly non-spherically symmetric model BU8 as a function of the number of cores. 
		The computational time per cycle reduces with increasing number of cores.
		The scaling close to the ideal scaling for the number of cores $\lesssim 100$, however it is not ideal with larger amount of cores.
		This is because the problem size is not larger enough to have good scaling.
		}
		}
	\label{fig:MG_BU8_alp_scaling}	
\end{figure}

{
Here, we assess the performance and scaling of our metric solver with the same setup mentioned above.
We measure the time per full multigrid (FMG) cycle by averaging over 500 cycles.
Figure~\ref{fig:MG_BU8_alp_scaling} shows the computational time per full multigrid (FMG) cycles when solving equation~\eqref{eq:alpha} of an highly non-spherically symmetric model BU8 as a function of the number of cores. 
The computational time per cycle reduces with increasing number of cores.
The scaling close to the ideal scaling for the number of cores $\lesssim 100$, however it is not ideal with larger amount of cores.
This is because the problem size is not larger enough to have good scaling.
For instance, in the uniform case ($N_x \times N_y \times N_z = 256 \times 256 \times 256$) with 416 cores, only about $32^3$ unknonws are involved.
}

\subsection{{Scaling of general-relativistic hydrodynamics}}
{
Finally, we assess the performance of \texttt{Gmunu} in general-relativistic hydrodynamics simulations. 
In this test, we again evolve the rapidly rotating neutron star BU8 (see section~\ref{sec:BU8_2D}) but this time in a \emph{full} three-dimensional setting.
For instance, the computational domain covers $[-200,200]$ for both $x$,$y$ and $z$, with the resolution {$N_x \times N_y \times N_z = 64 \times 64 \times 64$} and allowing 5 AMR level {(an effective resolution of $1024 \times 1024 \times 1024$)}.
The simulation box is decomposed into $8 \times 8 \times 8$ blocks with each block of $8 \times 8 \times 8$ cells.
Also, TVDLF Riemann solver and third-order accurate PPM limiter is used.
This system is evolved to $T_{\rm{final}} = 10\ \rm{ms}$, and output data at every 1 ms.
Note that, this test includes not only the hydrodynamics part (hyperbolic sector), but also the metric equations (elliptic sector).
Unlike in the case of divergence cleaning of magnetic field, where the elliptic equation needed to be solve is the trivial and well behave Poisson equation, the metric equations in extended CFC scheme are highly non-linear and include vectorial elliptic equations, the computational cost required by the metric solve may be different from time to time during the dynamical simulations.
}

\begin{figure}
	\centering
	\includegraphics[width=\columnwidth, keepaspectratio, angle=0]{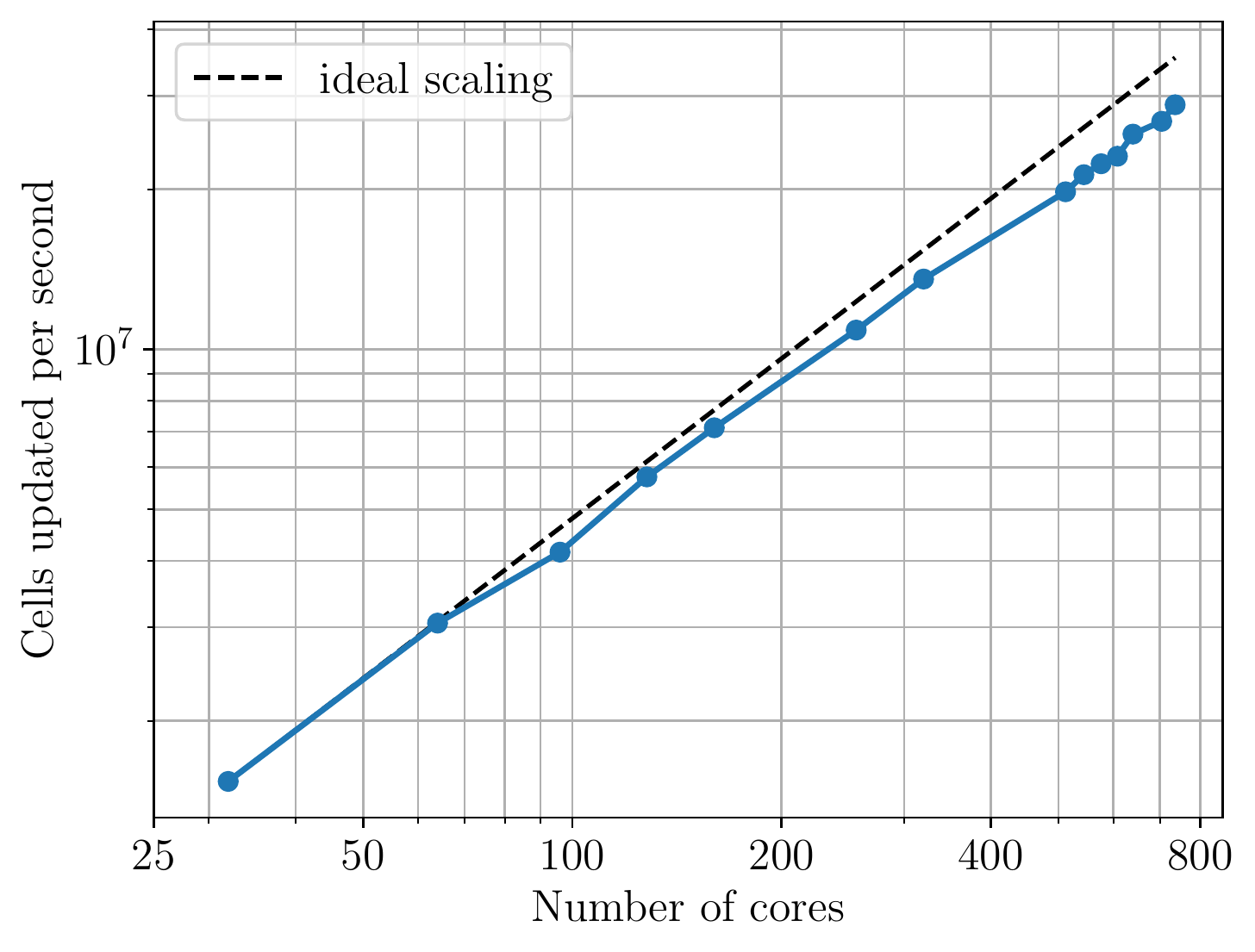}
	\includegraphics[width=\columnwidth, keepaspectratio, angle=0]{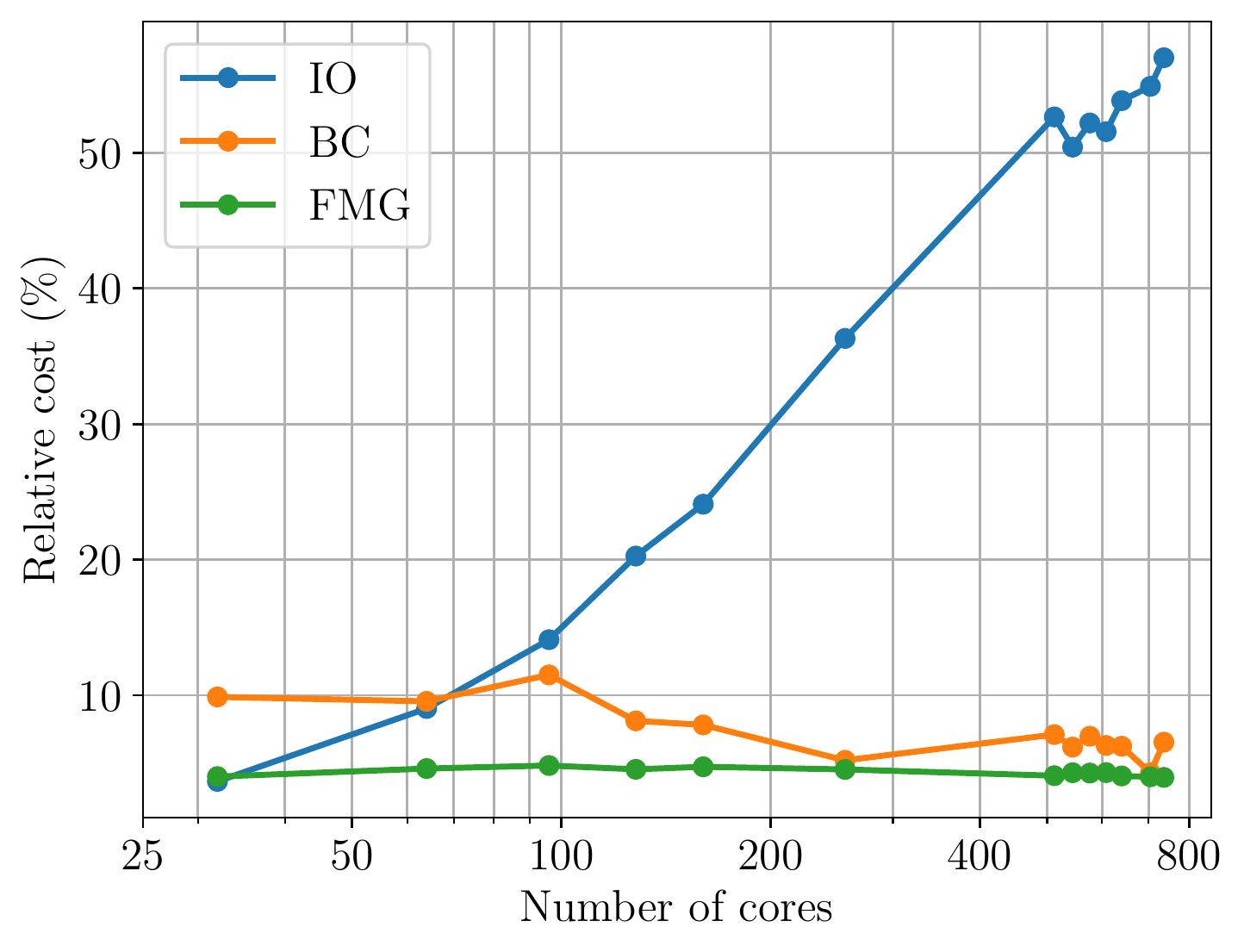}
	\caption{
		{Strong scaling of the evolution of full three-dimensional rapidly rotating neutron star BU8 in Cartesian coordinates.
		The computational domain covers $[-200,200]$ for both $x$,$y$ and $z$, with the resolution {$N_x \times N_y \times N_z = 64 \times 64 \times 64$} and allowing 5 AMR level {(an effective resolution of $1024 \times 1024 \times 1024$)}.
The simulation box is decomposed into $8 \times 8 \times 8$ blocks with each block of $8 \times 8 \times 8$ cells.
		The \emph{upper panel} shows the cells updated per second for increasing number of cores.
		The blue line shows the results obtained by \texttt{Gmunu} while the black dashed line is the ideal scaling.
		Note that even with the elliptic metric solver included, the scaling is closed to ideal scaling.
		The \emph{lower panel} shows the relative cost measure by time of different parts of the code.
		The cost of the metric solver (with full multigrid (FMG), green line) is slightly below 5\% in all cases we have tested, which is even below the cost required for updating boundary conditions (BC, orange line).
		The relative cost of output data (IO, blue line) gradually increase with increasing number of cores, the computational time of which is limited by the speed of writing/reading data to/from the hard disk drive (disk I/O speed).}
		}
	\label{fig:HD_BU8_3D_scaling}	
\end{figure}

{
Figure~\ref{fig:HD_BU8_3D_scaling} shows the strong scaling and the relative computational cost (measure by time) of this test problem.
As shown on the left panel, the scaling is closed to ideal scaling, even with the elliptic metric solver included.
Besides, as shown on the right panel, the relative cost of metric solver is slightly below 5\% in all cases we have tested, which is relatively low in the sense that even the cost of updating the boundary conditions (including ghost cells) requires more than the metric solver.
The relative cost of output data (IO) gradually increase with increasing number of cores, the computational time of which is limited by the speed of writing/reading data to/from the hard disk drive (disk I/O speed).
}

\section{Conclusions}\label{sec:conclusions}
We present the new methodology and implementation of \texttt{Gmunu}, a parallelised multi-scale multi-dimensional curvilinear general-relativistic magneto-hydrodynamics code with a cell-centred non-linear multigrid solver which is fully coupled with an adaptive mesh refinement modules.
The code has been designed to perform generic general relativistic (magneto-)hydrodynamical simulations in dynamical spacetime.
With the flexibility of choosing coordinates and the efficient block-based adaptive mesh refinement module, depending on the nature of the problems and the study interests, users can balance the computational cost and the accuracy of the results easily without changing to other codes.
For the divergenceless handling for the magnetic field, in this work, we present, to our knowledge, the first example of using elliptic divergence cleaning dynamically during the relativistic magneto-hydrodynamics simulations.
Currently, \texttt{Gmunu} is able to solve the elliptic-type metric equations in the extended conformally flat condition (xCFC) approximation to general relativity.

We have tested \texttt{Gmunu} with several benchmarking tests, from special-relativistic to general-relativistic (magneto-)hydrodynamics in one-, two- and three- dimensional {Cartesian}, cylindrical and spherical coordinates.
These tests include (i) SR(M)HD shock tubes, SRHD Riemann test, axisymmetric jet in SRHD, cylindrical blast wave and magnetic field loop advection in SRMHD, the evolution of rapidly/ differentially rotating, strongly magnetized neutron stars in GR(M)HD.
In the GRMHD tests, we demonstrate that the multigrid {algorithm} in \texttt{Gmunu} is able to solve CFC metric equations in {multiple dimensions} and in different coordinates with or without coupling with the AMR module.
In addition, the robust positivity preserving limiter and {conserved-to-primitive} variables conversions enable us to set the density of the ``atmosphere'' $\rho_\text{atmo}$ to the order of $\mathcal{O}(10^{-20})$ ({below} machine precision) even in the evolution of a rapidly rotating or strongly magnetized neutron star with good rest mass conservation and accurate results.

In the future, we will present the implementations and comparisons of various divergence-free treatments, i.e., elliptic cleaning, generalized Lagrange multiplier (GLM), constrained transport (CT) and the vector potential schemes.
Furthermore, we will implement radiation hydrodynamics for MHD also for neutrino physics.
We shall also extend \texttt{Gmunu} to a fully-constrained scheme in exact general relativity such as the formulation of \citet{2004PhRvD..70j4007B}.

\section*{Acknowledgements}
PCKC thanks David Yat-Tung Pong for setting up and providing technical support for CUHK-GW workstations. 
{
We acknowledge the support of the CUHK Central High Performance Computing Cluster, on which the scaling tests in this work have been performed.
}
This work was partially supported by grants from the Research Grants Council of the Hong Kong (Project No. CUHK24304317 and CUHK 14306419), the Croucher Innovation Award from the Croucher Fundation Hong Kong and by the Direct Grant for Research from the Research Committee of the Chinese University of Hong Kong.

\section*{Data Availability}
The data underlying this article are available in the article.
 



\bibliographystyle{mnras}
\bibliography{mybibfile}




\appendix
\section{Flat metric in 3D}
\label{appendix:coordinates}
The cell volume $\Delta V$, cell surface $\Delta A$ and the volume-average of the 3-Christoffel symbols $\left< \hat{\Gamma}^l_{ik}\right>$ which is contained in the geometrical are non-trivial when the reference metric $\hat{\gamma}_{ij}$ is chosen to be cylindrical or spherical.
Here we list out the relation we implemented in \texttt{Gmunu}.
\subsection{cylindrical coordinate}
The line element can be expressed as: $ ds^2 = dR^2 + dz^2 + R^2 d\varphi^2 $, with the reference metric $\hat{\gamma}_{ij}$:
\begin{align}
	\hat{\gamma}_{ij} &= \begin{bmatrix}
		1 & 0 & 0 \\
		0 & 1 & 0 \\
		0 & 0 & R^2 \\
       	\end{bmatrix}.
\end{align}
The associated 3-Christoffel symbols $\hat{\Gamma}^l_{ik}$ are:
\begin{align}
	\hat{\Gamma}^R_{ij} = &\begin{bmatrix}
		0 & 0 & 0 \\
		0 & 0 & 0 \\
		0 & 0 & -R \\
       	\end{bmatrix}, &
	\hat{\Gamma}^z_{ij} = &\begin{bmatrix}
		0 & 0 & 0 \\
		0 & 0 & 0 \\
		0 & 0 & 0 \\
       	\end{bmatrix}, &
	\hat{\Gamma}^\varphi_{ij} = &\begin{bmatrix}
		0  & 0& \frac{1}{R} \\
		0 & 0 & 0 \\
		\frac{1}{R} & 0 & 0 \\
       	\end{bmatrix}.
\end{align}
The geometrical source terms for the momentum equations are:
\begin{align}
	\hat{\Gamma}^l_{Rk}(f_{S_l})^k =& \hat{\Gamma}^\varphi_{R\varphi}(f_{S_\varphi})^\varphi ,\\
	\hat{\Gamma}^l_{zk}(f_{S_l})^k  =& 0  ,\\
	\hat{\Gamma}^l_{\varphi k }(f_{S_l})^k =& 0 .
\end{align}
Here we note that since $z$ and $\varphi$ do not explicitly enter into the reference metric $\hat{\gamma}_{ij}$, the corresponding geometrical source terms for the momentum equation $q_{S_j}$ are vanishing. 
In this formulations, the linear momentum $q_{S_z}$ and the angular momentum $q_{S_\varphi}$ are conserved to machine precision.

To work out the cell volume $\Delta V$, cell surface $\Delta A$ and the volume-average of the 3-Christoffel symbols $\left< \hat{\Gamma}^l_{ik}\right>$, we define the following notations: 
\begin{align}
&R_\pm \equiv R \pm \frac{1}{2}\Delta R,&
&z_\pm \equiv z \pm \frac{1}{2}\Delta z,& 
&\varphi_\pm \equiv \varphi \pm \frac{1}{2}\Delta \varphi ,&
\end{align}
where $(R,z,\varphi)$ are the location at the cell centre at some particular point while $(\Delta R,\Delta z,\Delta \varphi)$ are the corresponding grid sizes.
The cell surface $\Delta A$ and the cell volume $\Delta V$ can then be expressed as:
\begin{align}
\Delta A_R\Big|_{R_\pm} &= \left( R \pm \frac{\Delta R}{2} \right) \left( \Delta z \right) \left( \Delta \varphi \right) ,\\
\Delta A_z \Big|_{z_\pm} &= R \left( \Delta R \right) \left( \Delta \varphi \right) ,\\
\Delta A_\varphi \Big|_{\varphi_\pm} &= R \left( \Delta R \right) \left( \Delta z \right) ,\\
\Delta V &=  R \left( \Delta R \right) \left( \Delta z \right) \left( \Delta \varphi \right).
\end{align}
Finally, the non-vanishing volume-averaged 3-Christoffel symbols $\left< \hat{\Gamma}^l_{ik}\right>$ are:
\begin{align}
&\left<\hat{\Gamma}^{R}_{\varphi\varphi } \right> = - \frac{1}{R} \left(R^2 + \frac{1}{12}\left(\Delta R\right)^2\right), \\
&\left<\hat{\Gamma}^{\varphi}_{\varphi R} \right> = \left<\hat{\Gamma}^{\varphi}_{R \varphi } \right> = \frac{1}{R}.
\end{align}

\subsection{Spherical coordinates}
The line element can be expressed as: $ ds^2 = dr^2 + r^2 d\theta^2 + r^2\sin^2\theta d\phi^2 $, with the reference metric $\hat{\gamma}_{ij}$:
\begin{align}
	\hat{\gamma}_{ij} &= \begin{bmatrix}
		1 & 0 & 0 \\
		0 &r^2 & 0 \\
		0 & 0 &r^2 \sin^2\theta  \\
       	\end{bmatrix}.
\end{align}
The associated 3-Christoffel symbols $\left< \hat{\Gamma}^l_{ik}\right>$ are:
\begin{align}
	\hat{\Gamma}^r_{ij} &= \begin{bmatrix}
		0 & 0 & 0 \\
		0 &-r & 0 \\
		0 & 0 &-r \sin^2\theta  \\
       	\end{bmatrix}, \\
	\hat{\Gamma}^\theta_{ij} &= \begin{bmatrix}
		0 & \frac{1}{r} & 0 \\
		\frac{1}{r} & 0 & 0 \\
		0 & 0 & -\sin\theta\cos\theta  \\
       	\end{bmatrix}, \\
	\hat{\Gamma}^\phi_{ij} &= \begin{bmatrix}
		0 & 0 & \frac{1}{r}\\
		0 & 0 & \cot\theta \\
		\frac{1}{r} & \cot\theta & 0 \\
       	\end{bmatrix}.
\end{align}
The geometrical source terms for the momentum equations are:
\begin{align}
\hat{\Gamma}^l_{rk}(f_{S_l})^k =& \hat{\Gamma}^\theta_{r\theta}(f_{S_\theta})^\theta + \hat{\Gamma}^\phi_{r\phi}(f_{S_\phi})^\phi ,\\
\hat{\Gamma}^l_{\theta k }(f_{S_l})^k =& \hat{\Gamma}^r_{\theta \theta}(f_{S_r})^\theta + \hat{\Gamma}^\theta_{\theta r}(f_{S_\theta})^r + \hat{\Gamma}^\phi_{\theta \phi}(f_{S_\phi})^\phi ,\\ 
\hat{\Gamma}^l_{\phi k}(f_{S_l})^k =& 0.
\end{align}
Similarity, as in the cylindrical case, the angular momentum $q_{S_\phi}$ are conserved to machine precision.

To work out the cell volume $\Delta V$, cell surface $\Delta A$ and the volume-average of the 3-Christoffel symbols $\left< \hat{\Gamma}^l_{ik}\right>$, we define the following notations: 
\begin{align}
&r_\pm = r \pm \frac{1}{2}\Delta r ,&
&\theta_\pm = \theta \pm \frac{1}{2}\Delta \theta ,&
&\phi_\pm = \phi \pm \frac{1}{2}\Delta \phi ,&
\end{align}
where $(r,\theta,\phi)$ are the location at the cell centre at some particular point while $(\Delta r,\Delta \theta,\Delta \phi)$ are the corresponding grid sizes.
The cell surface $\Delta A$ and the cell volume $\Delta V$ can then be expressed as:
\begin{align}
\Delta A_r\Big|_{r_\pm} &= \left( r \pm \frac{\Delta r}{2} \right)^2 \left( 2 \sin \theta \sin\left(\frac{\Delta \theta}{2}\right) \right) \left( \Delta \phi \right) \\
\Delta A_\theta \Big|_{\theta_\pm} &=  \left( \left(r^2 + \frac{1}{12}\left(\Delta r\right)^2\right)\Delta r \right) \left( \sin \left( \theta \pm \frac{\Delta \theta}{2} \right) \right) \left( \Delta \phi \right) \\
\Delta A_\phi \Big|_{\phi_\pm} &=  \left( \left(r^2 + \frac{1}{12}\left(\Delta r\right)^2\right)\Delta r \right) \left( 2 \sin \theta \sin\left(\frac{\Delta \theta}{2}\right) \right)\\
\Delta V &= \left( \left(r^2 + \frac{1}{12}\left(\Delta r\right)^2\right)\Delta r \right) \left( 2 \sin \theta \sin\left(\frac{\Delta \theta}{2}\right) \right) \left( \Delta \phi \right)
\end{align}
Finally, the non-vanishing volume-averaged 3-Christoffel symbols $\left< \hat{\Gamma}^l_{ik}\right>$ are:
\begin{align}
&\left<\hat{\Gamma}^{\theta}_{r \theta} \right> = \left<\hat{\Gamma}^{\theta}_{ \theta r} \right>  = \left<\hat{\Gamma}^{\phi}_{r \phi} \right> = \left<\hat{\Gamma}^{\phi}_{\phi r} \right>
 = \frac{1}{\Delta V} \frac{1}{2} \left( \Delta A_r\Big|_{r_+} - \Delta A_r\Big|_{r_-} \right) \\
&\left<\hat{\Gamma}^{r}_{\theta \theta} \right> =  -\frac{1}{\Delta V} \frac{1}{4} \left( r_+^2\Delta A_r\Big|_{r_+} - r_-^2\Delta A_r\Big|_{r_-} \right) \\
&\left<\hat{\Gamma}^{r}_{\phi \phi} \right> = \frac{1}{\Delta V} \left( -\frac{1}{4} r^4 \Big|_{r_-}^{r^+} \right) \left(\frac{1}{3}\cos^3\theta - \cos\theta\right)\Big|_{\theta_-}^{\theta^+} \Delta \phi \\
&\left<\hat{\Gamma}^{\theta}_{\phi \phi} \right> = - \frac{1}{3} \frac{1}{\Delta V} \left( \sin^2 (\theta_+) \Delta A_\theta\Big|_{\theta_+} - \sin^2 (\theta_-) \Delta A_\theta \Big|_{\theta_-} \right) \\
&\left<\hat{\Gamma}^{\phi}_{\theta \phi} \right> = \left<\hat{\Gamma}^{\phi}_{\phi \theta } \right> = \cot \theta
\end{align}


\bsp	
\label{lastpage}
\end{document}